\documentclass[review,12pt]{elsarticle}

\usepackage{lineno,hyperref}
\modulolinenumbers[5]

\usepackage{graphicx}
\usepackage{amsmath,amssymb}
\usepackage{amsthm}
\usepackage{xcolor}
\usepackage{mathrsfs}
\usepackage{paralist}
\usepackage{multirow}
\usepackage[T1]{fontenc}
\usepackage[utf8]{inputenc}
\usepackage[english]{babel}
\usepackage{subfigure}
\DeclareMathAlphabet\mathbfcal{OMS}{cmsy}{b}{n}
\usepackage{caption}
\usepackage{lineno}
\usepackage{bbm}
\usepackage[normalem]{ulem}
\usepackage[percent]{overpic}
\usepackage{cleveref}
\usepackage{color}
\usepackage{tikz}
\usepackage{yhmath}
\usepackage{hyperref}

\newcommand{\ud}{\mathrm{d}}

\newcommand{\pd}[2]{\frac{\partial{#1}}{\partial{#2}}} 
\newcommand{\der}[2]{\frac{\ud{#1}}{\ud{#2}}} 

\usepackage[a4paper,bindingoffset=0.0in,%
            left=1.0in,right=1.0in,top=1.in,bottom=1.in,%
            footskip=.4in]{geometry}

\begin{document}

\begin{frontmatter}

\title{Generalisation of the Spectral Difference scheme for the diffused-interface five equation model}

\author[mymainaddress]{Niccol{\`o} Tonicello\corref{mycorrespondingauthor}}
\cortext[mycorrespondingauthor]{Corresponding author}
\ead{ntonicel@sissa.it}
\author[secondaddress]{Guido Lodato}
\author[thirdaddress]{Matthias Ihme}

\address[mymainaddress]{Scuola Internazionale Superiore di Studi Avanzati (SISSA), Italy}
\address[secondaddress]{Normandie Universit{\'e}, INSA et Universit{\'e} de Rouen, St.~Etienne du Rouvray, France}
\address[thirdaddress]{Department of Mechanical Engineering, Stanford University, Stanford, United States}

\begin{abstract}
The present work focuses on the generalisation of the Spectral Difference (SD) scheme to the reduced Baer-Nunziato system known as five-equation model for the simulation of two immiscible compressible fluids. This five equation model is considered with the additional Allen-Cahn regularisation to avoid both over-diffusion and over-thinning of the phase field representing the interface. Finally, in order to preserve contact discontinuities, in the reconstruction step of the spectral difference scheme, a change of variables from conservative to primitive is used. This approach is shown to be beneficial in avoiding pressure oscillations at material interfaces. An extensive series of numerical tests are proposed to assess accuracy and robustness of the present method. Both kinematic (Rider-Kothe vortex) and two-phase flow problems (Rayleigh-Taylor instability, shock-droplet interaction, Taylor-Green vortex) are considered.
\end{abstract}


\end{frontmatter}

\section{Introduction}
High-fidelity simulations of multi-phase flows represent a relevant tool in the design process of many different fields of engineering. The sudden change of density, viscosity and thermodynamic properties at the interface between two immiscible fluids poses a great challenge in terms of numerical modeling. The development of reliable, accurate and robust discretisation techniques to deal with these types of problems consequently represents an unavoidable step to handle simulations of challenging scenarios as the ones commonly encountered in the aeronautical industry.

Within the framework of interface capturing techniques, the main approaches to deal with interfaces between two phases can be categorised in sharp and diffused interface techniques. In the former case, the interface is modelled as an infinitesimally thin layer separating the two phases. Among the many methods falling into this category, the Volume of Fluid (VOF) ~\cite{debar1974fundamentals,nichols1975methods} and level-set method~\cite{osher1988fronts,sussman1994level} are the most common. A different approach, instead, relies on a diffused interfaces where the transition from one phase to the other takes place within a thin interface of finite width. Relevant examples of these models are the Cahn-Hilliard~\cite{cahn1958free} and Allen-Cahn~\cite{allen1976mechanisms} models, which can be included in what are commonly called \emph{phase-field} methods. 
This work will focus on the latter approach. In particular, the conservative Allen-Cahn equations by Chiu et Lin~\cite{chiu2011conservative} are considered in this work for the simulation of compressible two-phase flows.

Once the mathematical model is properly defined, the discretisation technique can be derived. Over the last decades, a significant interest in high-order discretisation techniques spreads in the computational fluid dynamics community. A substantial amount of research has been dedicated to the development of such schemes for increasingly complex applications. In particular, spectral element methods such as the Discontinuous Galerkin method~\cite{hesthaven:book,cockburn:98,cockburn:98b,CANTWELL2015205}, the Flux Reconstruction scheme~\cite{huynh2007flux,vincent2011new,witherden2014pyfr}  and the Spectral Difference method~\cite{kopriva1996conservative,karniadakis2013spectral} have shown great promise in the simulation of complex fluid dynamics simulations~\cite{lv2014discontinuous,mengaldo2021industry,tonicello2022turbulence,ching2019shock,lv2021discontinuous,dzanic2022positivity,ferrer2023high}. The Discontinuous Galerkin scheme has been recently applied to different types of diffused interface models for the simulation of two-phase flows~\cite{tonicello2023high,ntoukas2022entropy,manzanero2020entropy,cheng2020quasi,orlando2023implicit}, as the Flux Reconstruction method~\cite{al2021high}. The same schemes are also growing into the multi-component flow simulations field such as combustion~\cite{lv2014discontinuous,lv2017high,bando2020comparison,du2022high,marchal2023extension} (see~\cite{lv2023recent} for a recent review on high-order schemes for multi-phase and multi-component flows).

With respect to the use of spectral element high-order schemes for single-phase simulation, the addition of multiple phases greatly influences how the numerical method should be designed, which type of conservation or structures it should preserve. One particularly challenging task in compressible multi-phase flows is avoiding pressure oscillations close to material interfaces~\cite{abgrall1996prevent}. These oscillations are caused by the sudden change of thermodynamic parameters (e.g. polytropic coefficient) and  in the two fluids, leading, from a mathematical point of view, to an inherently non-linear relation between pressure and internal energy. Similar problems are, in fact, encountered also in the simulation of multicomponent flows~\cite{karni1994multicomponent,billet2003adaptive,lv2014discontinuous}.
In this work, the spectral difference scheme is modified in order to deal with such issues. In particular, in order to avoid the interpolation of sharply varying quantities across the interface such as internal energy and momentum, pressure and velocity are instead used. In the relaxed Baer-Nunziato system, which is employed in this work (the five equation model~\cite{allaire2002five}), these quantities are in fact varying smoothly across the interface. After the primitive variables are interpolated to compute the flux at the flux points, it is possible to pass back to the conservative variables and compute the flux's divergence. This additional step can significantly improve the fulfilment of the interface equilibrium condition with, at the same time, a considerably small modifications of the main core of the single-phase implementation.

The present works aims at generalising the spectral difference scheme to deal with two-phase flows. In particular, the way the scheme handles the flux reconstruction is accordingly modified in order to deal with sharp variations of variables across the interface.

The paper will be structured as follows. Section \ref{sec:5eq} will introduce the the five equation model used to simulate compressible two-phase flows. Section \ref{sec:SD} will briefly introduce the spectral difference scheme for a general conservation law and subsequently show how to generalise the SD scheme for the five equation model. Finally, section \ref{sec:results} will be dedicated to numerical tests of the proposed approach. Numerical tests will be divided in kinematic and two-phase tests. In the former, the velocity field is imposed and the accuracy of the interface capturing technique is assessed. In the latter, instead, the velocity field itself is solved for.
\section{Five equation model} \label{sec:5eq}
The five equation model first proposed by Allaire et al.~\cite{allaire2002five} is a commonly used model for simulating compressible two-phase flows. The present formulation differs from the original model by considering an additional equation for an advected function $\psi$, which is used to compute the interface normal vectors similar to the work by Al-Salami et al.~\cite{al2021high}. This model was also considered in the context of a high-order Discontinuous Galerkin discretisation~\cite{tonicello2023high}.

The five equation model, including the contributions of viscous and gravitational forces reads:
\begin{align}
 \frac{\partial \psi}{\partial t} + \textbf{u} \cdot \nabla \psi &= 0, \label{eq:5eq_finalPC2}\\
 \frac{\partial \phi_{1}}{\partial t} + \textbf{u} \cdot \nabla \phi_{1} &= \nabla \cdot \textbf{a}_{1}, \\
 \frac{\partial (\rho_{1} \phi_{1})}{\partial t} + \nabla \cdot (\rho_{1} \phi_{1} \textbf{u}) &= \nabla \cdot \textbf{R}_{1}, \\
 \frac{\partial (\rho_{2} \phi_{2})}{\partial t} + \nabla \cdot (\rho_{2} \phi_{2} \textbf{u}) &= \nabla \cdot \textbf{R}_{2}, \\
 \frac{\partial (\rho \textbf{u})}{\partial t} + \nabla \cdot (\rho \textbf{u}  \otimes \textbf{u} + p \mathbb{I}) &= \nabla \cdot (\textbf{f} \otimes \textbf{u}) + \nabla \cdot \boldsymbol{\tau} + \rho \textbf{g}, \\
 \frac{\partial (\rho E)}{\partial t} + \nabla \cdot [(\rho E + p) \textbf{u}] &= \nabla \cdot (\textbf{f} k) + \sum_{l=1}^{2} \nabla \cdot ( \rho_{l} H_{l} \textbf{a}_{l}) + \nabla \cdot (\boldsymbol{\tau} \cdot \textbf{u}) + \rho \textbf{g} \cdot \textbf{u},
\label{eq:5eq2}
\end{align}
where $\phi$ is the phase field, $\rho$ is the density, $\psi$ is the additional level-set function, $\rho \textbf{u}$ is the total momentum, $p$ is the pressure, $\rho E = \rho e + \frac{1}{2}\rho \|\textbf{u}\|^{2}$ is the total energy (as a sum of internal and kinetic energy) and $\textbf{g}$ is the gravitational acceleration. The subscript $l$ denotes the $l^{\mathrm{th}}$ phase.

In addition,
\begin{align}
& \textbf{a}_{l} = \Gamma[ \epsilon \nabla \phi_{l} - \phi_{l} (1- \phi_{l})\widehat{\textbf{n}}_{l}], \quad \widehat{\textbf{n}}_{l} = \nabla \psi/\|\nabla \psi\|, \\
&\textbf{R}_{l} = \rho_{l}^{(0)} \textbf{a}_{l}, \\
&\textbf{f} = \sum_{l=1}^{2} \textbf{R}_{l}, \quad k=\frac{1}{2} \|\textbf{u}\|^{2},
\end{align}
and $H_{l}$ is the specific enthalpy of the $l$-th phase. The parameters of the Allen-Cahn terms, $\epsilon$ and $\Gamma$, represent the thickness of the interface and the magnitude with respect to advection, respectively. Consequently, $\epsilon$ is commonly chosen to be proportional to the grid size and $\Gamma$ to be proportional to the maximum velocity magnitude in the fluid domain. The viscous stress reads: $\boldsymbol{\tau}= 2 \mu (\mathbb{S} - 1/3(\nabla \cdot \textbf{u}) \mathbb{I})$ with $\mu$ the dynamic viscosity of the mixture evaluated as $\mu = \mu_{1} \phi_{1} + \mu_{2} \phi_{2}$, $\mathbb{S} = [\nabla \textbf{u} + (\nabla \textbf{u})^{\intercal}]/2$ is the strain-rate tensor.
The system is then closed by relating internal energy with the pressure field using an equation of state (EOS). A classical choice is the stiffened-gas EOS~\cite{harlow1971fluid}:
\begin{equation}
p = \frac{\rho e - \bigg ( \frac{ \gamma_{1} p^{\infty}_{1}}{\gamma_{1}-1} \phi_{1} + \frac{ \gamma_{2} p^{\infty}_{2}}{\gamma_{2}-1} \phi_{2} \bigg)}{ \bigg ( \frac{\phi_{1}}{\gamma_{1}-1} +   \frac{\phi_{2}}{\gamma_{2}-1} \bigg)},
\end{equation}
where $\gamma_{l}$ and $p^{\infty}_{l}$ are the parameters of the EOS. 
From the stiffened-gas equation of state it is possible to write the speed of sound and specific enthalpy of each phases as
\begin{equation}
c_{l} = \sqrt{\gamma_{l}\bigg( \frac{p + p^{\infty}_{l}}{\rho_{l}} \bigg)} \quad \mathrm{and} \quad H_{l} = \frac{(p + p^{\infty}_{l})\gamma_{l}}{\rho_{l} (\gamma_{l} -1)} \quad \mathrm{for} \quad l =1,2.
\end{equation}
Finally, for completeness, the following mixture relations apply:
\begin{align}
\phi_{2} = & 1- \phi_{1},   \\
\rho = & \rho_{1}\phi_{1} + \rho_{2} \phi_{2},  \\
\frac{1}{\gamma-1} = & \phi_{1} \frac{1}{\gamma_{1}-1} + \phi_{2} \frac{1}{\gamma_{2}-1}, \\
p^{\infty} \frac{\gamma}{\gamma-1} =  & \phi_{1} \frac{\gamma_{1} p_{1}^{\infty}}{\gamma_{1}-1} + \phi_{2} \frac{\gamma_{2} p_{2}^{\infty}}{\gamma_{2}-1}.
\end{align}
%
\section{The spectral difference scheme} \label{sec:SD}
In this section the Spectral Difference scheme is introduced for a general conservation law. The application of this approach to the five equation model will naturally arise from the discussion. Finally, we will present the additional modification needed to allow the SD to preserve the interface equilibrium condition (IEC). 

\begin{figure}[t]
\centering
\includegraphics[width=0.6\textwidth]{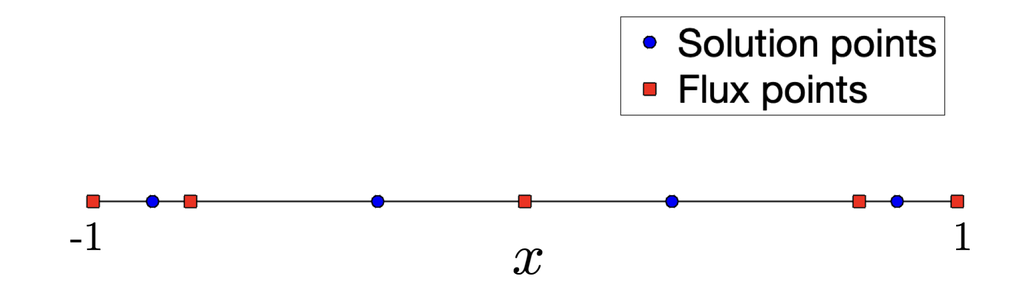}
\caption{Solution and flux points of SD discretisation in the $[-1,1]$ element for a $4^{\mathrm{th}}$-order approximation.}
\label{nodes}
\end{figure}

The SD method~\cite{kopriva1996conservative,liu:06a,wang:07}  solves the strong form of the differential equation using element-wise continuous functions as approximation space.
Consequently, the solution is assumed to be discontinuous at element interfaces.
In order to have a consistent discretisation, the solution is interpolated using a polynomial of degree $k$ while the flux, which is connected to the conservative variables via a divergence operator, is approximated with a polynomial of degree $k+1$.
The most important ingredient of the SD discretisation is the definition of two different sets of points that are associated to solution and flux points. 
The numerical solution is defined on the nodes $x_i^s$ with $i$=0 to $K$.
Fluxes, instead, are defined on a different set of nodes $x_{i}^f$,  with $i$=0 to $K+1$, among which boundary points are included. 
In the present study, the solution points are set as the Gauss-Legendre points of order $K+1$, a sensible choice to minimize aliasing errors in the nonlinear case while defining a well conditioned basis set for the solution interpolation~\cite{jameson:12}, whereas the flux points are set as the Gauss-Legendre points of order $K$ plus the two end points -1 and 1 to ensure linear stability~\cite{jameson:10}.
An example of solution and flux points for a $4^{\mathrm{th}}$-order approximation is shown in figure~\ref{nodes}.
Let us consider a general one dimensional, scalar conservation law
\begin{equation}
\frac{\partial w}{\partial t} + \frac{\partial f(w)}{\partial x}.
\end{equation}
The solution is approximated with a polynomial of degree $K$:
\begin{equation}
\hat{w}(\hat{x})=\sum_{i=0}^{K}w_{i}l^{s}_{i}(\hat{x}).
\label{solSD}
\end{equation}
where $l^{s}_{i}(\hat{x})$ are lagrange polynomials defined on the point $\hat{x}^{s}_{i}$.
Subsequently, the values of the solution are extrapolated at the flux points
\begin{equation}
\hat{w}(\hat{x}^{f}_{j})=\sum_{i=0}^{K}w_{i}l^{s}_{i}(\hat{x}^{f}_{j}), \qquad j=0,...,K+1,
\end{equation}
and then used to compute fluxes on the same collocation basis:
\begin{equation}
f_{j}= \hat{f}(\hat{x}^f_{j})=\hat{f}(\hat{w}(\hat{x}^{f}_{j})).
\end{equation}
Then, a continuous flux polynomial of degree $K+1$ is constructed, by Lagrange interpolation, using the fluxes evaluated from the interpolated solution at the interior flux points and the numerical fluxes at the element interfaces (denoted as $\hat{f}^{I}_{L/R}$):
\begin{equation}
\hat{f}(\hat{x})= \hat{f}^{I}_{L}l^f_{0}(\hat{x}) +\sum_{j=1}^{k}f_{j}l^f_{j}(\hat{x}) +\hat{f}^{I}_{R}l^f_{k+1}(\hat{x}).
\end{equation}
In other words, the interpolated values of the flux at elements extrema are substituted by the interface numerical fluxes $\hat{f}^{I}_{L}$ and $\hat{f}^{I}_{R}$.
The flux divergence is then evaluated at the solution points,
\begin{equation}
\der{\hat{f}}{\hat{x}}(\hat{x}^s_{i})= \hat{f}^{I}_{L}\der{l^f_{0}}{\hat{x}}(\hat{x}^{s}_{i}) +\sum_{j=1}^{k}f_{j}\der{l^f_{j}}{\hat{x}}(\hat{x}^{s}_{i}) +\hat{f}^{I}_{R}\der{l^f_{k+1}}{\hat{x}}(\hat{x}^{s}_{i}),
\label{finalSD}
\end{equation}
and, finally, the numerical solution can be advanced in time using a suitable time integration scheme discretising the following equation:
\begin{equation}
\frac{\ud{\hat{w}}}{\textrm{d} t}=-\der{\hat{f}}{\hat{x}}(\hat{x}^s_{i}).
\label{fullSD}
\end{equation}

Now that the spectral difference scheme was introduced on a general one dimensional, scalar conservation law, let us consider the one-dimensional form of the five equation model that will be used in this work:

\begin{equation}
\frac{\partial \textbf{w}}{\partial t} + \frac{\partial (\textbf{F}(\textbf{w}))}{\partial x} = \textbf{S}(\textbf{w})
\end{equation}

with 

\begin{equation}
\textbf{w}=
\begin{pmatrix}
\phi \\
\psi \\
\rho_{1} \phi_{1} \\
\rho_{2} \phi_{2} \\
\rho u  \\
\rho E \\
\end{pmatrix},
\quad 
\textbf{F} =
\begin{pmatrix}
a_{1} \\
0 \\
\rho_{1} \phi_{1} u + R_{1} \\
\rho_{2} \phi_{2} u + R_{2} \\
\rho u^{2}+p + f u  \\
\rho (E +p)u + f k \\
\end{pmatrix},
\quad
\mathrm{and}
\quad 
\textbf{S} =
\begin{pmatrix}
-u \frac{\partial \phi}{\partial x} \\
-u \frac{\partial \psi}{\partial x}  \\
0\\
0 \\
0 \\
0 \\
\end{pmatrix}.
\end{equation}

including the Allen-Cahn regularisation terms.

Based on the physical assumption in the mathematical model itself, variables will either vary smoothly across the interface or follow a hyperbolic tangent profile as the one characterizing the interface. If we consider the simulation of a single droplet in the previously written conservation law phase field and partial densities will rapidly vary in the interface region, whereas pressure and velocity should be exactly constant. A desirable property of numerical schemes for the simulation of compressible two-phase flows is to intrinsically satisfy such conditions. This is commonly referred to as IEC or Abgrall condition\cite{abgrall1996prevent}. As an example, centered finite differences equipped with proper corrective terms in the differential equations of the full system can preserve this property \cite{jain2020conservative}, writing the system in a quasi-conservative form \cite{abgrall1996prevent,cheng2020quasi} or even considering specific Riemann solvers for each equation of the system \cite{tonicello2023high} are shown to obey IEC.

Due to the specific way fluxes are computed in the spectral difference scheme, we propose an additional interpolation step in the overall strategy. Instead of interpolating the conservative variables, which vary abruptly across the interface (making them prone to numerical errors in the interpolation step), the primitive variables that are smooth across the interface are used. Specifically, instead of using the vector $\textbf{w}$ we interpolate the primitive variables $\textbf{v}=(\phi, \psi, \rho_{1}\phi_{1},  \rho_{2}\phi_{2}, u, p)^{T}$. These, we evaluate the discrete fluxes as
\begin{equation}
\textbf{F}(\hat{x}^{f}_{j}) =\textbf{F}(\textbf{v}(\hat{x}^{f}_{j}))=\textbf{F} \bigg (\sum_{i=0}^{k}\textbf{v}_{i}l^{s}_{i}(\hat{x}^{f}_{j}) \bigg)
\end{equation}
with $\textbf{v}=\mathcal{G}(\textbf{w})$ the transformation from conservative to primitive variables. The remaining building blocks of the spectral difference scheme are exactly the same. This modification is consequently very localised and can be applied in many different nodal spectral element methods without the need of substantial change in the solver. It is intuitive to believe that such change of variables in the interpolation step should mitigate spurious oscillations in proximity of a material interfaces where pressure and velocity are supposed to be exactly constant.

The remaining implementational details regarding the discretisation follow similar concepts to the work by Tonicello and Ihme~\cite{tonicello2023high}. The numerical fluxes are based on the Lax–Friedrichs Riemann solver, the transport terms in the phase-field and level-set equations are treated as source terms and the sharpening/diffusive terms are discretised using a Local Discontinuous Galerkin (LDG) approach~\cite{cockburn:98}. The re-initialization of the level-set function is based on the same equation as in~\cite{al2021high,tonicello2023high}. The diffusive term is based on the LDG flux and the Hamiltionian term is implemented as a source term. Further details can be found in~\cite{tonicello2023high}. Finally, it is worth highlighting that in~\cite{tonicello2023high} a special discretisation of the phase field was used in order to exactly preserve the interface equilibrium condition. In this work, the interpolation of the primitive variables is proposed as a less restricting strategy to avoid pressure oscillations in proximity of the interface. Even if a Lax–Friedrichs flux is considered in this work, the present formulation allows for the implementation of more elaborate Riemann solvers such as Roe and HLLC fluxes in a relatively straightforward manner, provided that suitable corrections are introduced in their original formulation to account for the particular EOS herein considered. 
\section{Numerical results} \label{sec:results}
To assess the method we consider a series of test cases. The numerical test cases herein proposed will be sorted in kinematic tests and two-phase flow tests. For the former case, the velocity field is represented by a given analytical function which will be used in the phase field and level-set equations only. In this way, it is possible to assess the performance of the interface capturing technique. The latter set of tests consider the full  system of equations, which are used to describe the flow of compressible two-phase flows. Notice that all the settings herein proposed are adimensional. Therefore, the units of the all the physical quanties have been omitted.
\subsection{Kinematic tests}
\subsubsection{Boundedness} 
One of the most important aspect of numerical schemes for advecting interfaces is the capability of obtaining bounded solutions of the phase field.
Therefore, we first consider the advection of a single droplet in $2$D and investigate numerically the parameter space $\Gamma$-$\epsilon$ in order to estimate the regions where the overall approach is bounded and where it is not. We limit ourselves to a fixed time-integration scheme (RK$4$), fixed number of steps before re-initialization ($1000$) and fixed amount of viscosity in the re-initialization equation ($\nu = 0.25 \epsilon/\Delta x$) with $\Delta x$ the grid size. In the framework of high-order methods, $\Delta x$ is commonly evaluated as the size of the element divided by the order of approximation.
At the same time, the parameters $\Gamma$ and $\epsilon$ are changed along with different orders of approximation to evaluate the boundedness of the scheme.

\begin{figure}[h!]
\centering
\includegraphics[width=.75\textwidth]{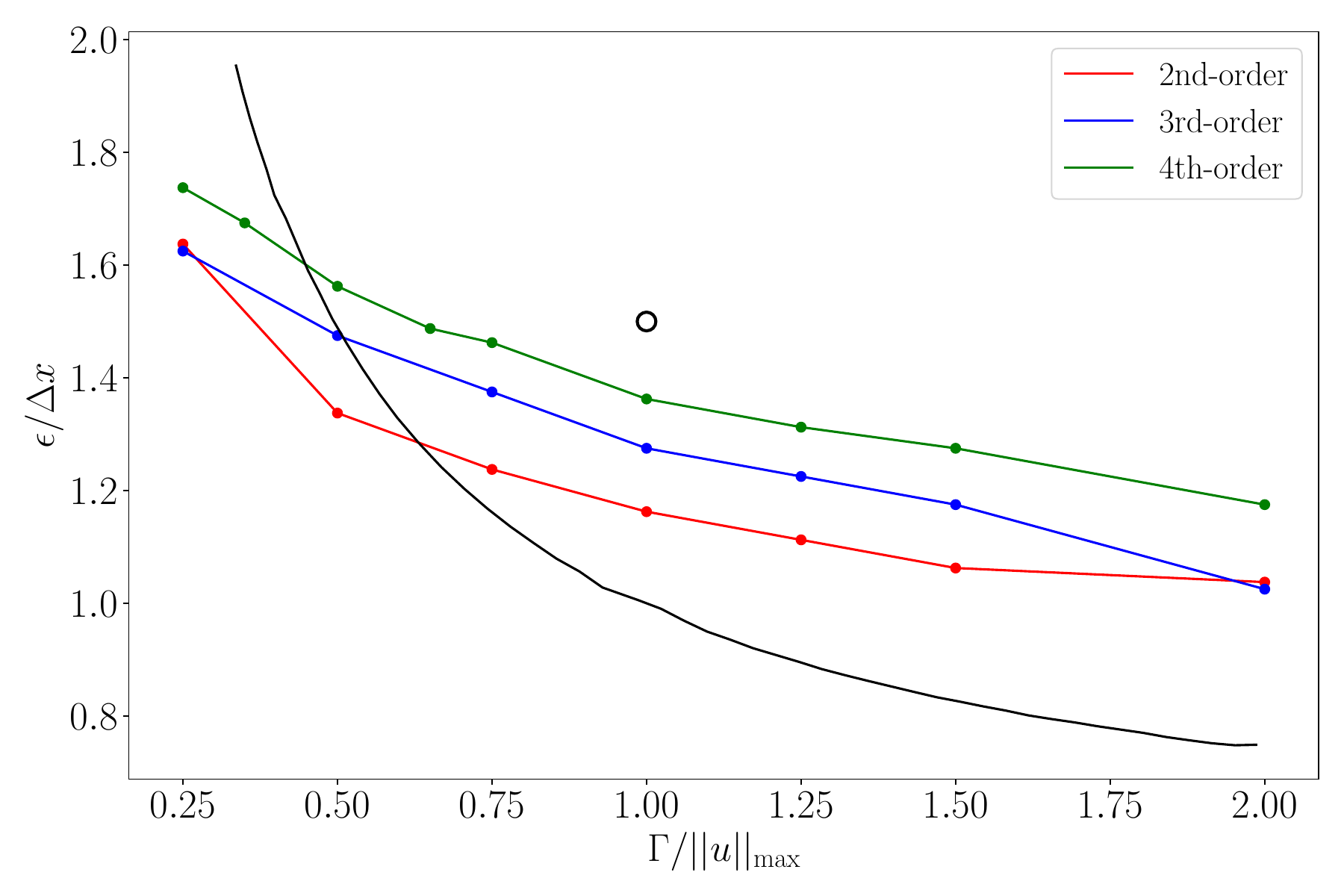}
\caption{Boundedness limits for different orders of approximation in the $\Gamma$-$\epsilon$ space. Red line, $3^{\mathrm{rd}}$-order; blue line, $4^{\mathrm{th}}$-order; $5^{\mathrm{th}}$-order. The black line denotes the boundedness limit presented by Mirjalili et al.~\cite{mirjalili2020conservative} for the same phase field equation. The single circle is located at $(\Gamma/\|u\|_{\mathrm{max}},\epsilon/\Delta x)=(1.0,1.6)$ which represents the standard value used in the rest of the paper (if not differently stated in the text).}
\label{fig:boundedness}
\end{figure}

In figure \ref{fig:boundedness} we show the approximated regions of boundedness of the proposed scheme. In particular, regions located above the different curves are characterised by bounded results, whereas below those lines, the phase field is, at times, unbounded.  We can observe that the relationship defining the boundedness limit is almost linear and, for $\Gamma/\|u\|_{\mathrm{max}}$ in the range $(0.25,2.00)$ the parameter $\epsilon/\Delta x$ varies approximately between $1.1$ and $1.7$. We can also observe that increasing the order of approximation further restricts the parameters giving bounded solutions. For example, at $\Gamma/\|u\|_{\mathrm{max}}=1$, for a $3^{\mathrm{rd}}$-order approximation, values of $\epsilon/\Delta x$ larger than $1.2$ leads to bounded solutions, whereas for $5^{\mathrm{th}}$-order discretisation, this value increases to $1.4$.
The single circle located at $(\Gamma/\|u\|_{\mathrm{max}},\epsilon/\Delta x)=(1.0,1.6)$ represents the standard value used in the rest of the paper. For this value we never observed unbounded solutions of the phase field in any of the following test cases.
\subsubsection{Rider-Kothe vortex}
Another test case for assessing interface capturing techniques consists in the deformation of a circular bubble by a shear flow~\cite{rider1998reconstructing}. In particular, a $2$D droplet of radius $R=0.15$ centered at $(0.5,0.25)$ in a $[0,1] \times [-0.5,0.5]$ periodic domain is advected by the following divergence-free velocity field:
\begin{align}
u_{1} = &-\sin^{2} (\pi x_{1}) \sin [2 \pi (x_{2}+0.5)] \cos \bigg( \frac{\pi t}{T}\bigg) \\
u_{2} = &\sin ( 2 \pi x_{1}) \sin^{2} [\pi (x_{2}+0.5)] \cos \bigg( \frac{\pi t}{T}\bigg),
\end{align} 
where $T$ denotes the characteristic period of the shear flow. The classical value of $T=4$ is chosen for this particular case. Under the action of this velocity field, in the first half of the period, the initial circular bubble is strongly deformed into a thin filament. After velocity reversal, the filament is stretched back to the initial condition. 
Relevant analyses for this problem can be sought in the evaluation of both $L_{1}$ and mass conservation errors of the scheme. Since the overall system is written in a quasi-conservative form, even if the regularisation terms are conservative, mass conservation errors are expected from the discretisation of the transport term. 

\begin{figure}[h!]
\centering
\subfigure[$3^{\mathrm{rd}}$-order.]{\includegraphics[width=0.32\textwidth]{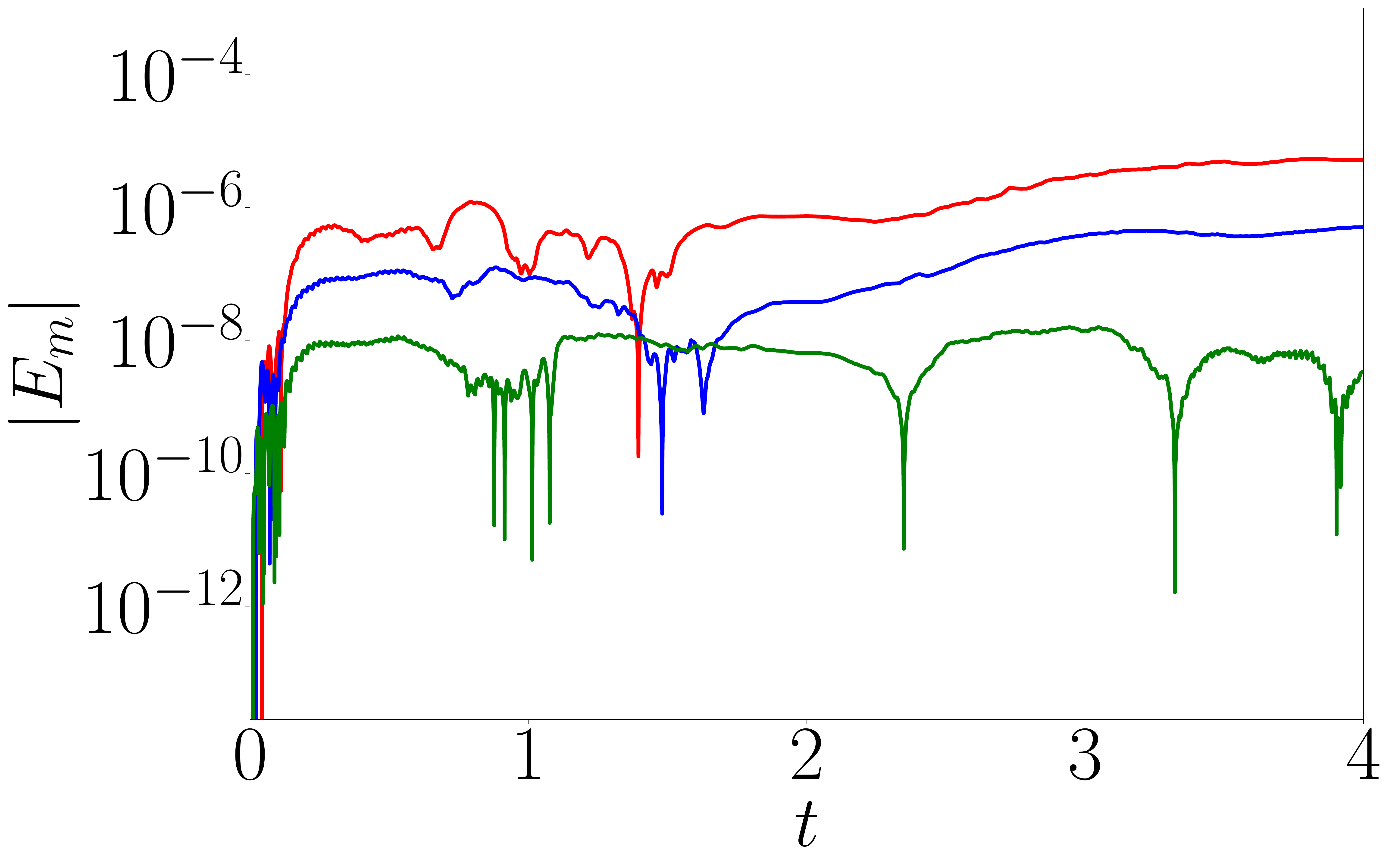}}
\subfigure[$4^{\mathrm{th}}$-order.]{\includegraphics[width=0.32\textwidth]{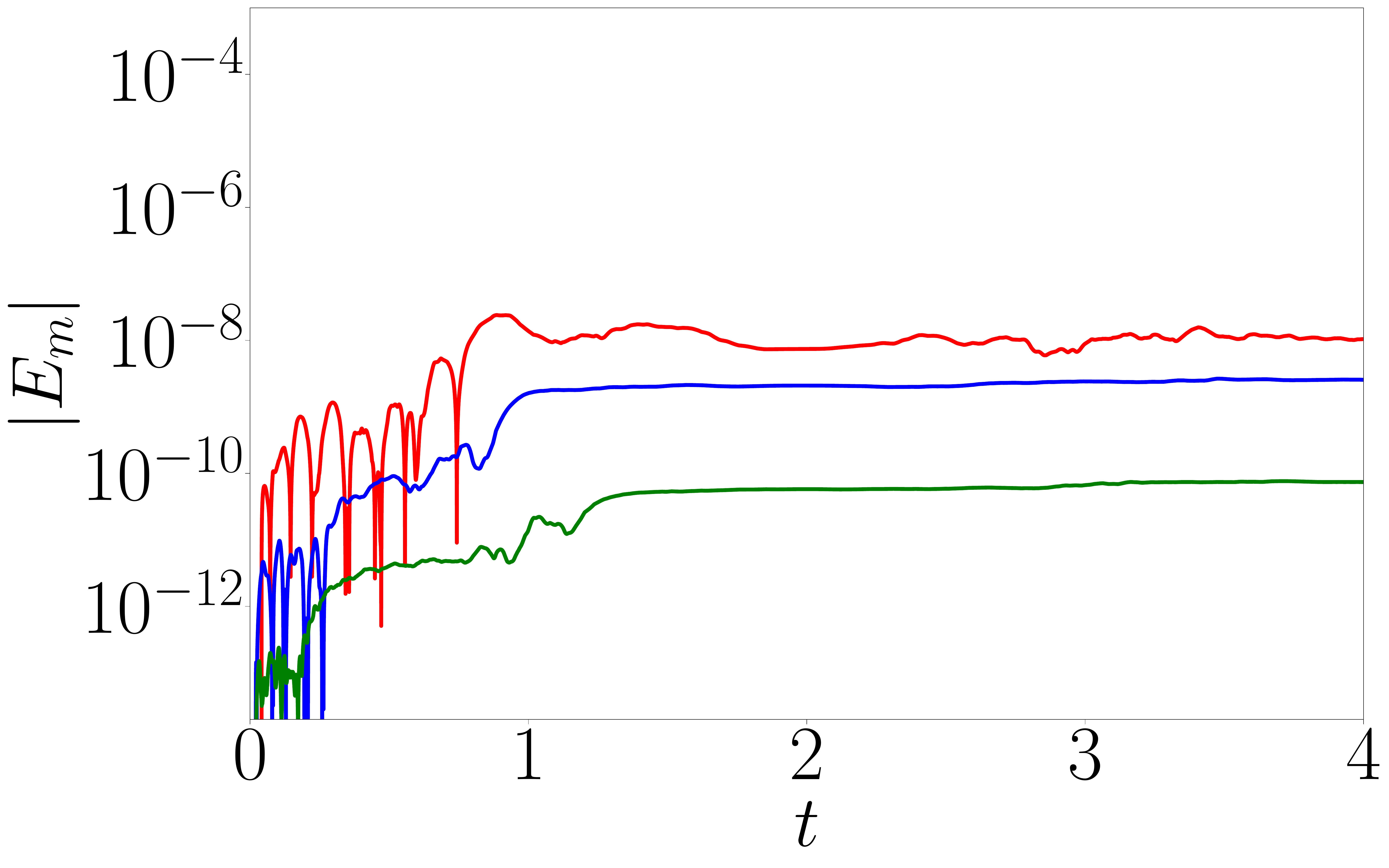}}
\subfigure[$5^{\mathrm{th}}$-order.]{\includegraphics[width=0.32\textwidth]{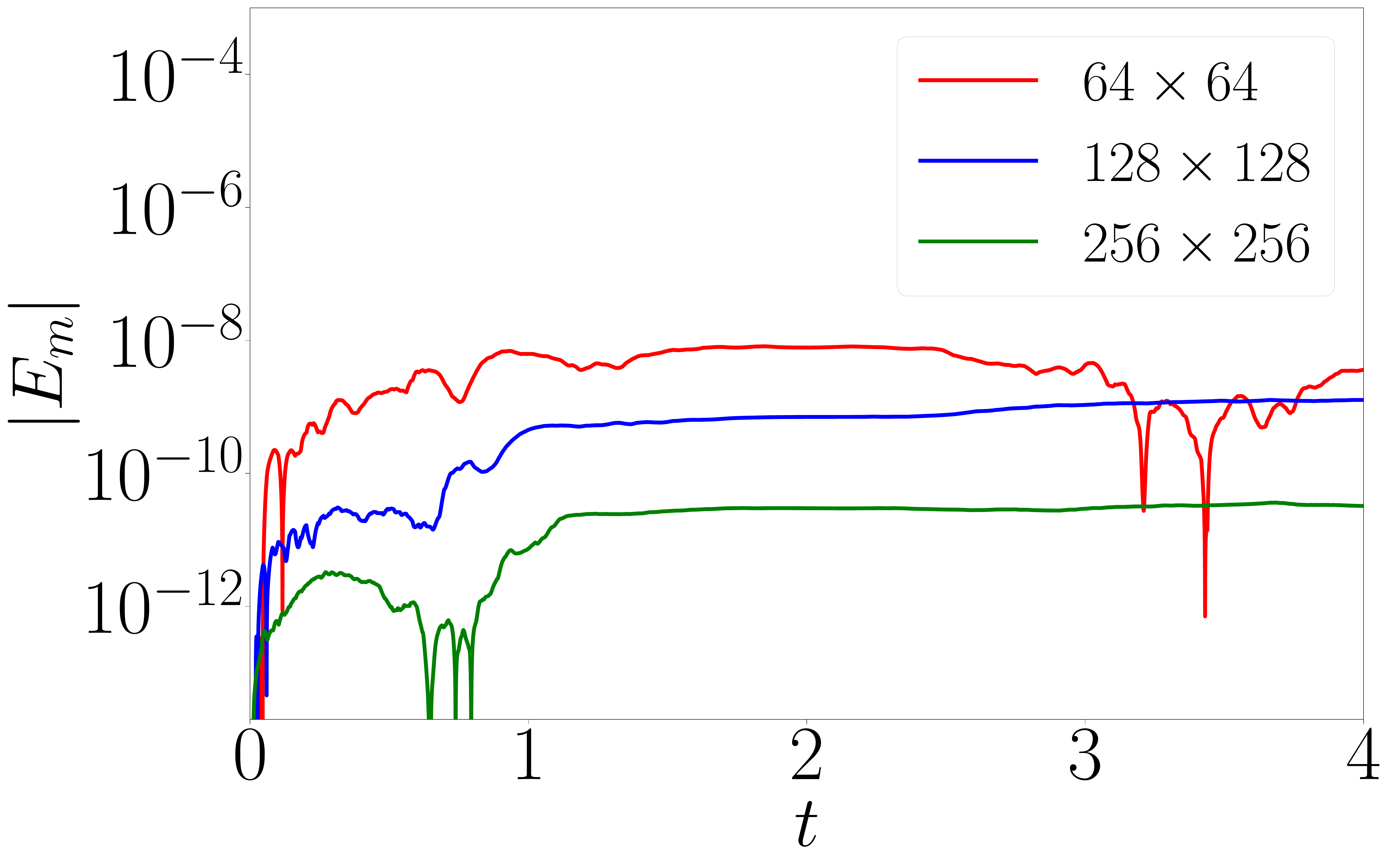}}
\caption{Mass conservation errors as a function of time for different order of approximation}
\label{fig:mass_order}
\end{figure}

In figure \ref{fig:mass_order} we are showing the mass conservation error, defined as
\begin{equation}
E_{m} = \frac{\int (\phi(\textbf{x},t) - \phi(\textbf{x},0)) d \Omega}{\int \phi(\textbf{x},0) d \Omega }.
\end{equation}
for different orders of approximation and different spatial resolutions. 

\begin{figure}[h!]
\centering
\subfigure[$64^{2}$ DoF.]{\includegraphics[width=0.32\textwidth]{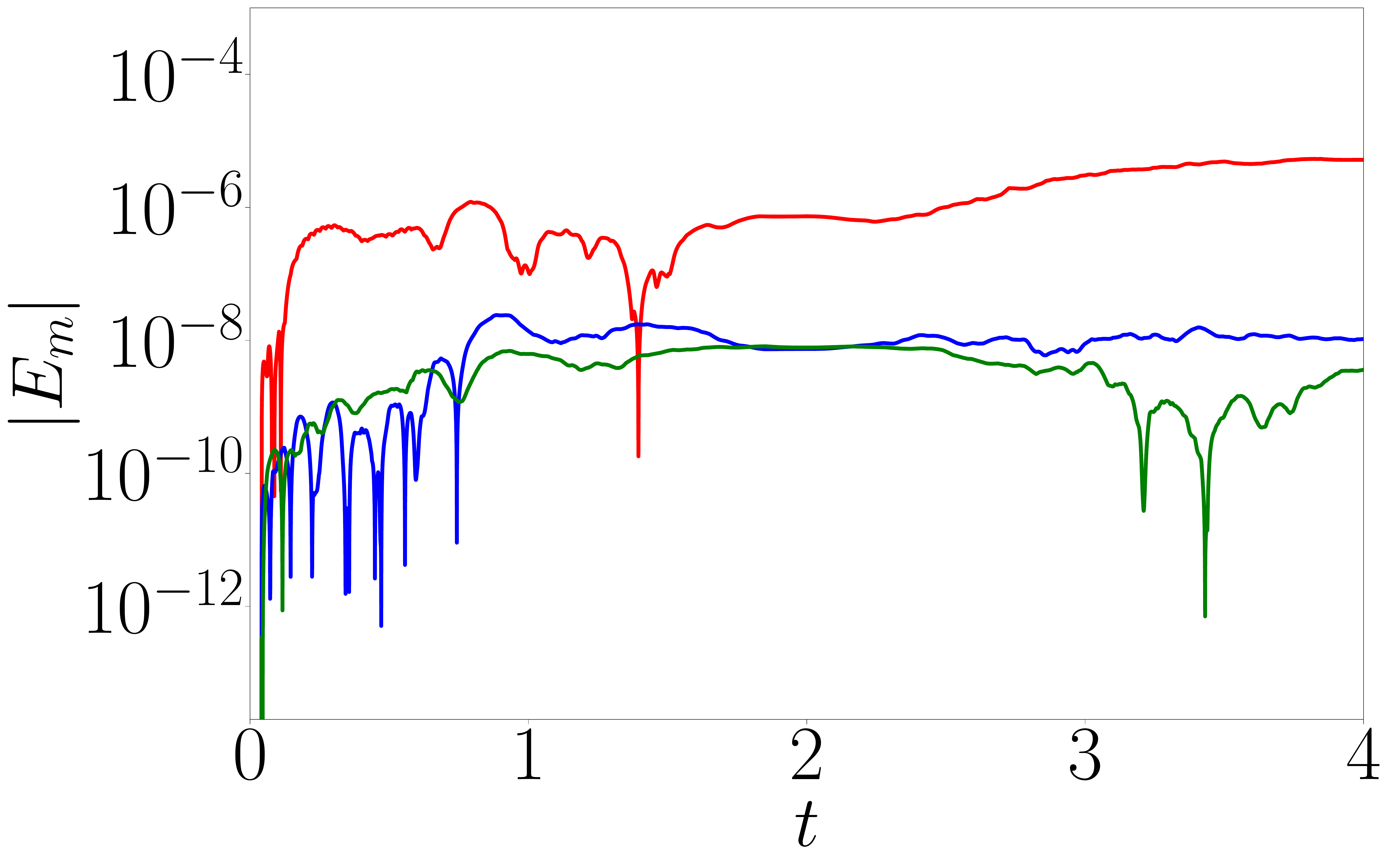}}
\subfigure[$128^{2}$ DoF.]{\includegraphics[width=0.32\textwidth]{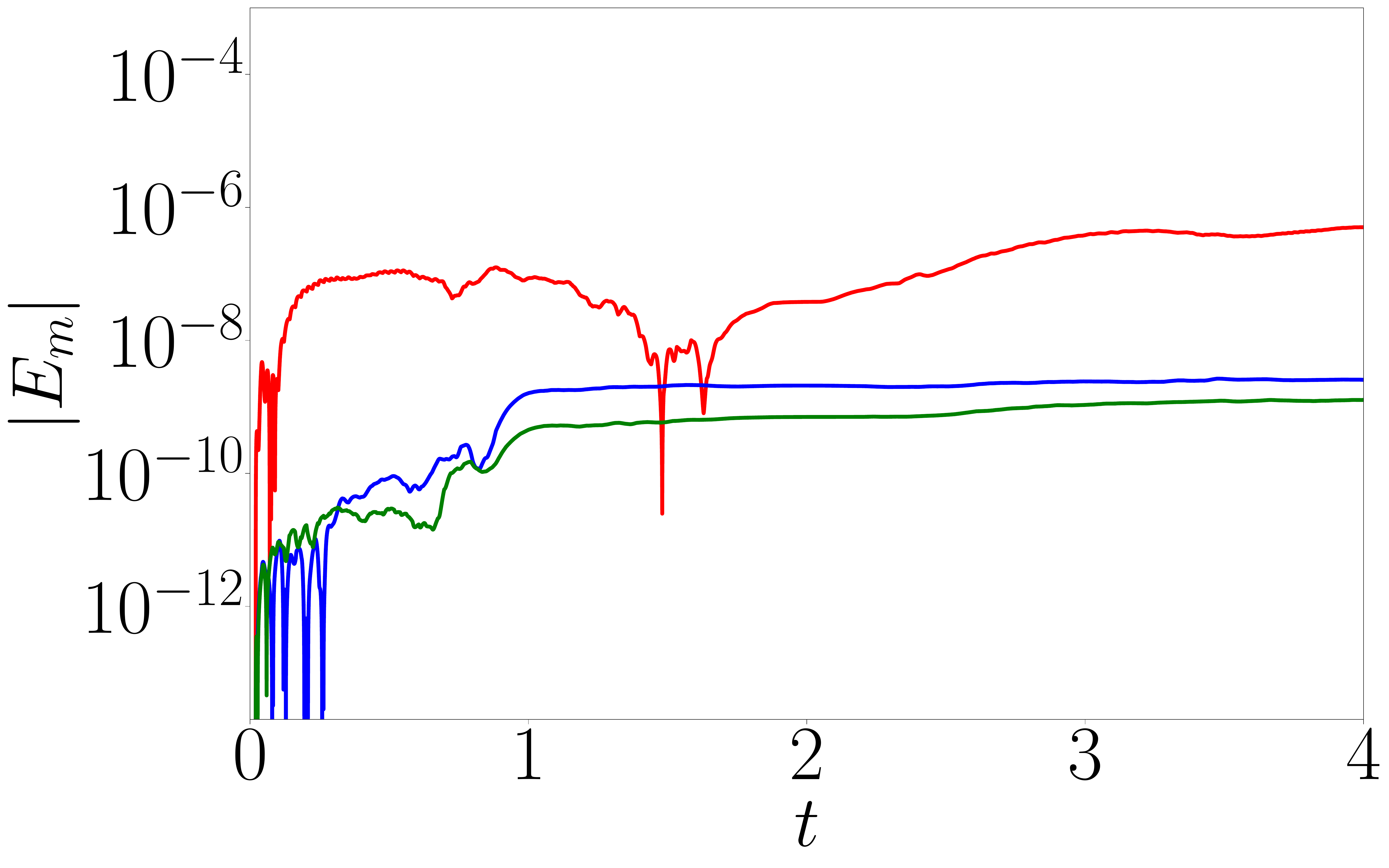}}
\subfigure[$256^{2}$ DoF.]{\includegraphics[width=0.32\textwidth]{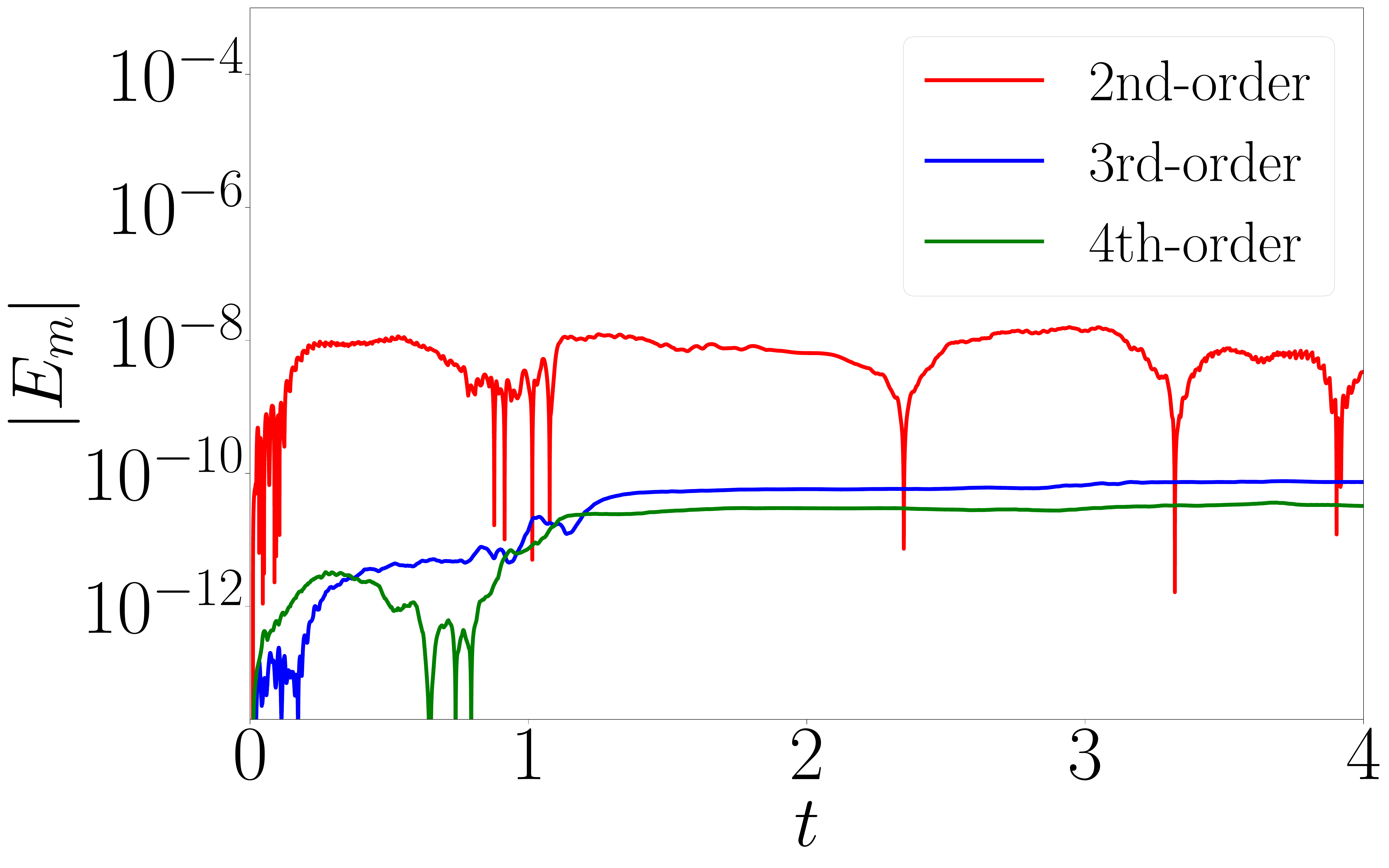}}
\caption{Mass conservation errors as a function of time for different grids.}
\label{fig:mass_res}
\end{figure}

\begin{figure}[htpb!]
\centering
\subfigure[]{\includegraphics[width=0.40\textwidth]{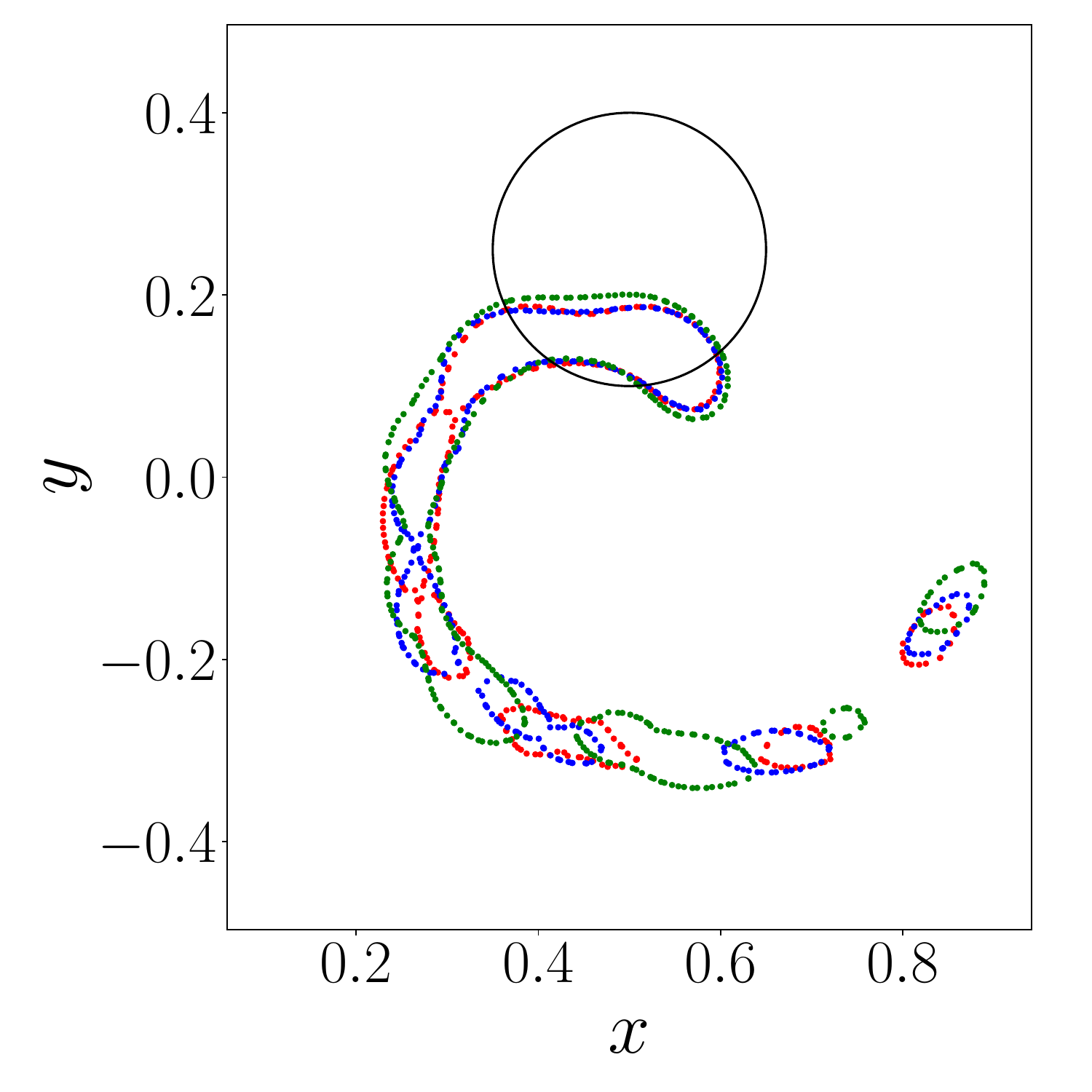}}
\subfigure[]{\includegraphics[width=0.40\textwidth]{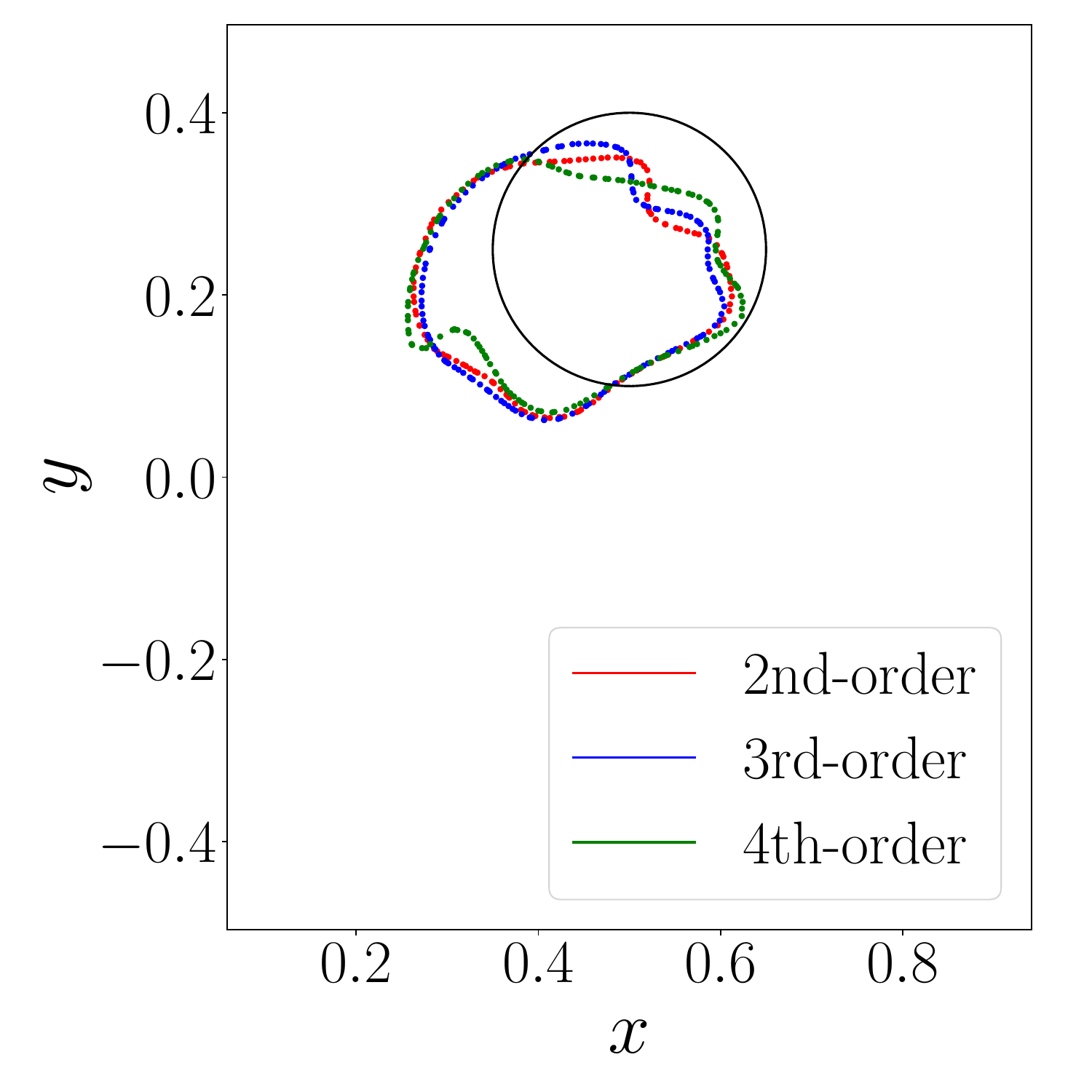}}
\subfigure[]{\includegraphics[width=0.40\textwidth]{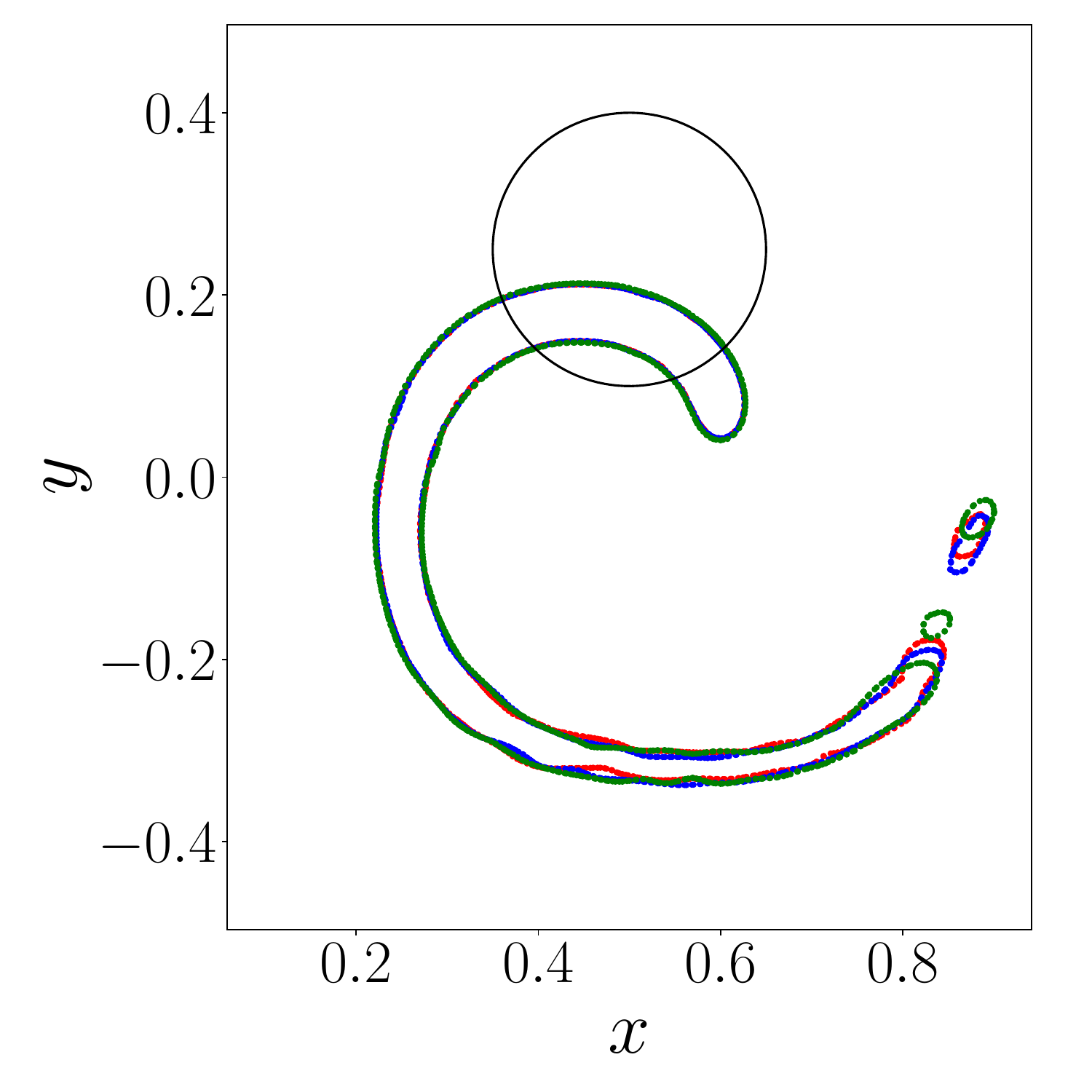}}
\subfigure[]{\includegraphics[width=0.40\textwidth]{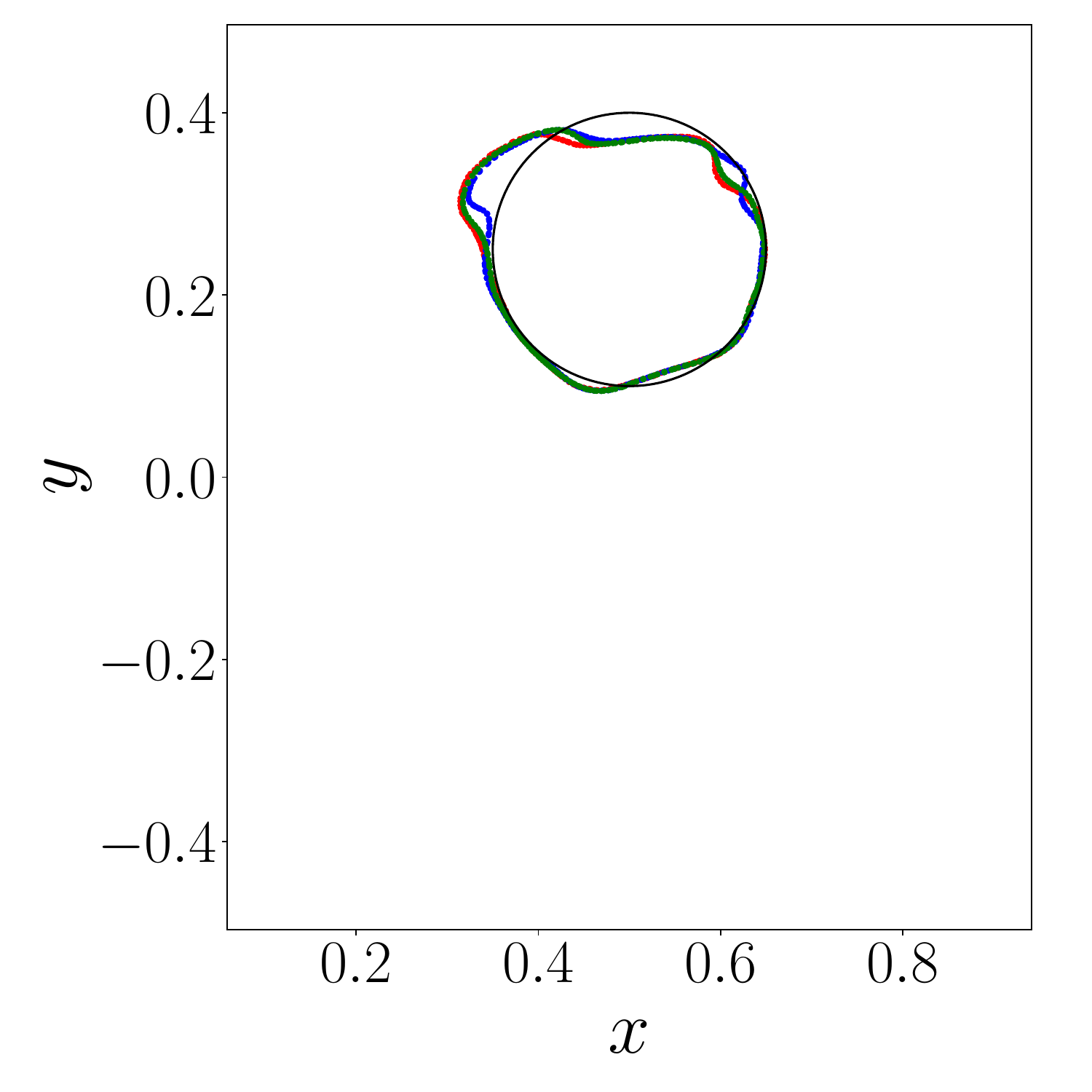}}
\subfigure[]{\includegraphics[width=0.40\textwidth]{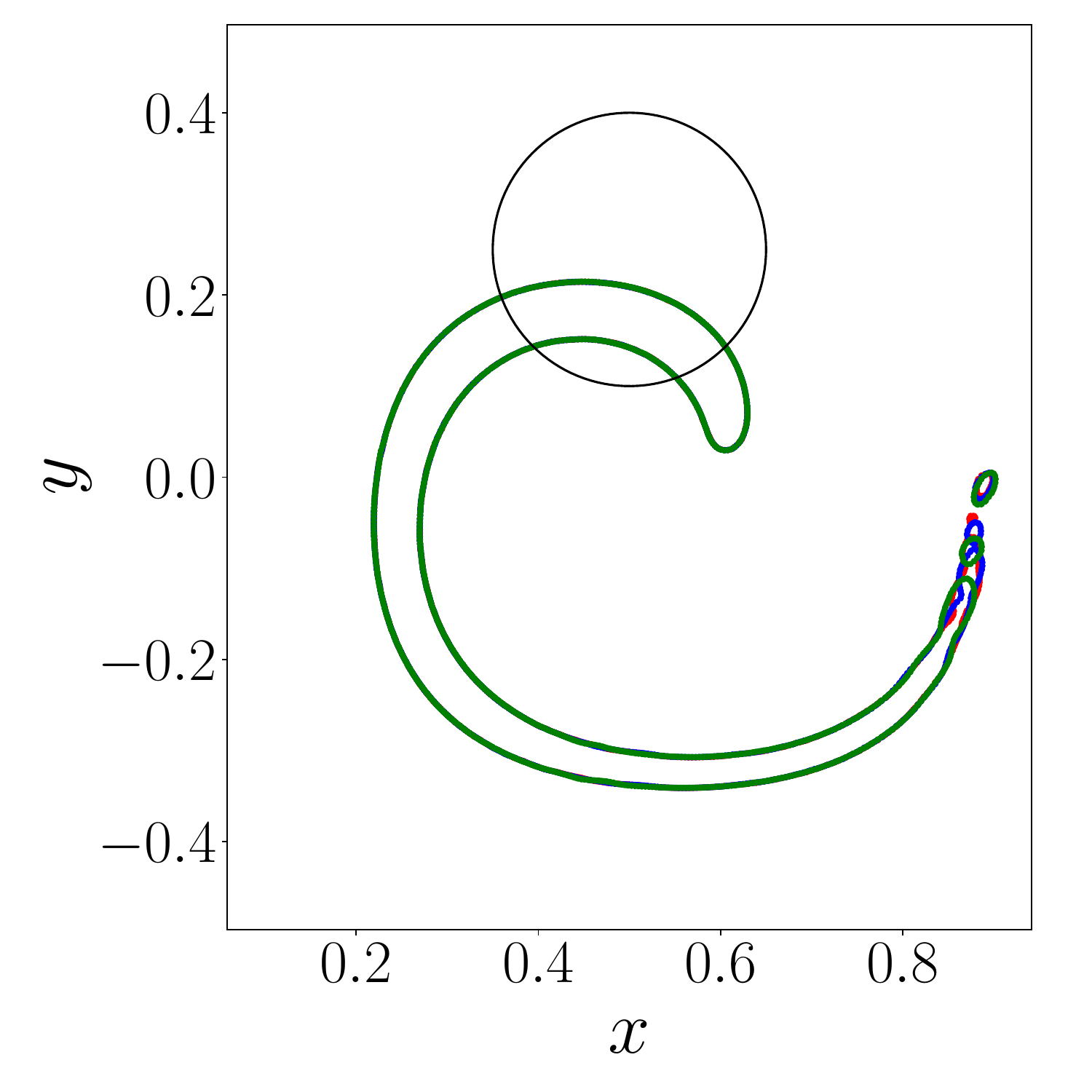}}
\subfigure[]{\includegraphics[width=0.40\textwidth]{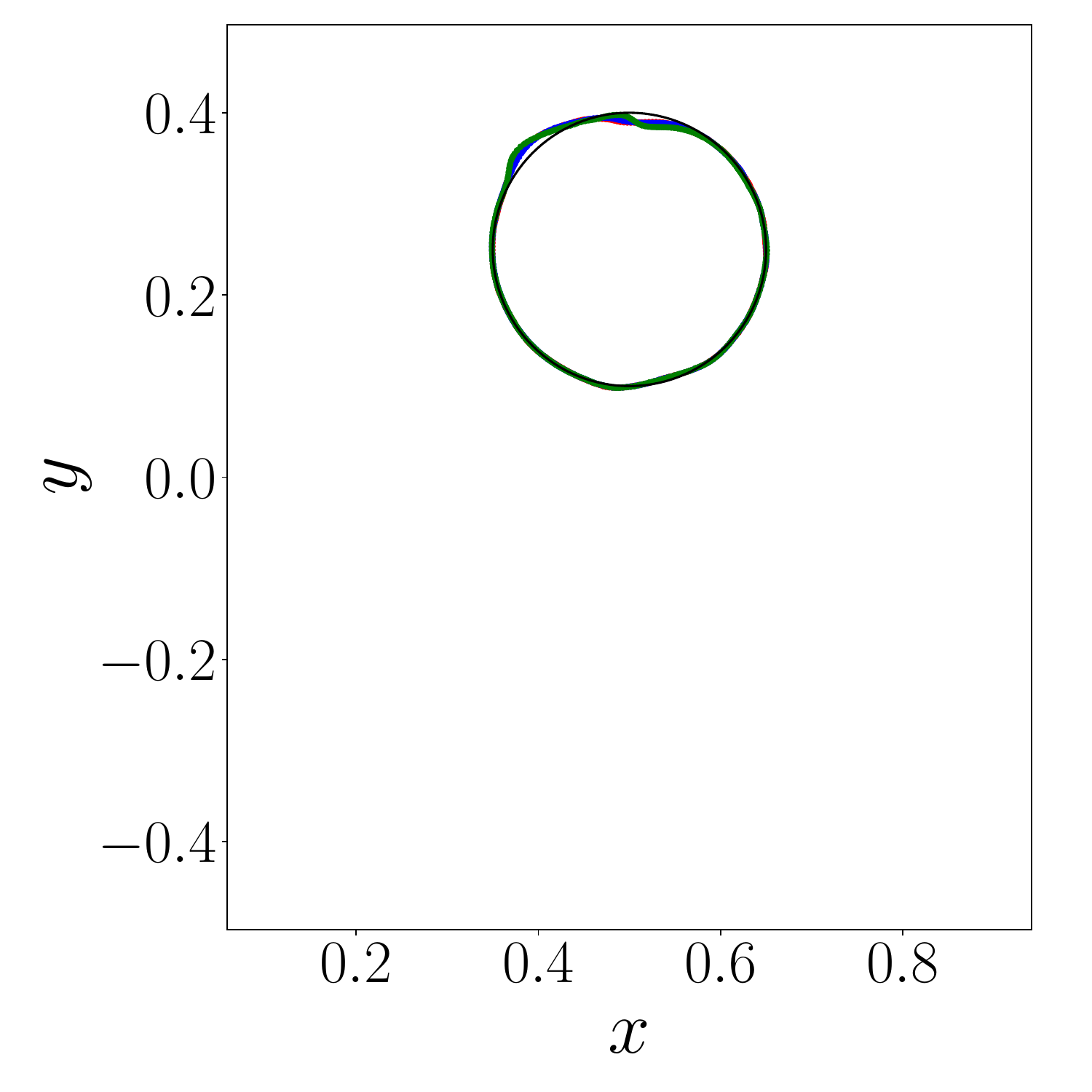}}
\caption{Interface location after half period (left) and full period (right). From top to bottom: $64^{2}$, $128^{2}$ and $256^{2}$ DoF. The thin black line represents the initial location of the interface.}
\label{fig:interfaceRK}
\end{figure}

\begin{figure}[h!]
\centering
\includegraphics[width=.75\textwidth]{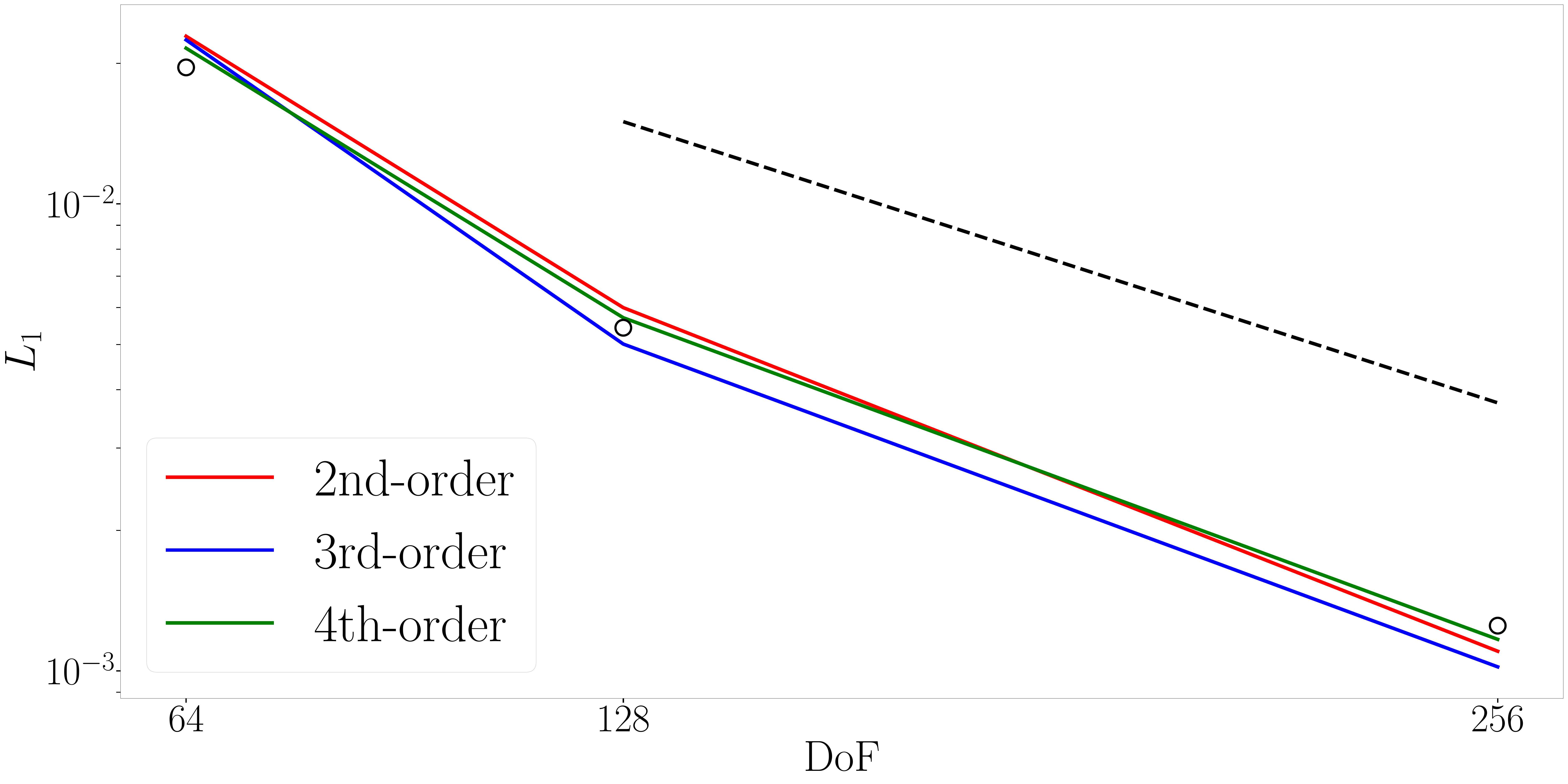}
\caption{$L_{1}$ error. Circles represent the errors reported by Mirjalili et al.~\cite{mirjalili2020conservative}. The dashed black line represents first-order rate.}
\label{fig:L1}
\end{figure}

We can immediately notice that, for a fixed order of approximation, mass conservation errors decrease with mesh refinement. More interestingly, we can observe that increasing the order of approximation (for fixed total number of DoF) provides significantly smaller mass conservation errors. This is further highlighted in figure \ref{fig:mass_res} where the different orders are compared for fixed total number of DoF. We can observe that the jump between $3^{\mathrm{rd}}$-order  and $4^{\mathrm{th}}$-order is substantial, whereas further increasing the order to $5$ gives a much less significant improvement (even if still present). 
From a more qualitative point of view, we show in figures \ref{fig:interfaceRK} the location of the interface at half period and full period for different orders and different resolutions. We can observe that the different orders provide similar results. This is further confirmed by the evaluation of the $L_{1}$ error shown in figure \ref{fig:L1}. Notice that the similarity between the different orders is expected: since we are solving for a function which should ideally converge towards an exact discontinuity, the numerical solution cannot converge faster than first order in $L_{1}$-norm. 

It is worthwhile highlighting that in the resolution of the full system of equations, even if the phase field and partial densities cannot converge faster than first order, the smoothness of velocity and pressure across the interface can significant benefit from high-order discretisations.
%
\subsection{Two-phase flows}
Here we will consider a series of validation tests involving the simulation of two-phase flows using the five equation model presented in section \ref{sec:5eq}. 
\subsubsection{Droplet advection}
Before proceeding to more complex configurations, we want to assess the capability of the proposed scheme to fulfil the IEC. 

The classical test case for this purpose is the linear advection of a droplet in a periodic domain at constant velocity. Notice that with respect to the previous case used for the boundedness study, here we solve the fully-coupled system of equations.

A bubble of radius $R=25/89$ is located at the center of a $[0,1]^{2}$ periodic domain. 
The material properties of the air medium for this test case are $\gamma_{1}=1.4$, $\rho_{1} = 1 \times 10^{-3}$ and $p_{1}^{\infty} = 0 $, whereas for the water medium the fluid properties are $\gamma_{2}=4.4$, $\rho_{2} = 1$ and $p^{\infty}_{2} = 6 \times 10^{3}$. 

\begin{figure}[h!]
\centering
\includegraphics[width=.75\textwidth]{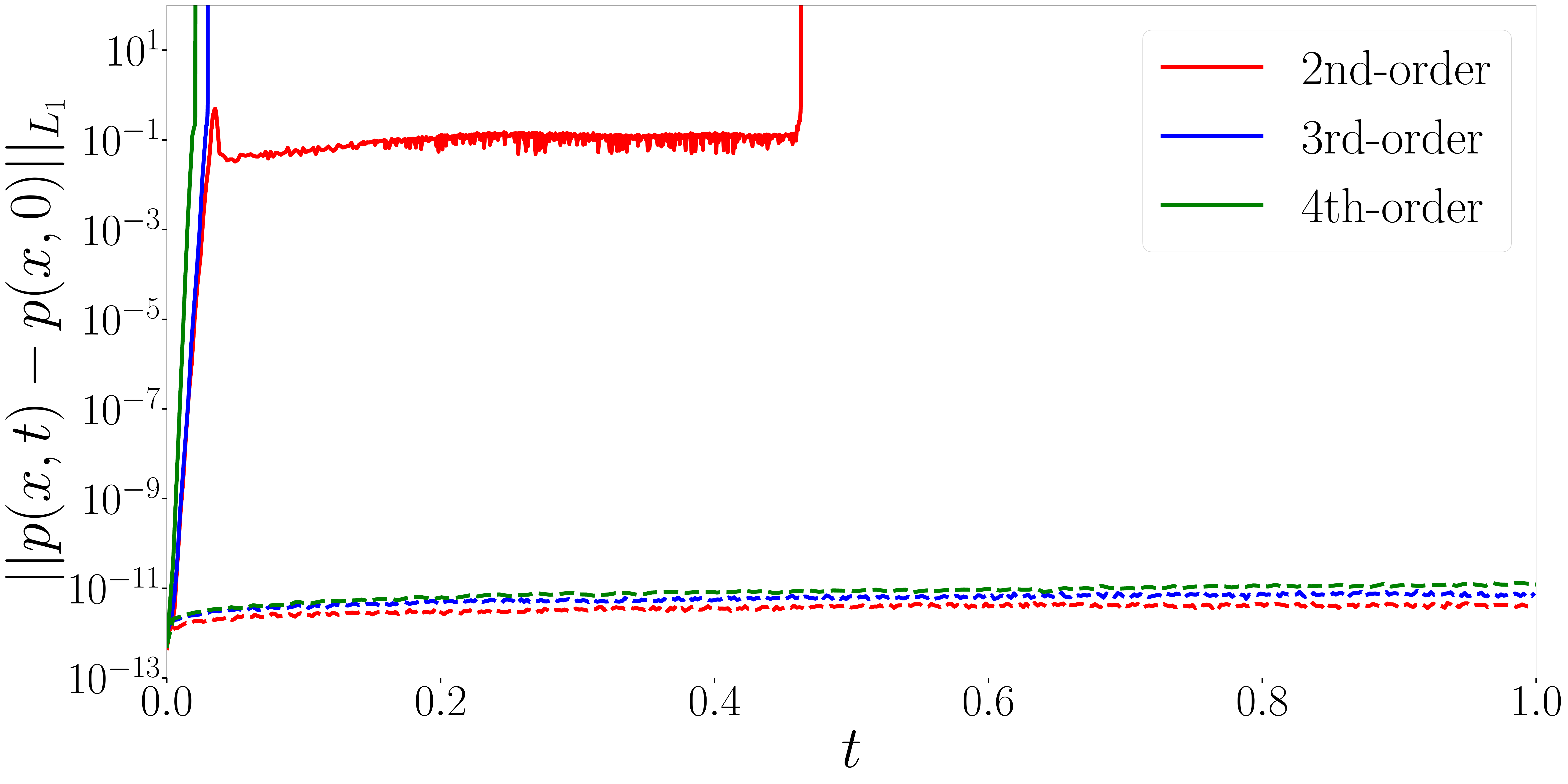}
\caption{Pressure error as a function of the time. Solid lines, conservative approach; dashed lines, primitive approach.}
\label{fig:pressure_L1}
\end{figure}

The initial conditions read:
\begin{equation}
\textbf{u} = (5,5), \quad p=1, \quad \phi_{1} = \frac{1}{2}\bigg(1 + \tanh \bigg( \frac{r-R}{2\epsilon}\bigg) \bigg) \quad \mathrm{and} \quad \rho = \rho_{2} + (\rho_{1}-\rho_{2}) \phi_{1}.
\end{equation}
In figure \ref{fig:pressure_L1} the $L_{1}$ error of the pressure is shown as it evolves in time. In particular, we compare different orders of approximation and the two different approaches (interpolating either primitive or conservative variables).

\begin{figure}[h!]
\centering
\includegraphics[width=.75\textwidth]{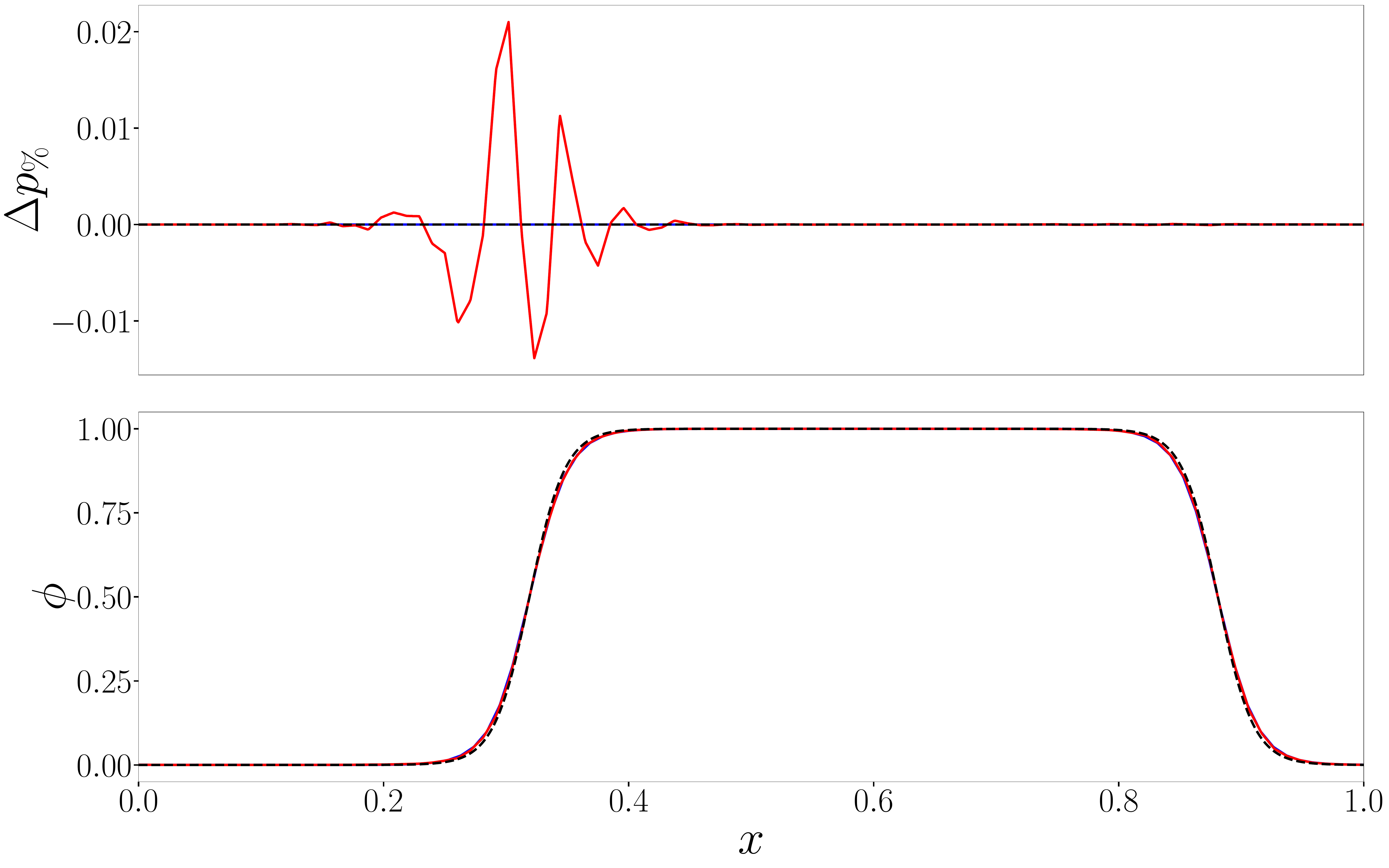}
\caption{Comparison of percentile pressure variation (top) and phase-field (bottom) along the $x$ direction for conservative (red line) and primitive (blue line) approaches. The dashed black line denotes the exact solution of the problem.}
\label{fig:pressure_line}
\end{figure}

\begin{figure}[h!]
\centering
\includegraphics[trim=0 100 0 100,clip, width=.95\textwidth]{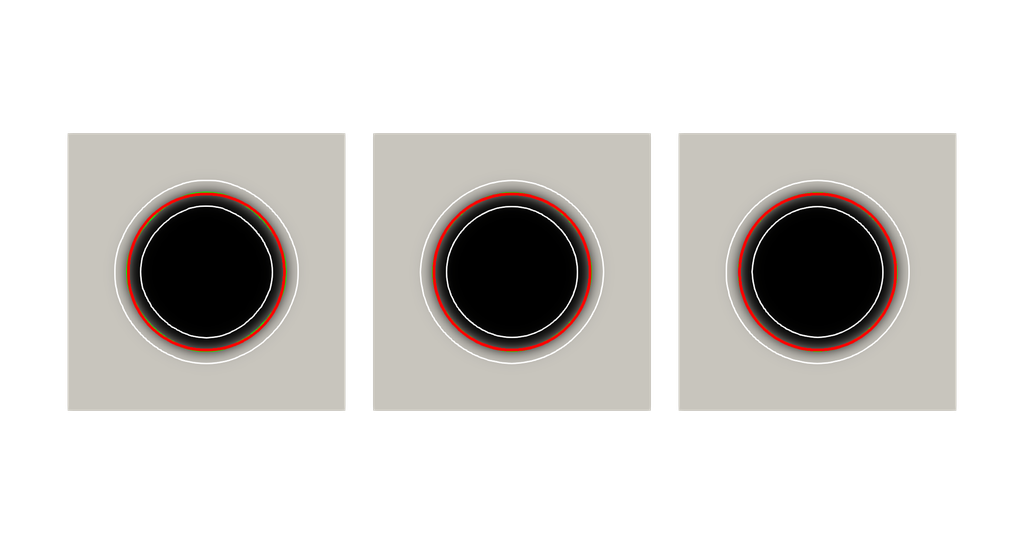}
\caption{Comparison of the phase field after five periods for different orders of approximation. From left to right: $2^{\mathrm{nd}}$, $3^{\mathrm{rd}}$ and $4^{\mathrm{th}}$ order simulations.}
\label{fig:shape}
\end{figure}

We can observe that the interpolation using the primitive variables greatly improves the stability of the problem. By using the conservative variables, even for low-order approximation ($2^{\mathrm{nd}}$), the simulation becomes unstable due to the pressure oscillations close to the interface (see figure \ref{fig:pressure_line}). Instead, all  simulations using the primitive variables are stable and no oscillations are observed in proximity of the interface.
Finally, in figure \ref{fig:shape} we can observe the improvement, in particular passing from $2^{\mathrm{nd}}$- to $3^{\mathrm{rd}}$-order, in the shape of the droplet. In the $2^{\mathrm{nd}}$-order simulation we can observe a mild tendency of the droplet to deform in a diamond-like shape, whereas the higher-order computations are able to preserve the circularity of the droplet over long-time integration.
%
\subsubsection{Rayleigh-Taylor instability}
The Rayleigh–Taylor instability occurs when an interface between two fluids with different densities experiences a pressure gradient opposing the density gradient. The domain consists in a $[0, d] \times [0, 4d]$ with $d=1$. The interface is initially defined as the curve $y(x) = 2d + 0.1d \cos{(2\pi x)}$. The Rayleigh-Taylor instability is characterised by the Reynolds number $\mathrm{Re} = (\rho_{1}d^{3/2} \|\textbf{g}\|^{1/2})/\mu$ and the Atwood number $\mathrm{At} = (\rho_{1} - \rho_{2})/(\rho_{1} + \rho_{2})$ which are respectively set to $3000$ and $0.5$.
The top boundary is treated as a Riemann-invariant boundary condition with zero velocity and constant pressure, the bottom boundary is a no-slip wall whereas slip wall boundary conditions are prescribed at the lateral sides of the domain. 

\begin{figure}[h!]
\centering
\includegraphics[width=.75\textwidth]{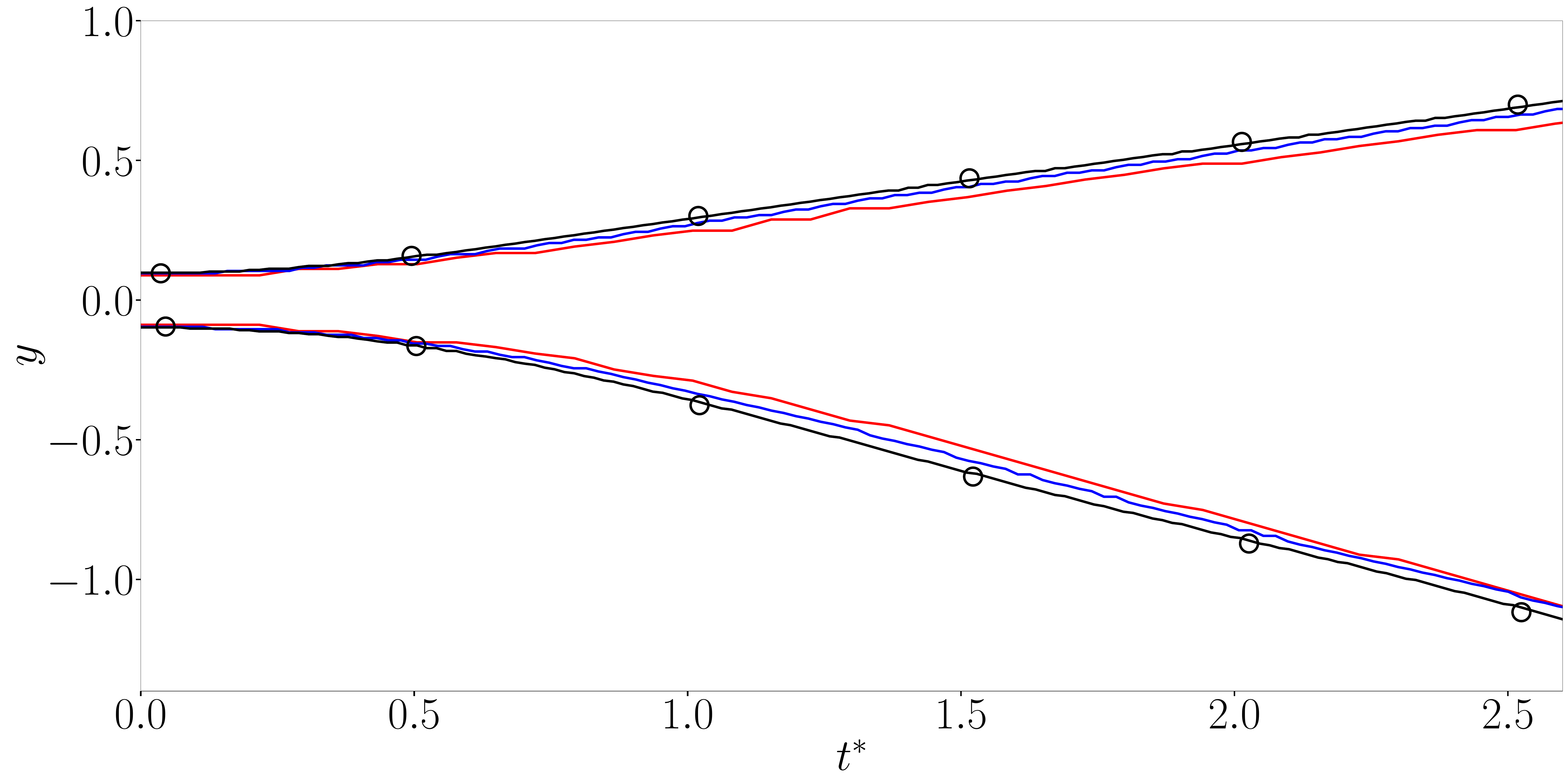}
\caption{Evolution of the lower and upper location of the interface for the Rayleigh-Taylor instability versus non-dimensional time for the $2^{\mathrm{nd}}$-order simulation. Red line, $200\times800$ DoF; blue line $400\times1600$ DoF; black line $800\times3200$ DoF. Symbols indicated the reference solution by Chiu \& Lin~\cite{chiu2011conservative}.}
\label{fig:plumes}
\label{fig:RTp1}
\end{figure}

\begin{figure}[h!]
\centering
\includegraphics[width=.75\textwidth]{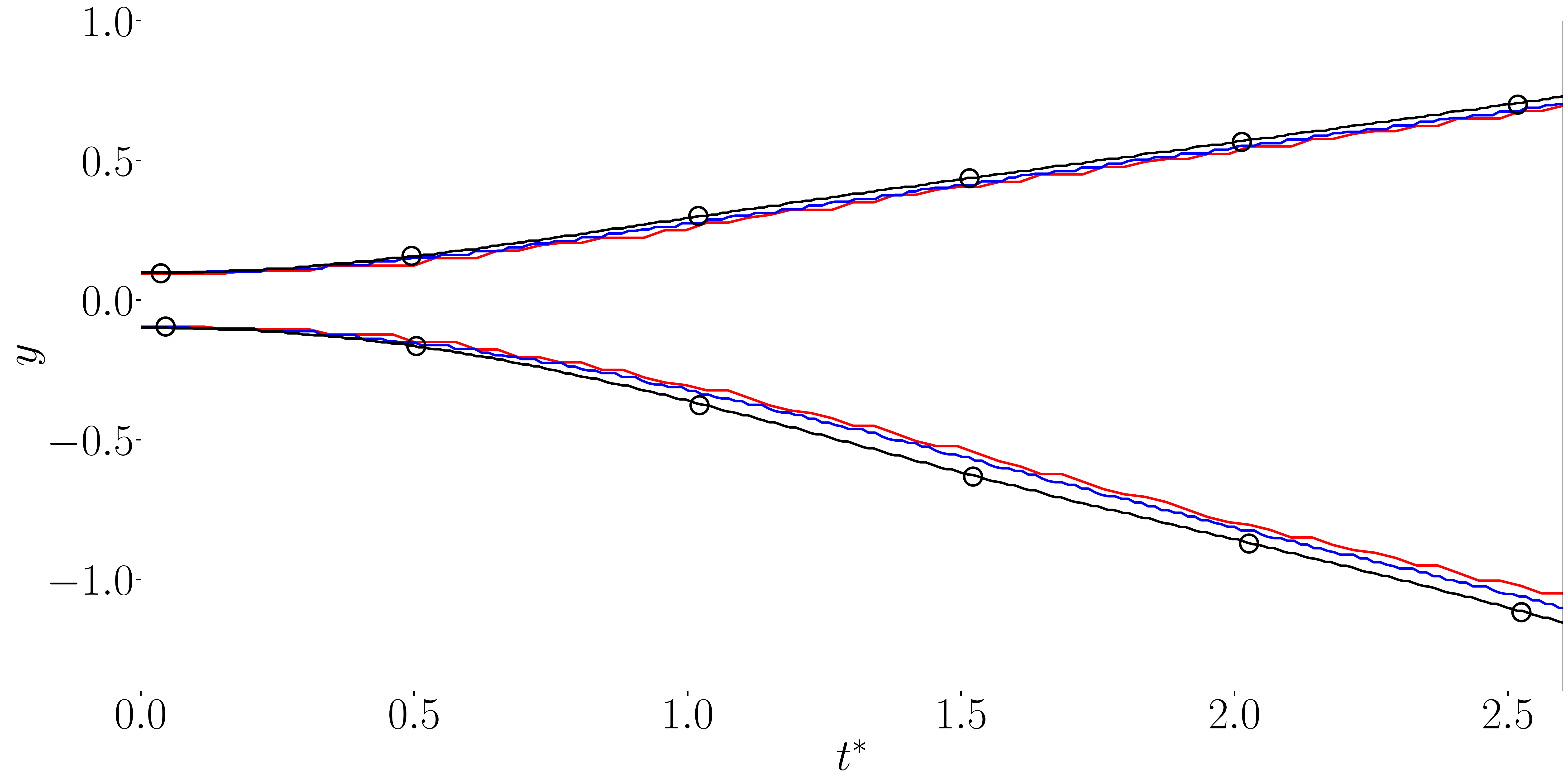}
\caption{The evolution of the top and bottom of the interface for the Rayleigh-Taylor instability versus non-dimensional time for the $5^{\mathrm{th}}$-order simulation. Red line, $200\times800$ DoF; blue line $400\times1600$ DoF; black line $800\times3200$ DoF.Symbols indicated the reference solution by Chiu \& Lin~\cite{chiu2011conservative}.}
\label{fig:RTp4}
\end{figure}

\begin{figure}[h!]
\centering
\subfigure[$t^{*}=1.65$.]{\includegraphics[trim=300 0 300 0,clip,width=0.32\textwidth]{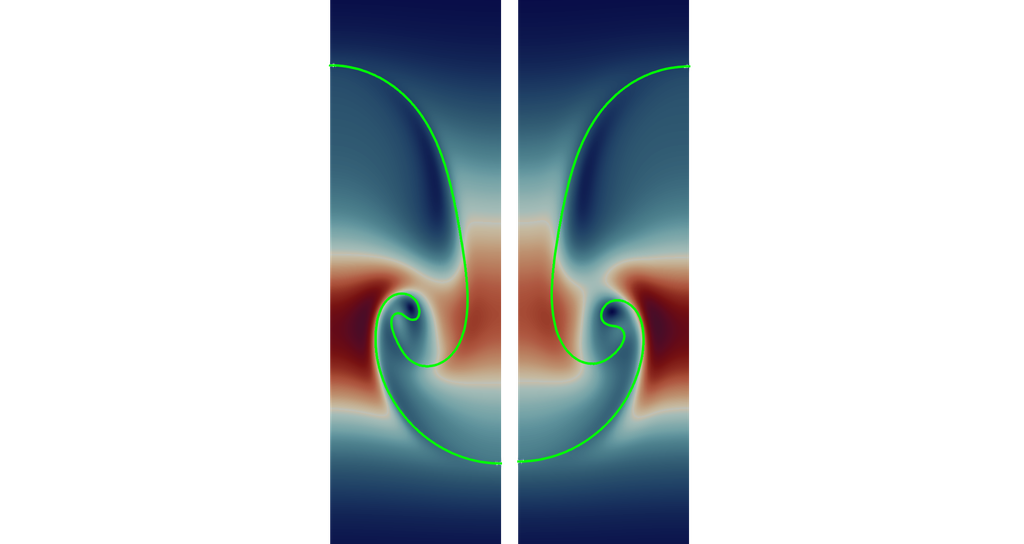}}
\subfigure[$t^{*}=1.75$.]{\includegraphics[trim=300 0 300 0 ,clip,width=0.32\textwidth]{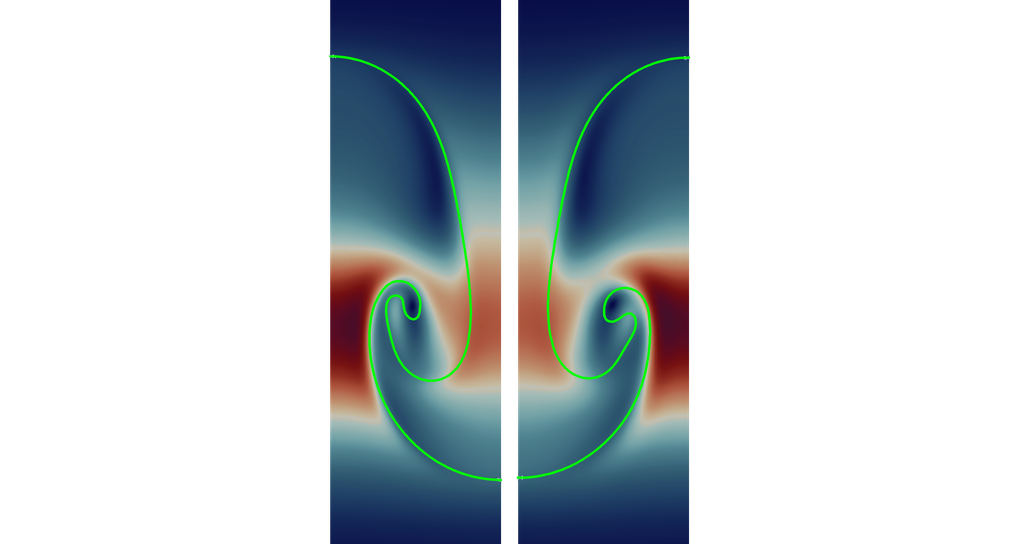}}
\subfigure[$t^{*}=1.85$.]{\includegraphics[trim=300 0 300 0 ,clip,width=0.32\textwidth]{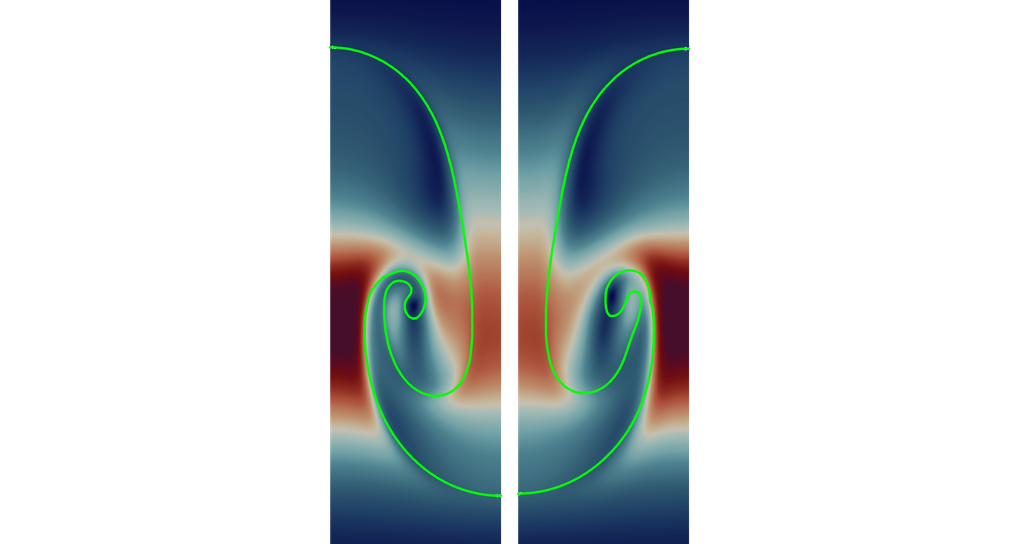}}
\caption{Velocity magnitude at different times for different orders of approximation. In each subfigure: left, $5^{\mathrm{th}}$-order simulation; right $2^{\mathrm{nd}}$-order simulation. The green line represents the instantaneous location of the interface.}
\label{fig:snap1}
\end{figure}

\begin{figure}[h!]
\centering
\subfigure[$t^{*}=2.90$.]{\includegraphics[trim=400 0 400 0,clip,width=0.32\textwidth]{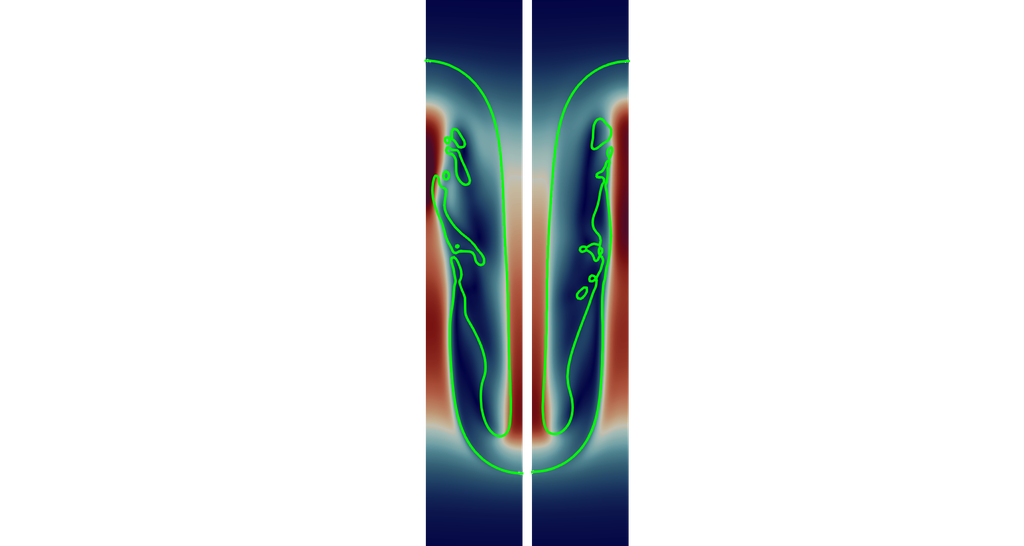}}
\subfigure[$t^{*}=3.00$.]{\includegraphics[trim=400 0 400 0 ,clip,width=0.32\textwidth]{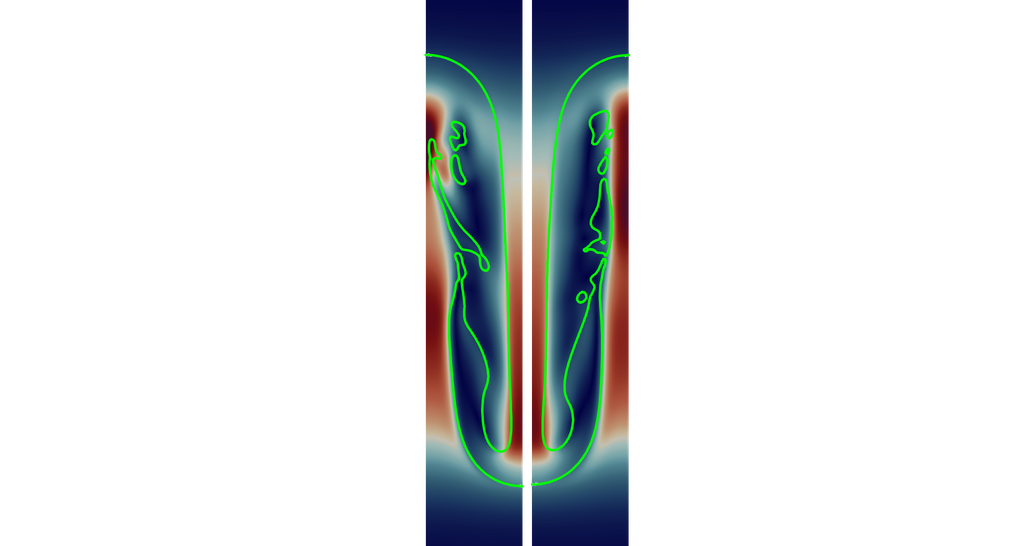}}
\subfigure[$t^{*}=3.10$.]{\includegraphics[trim=400 0 400 0 ,clip,width=0.32\textwidth]{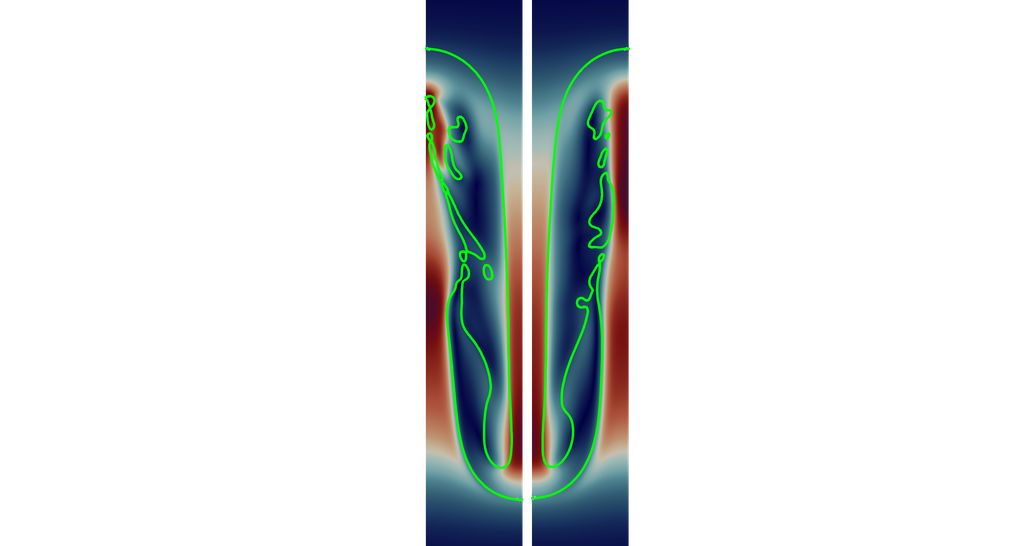}}
\caption{Velocity magnitude at different times for different orders of approximation. In each subfigure: left, $5^{\mathrm{th}}$-order simulation; right $2^{\mathrm{nd}}$-order simulation. The green line represents the instantaneous location of the interface.}
\label{fig:snap2}
\end{figure}

First, as a verification test, we perform a grid convergence study for two different orders of approximation ($2^{\mathrm{nd}}$ and $5^{\mathrm{th}}$). For this problem, a good indicator of convergence is the tracking of the interface lower and upper locations over time. In figures \ref{fig:RTp1} and \ref{fig:RTp4} we show these locations for different grids and compare them with the solution by Chiu \& Lin~\cite{chiu2011conservative}.
In figure \ref{fig:snap1} we compare the two different simulations on the medium grid ($400\times1600$ DoF) at different adimensional times ($t^{*} = t /\sqrt{d/(\|\textbf{g}\| \mathrm{At})}$). As mentioned in previous sections of the paper, we do not expect a great improvement by increasing the order of accuracy if the velocity field is analytically given. However, if the velocity field is an unknown to be resolved and it is sufficiently smooth (as it is expected in the five equation model) we can expect some improvement by increasing the order of accuracy. We can observe in figure \ref{fig:snap1} that before primary break-up, when the flow field starts to become more complex, some differences between the two orders of accuracy start to emerge. In particular, we can observe that the  $5^{\mathrm{th}}$-order simulation provides a thinner stretching of the interface. This is consequence of the better resolved velocity field which defines the dynamics of the interface. The velocity magnitude, in fact, is over-smoothed for the $2^{\mathrm{nd}}$-order simulation due to the excessive amount of numerical dissipation.
By letting the system evolve for longer times, other differences in the break-up region emerge: in particular, in figure \ref{fig:snap2}, we can observe that in the $2^{\mathrm{nd}}$-order simulation there is a tendency in agglomerating droplets whereas finer structures are preserved in the $5^{\mathrm{th}}$-order case. Once again, this is not due to the improvement in the resolution of the phase field equation, but rather the consequence of better resolved velocity fields.
%
\subsubsection{Shock-bubble interaction}
This problem consists in the interaction between a $2$D helium bubble with a plane shock wave in air. This setup has been extensively studied to assess both numerical methods and mathematical models to deal with two-phase flows for supersonic regimes \cite{terashima2009front,shukla2010interface,johnsen2006implementation,shyue2014eulerian,jain2023assessment}. 

The problem is defined on a $[-2,4]\times[0,1]$ rectangular domain, with periodic boundary conditions in the $y$ direction. Inlet and outlet boundary conditions are respectively prescribed at the left and right sides. The bubble has a radius of $25/89$ and it is initially located at $(0,0.5)$. The material properties for the air phase are $\gamma_{1}=1.4$, $\rho_{1}=1.0$, $p^{\infty}_{1}=0$ and $\mu_{1}=0$. The same properties for the helium medium, instead, are $\gamma_{2}=1.67$, $\rho_{2}=0.138$, $p^{\infty}_{2}=0$ and $\mu_{2}=0$.

\begin{figure}[h!]
\centering
\includegraphics[width=.95\textwidth]{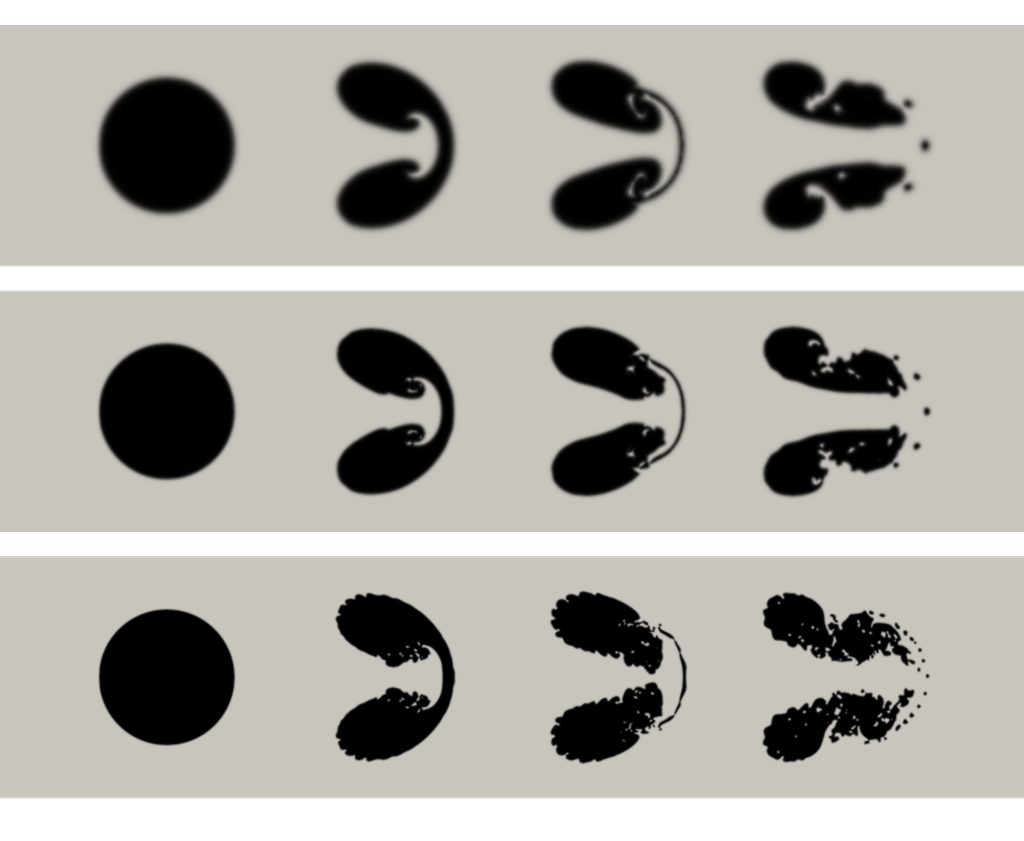}
\caption{Comparison of different refinements for a $4^{\mathrm{th}}$-order simulation: top, $1200\times200$ DoF; middle, $2400\times400$ DoF; bottom $4800\times800$ DoF.}
\label{fig:SD}
\end{figure}

The physical domain is discretised using a cartesian uniform grid and a $4^{\mathrm{th}}$-order approximation. In particular, different mesh refinements have been considered, including grids involving  $1200\times200$, $2400\times400$ and $4800\times800$ total number of DoF. 
After the shock wave hits the bubble strong deformations occur, leading to the breakup of the droplet. Primary breakup triggers the development of smaller and smaller structures from the main bubble (see figure \ref{fig:SD}).

\begin{figure}[h!]
\centering
\includegraphics[width=.95\textwidth]{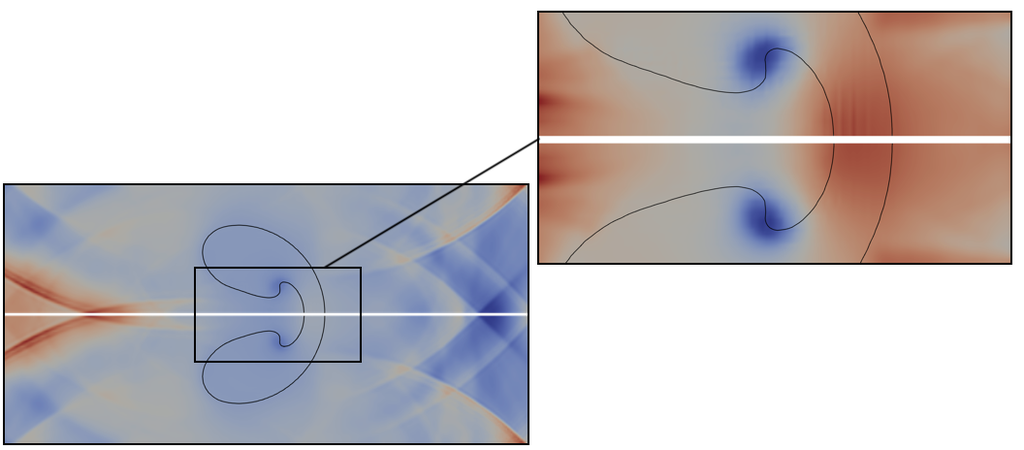}
\caption{Comparison of pressure field between conservative (top) and primitive (bottom) simulations.}
\label{fig:SD_pressure}
\end{figure}

With respect to the previous cases, compressibility effects are much more relevant in this problem. The presence of shock waves requires the use of artificial viscosity in order to keep the simulation stable. In this case we considered a physics-based artificial viscosity shock capturing technique \cite{tonicello2020entropy} in combination with the Pearsson \& Peraire discontinuity sensor \cite{persson2006sub}.
In figure \ref{fig:SD_pressure}, we compare conservative and primitive approaches for this problem. In particular, we can observe that even if the conservative approach is stable, pressure oscillations can be observed close to the interface, whereas the primitive approach provides smoother results.
Based on the specific relaxation approach of  the five equation model it is reasonable to base the shock detection technique on pressure which is supposed to be continuous across the material interface. In this way, ideally, the shock capturing technique should be active only at shock waves and not at material interfaces, whose regularisation is instead determined by the diffusion/sharpening.
\begin{figure}[h!]
\centering
\subfigure{\includegraphics[trim=0 200 0 150 ,clip,width=0.32\textwidth]{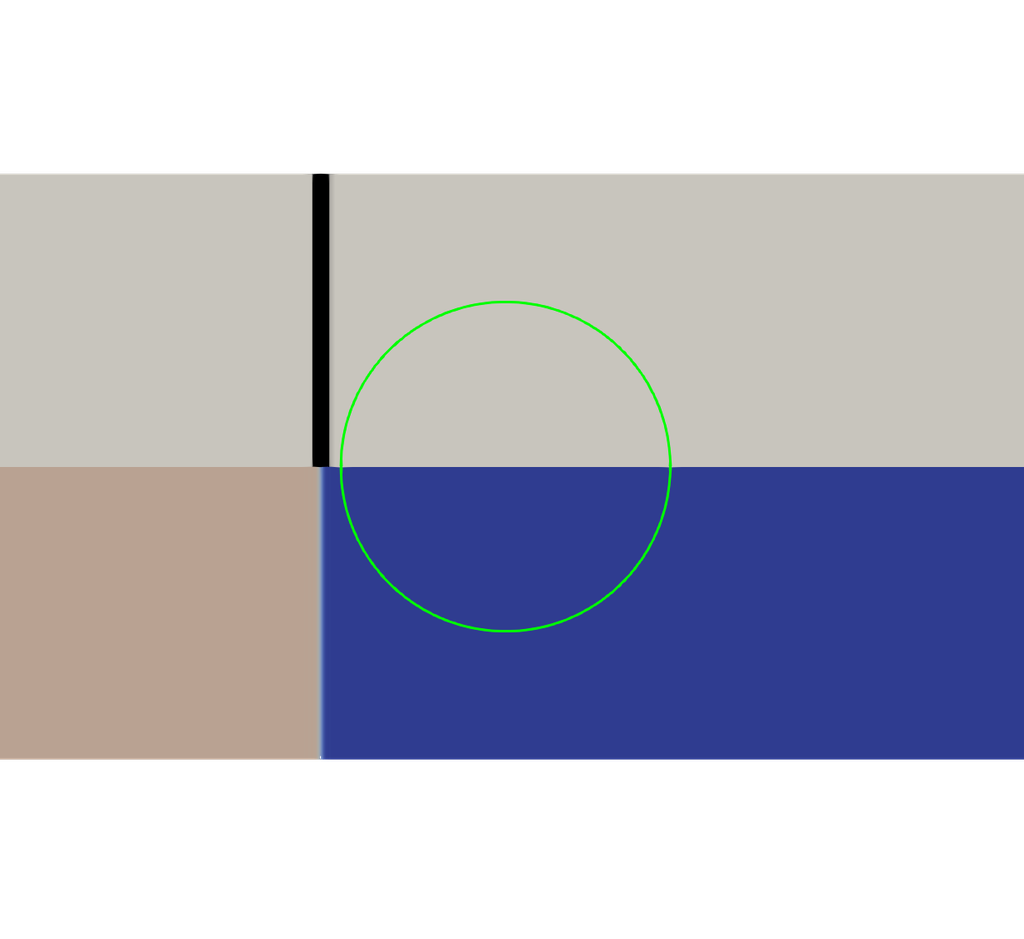}}
\subfigure{\includegraphics[trim=0 200 0 150 ,clip,width=0.32\textwidth]{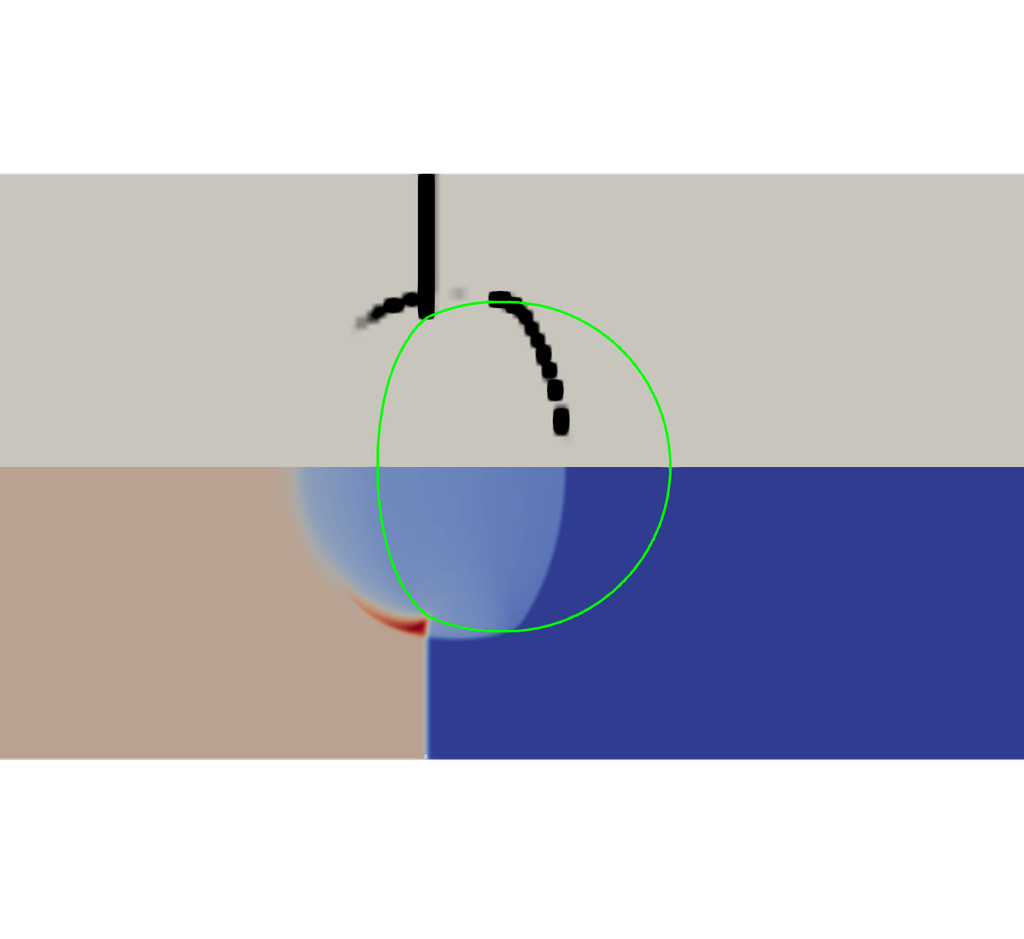}}
\subfigure{\includegraphics[trim=0 200 0 150 ,clip,width=0.32\textwidth]{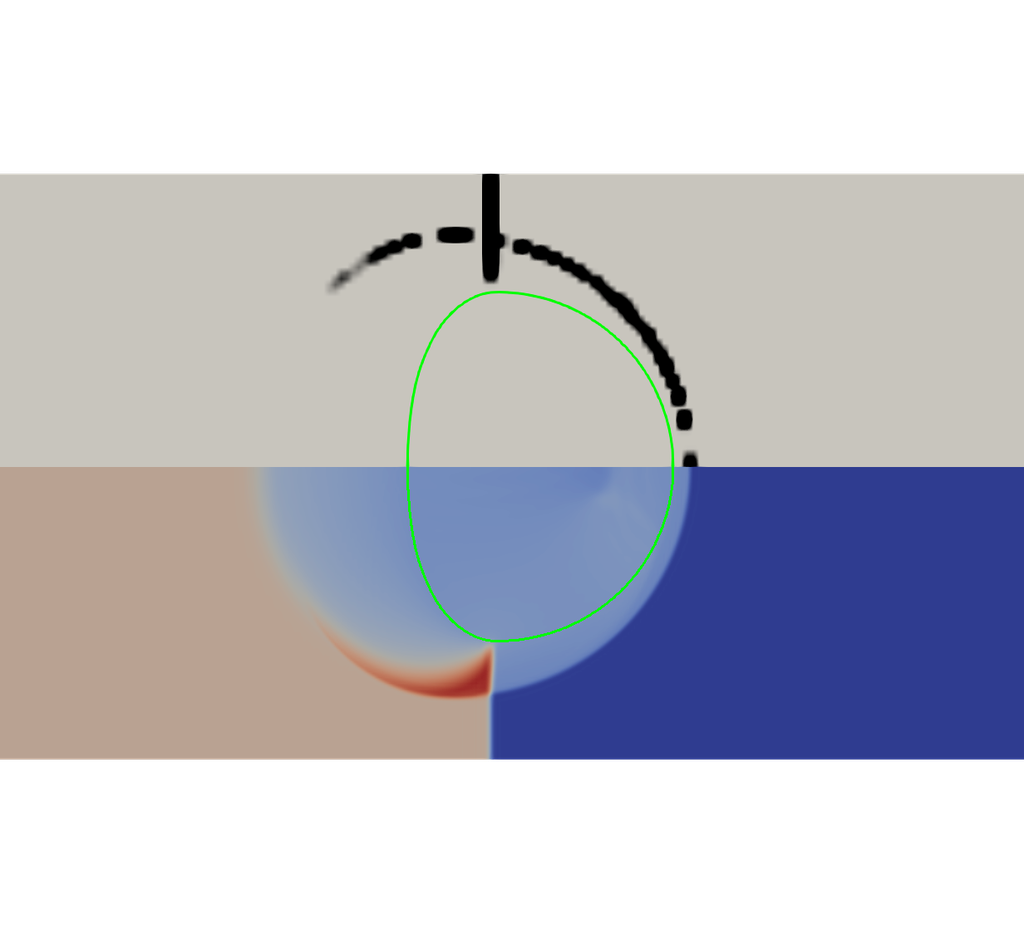}}
\caption{Artificial viscosity (top) and pressure field (bottom) at times close to the shock's impact on the droplet. The green line represents the instantaneous location of the interface.}
\label{fig:SD_AV}
\end{figure}
To further validate that the present approach avoids undesirable pressure oscillations close to the interface, we compare in figure \ref{fig:SD_AV} the activation of the shock sensor on top of the phase field at different time steps. We can observe that the shock detection correctly tracks the shock wave while being inactive in proximity of the material interface. It is interesting that the combined model (interface regularisation and artificial viscosity) is able to detect and treat different types of sharp features simultaneously: the Allen-Cahn regularisation terms are used only in proximity of the interface while the shock capturing technique only acts on the shock and not at the interface.

\begin{figure}[h!]
\centering
\subfigure[$t^{*}=2.00$.]{\includegraphics[trim=0 200 0 150,clip,width=0.32\textwidth]{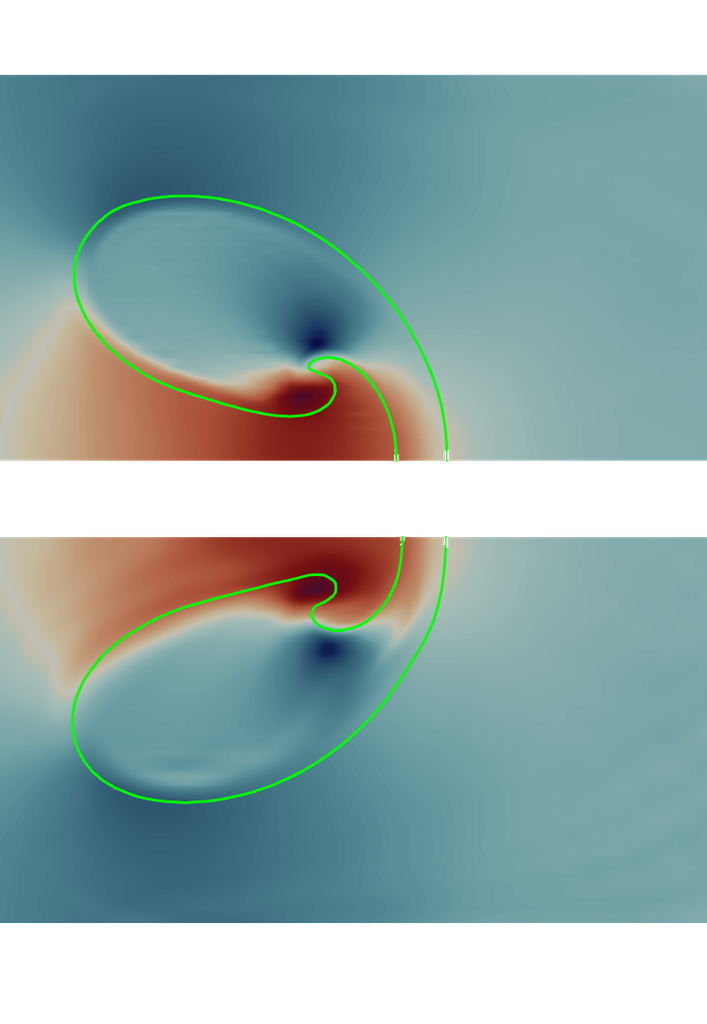}}
\subfigure[$t^{*}=2.25$.]{\includegraphics[trim=0 200 0 150 ,clip,width=0.32\textwidth]{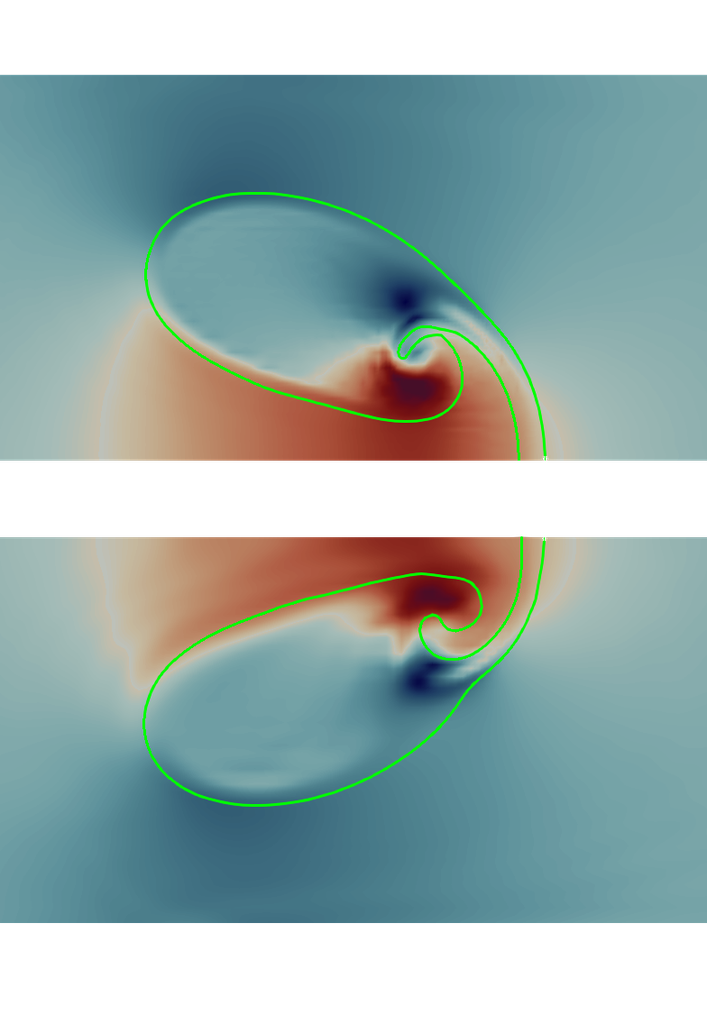}}
\subfigure[$t^{*}=2.50$.]{\includegraphics[trim=0 200 0 150 ,clip,width=0.32\textwidth]{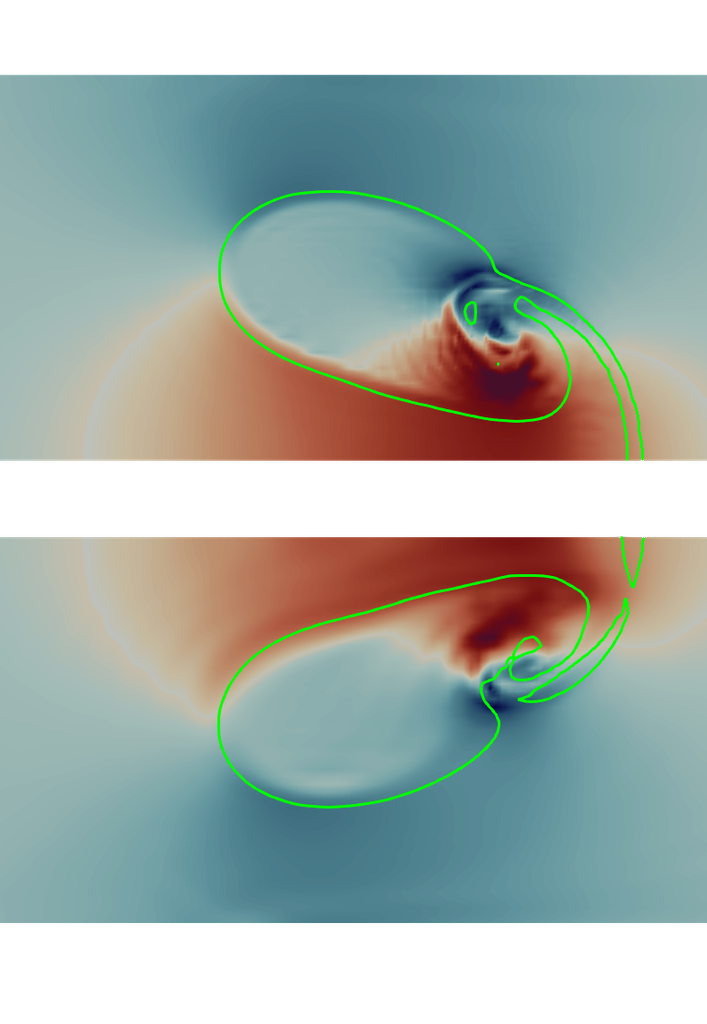}}
\caption{Velocity magnitude at different times for different orders of approximation. In each subfigure: top, $4^{\mathrm{th}}$-order simulation; bottom $2^{\mathrm{nd}}$-order simulation. The green line represents the instantaneous location of the interface.}
\label{fig:p1vsp3}
\end{figure}

In figure \ref{fig:p1vsp3} we compare the same simulation with different orders of approximation. Similarly with respect to the Rayleigh-Taylor instability, we notice some benefits in using a high-order discretisation: the plumes arising from the center line jet are much thinner using a $4^{\mathrm{th}}$-order approximation. The second order approximation, instead is characterised by larger, more dissipated structures.
%
\subsubsection{Two-phase Taylor-Green vortex: $\mathrm{Re}=500$}
The Taylor-Green vortex is a popular test case to evaluate the role played by the numerical scheme in under-resolved flows. Classically, this is done in a compressible framework with different values of the characteristic Mach number. Here we consider a modification of it for two-phase flows proposed by~\cite{jain2022kinetic}. In this problem the parameters of the two fluids are exactly the same except for the respective densities.

The initial conditions read:
\begin{equation}
\left\{
\begin{aligned}
\phi & =1 + \frac{1}{2} \Bigg [ 1 + \tanh \Bigg \{ \frac{\sqrt{(z-\pi)^{2}} - L}{2 \epsilon} \Bigg \} \Bigg], \\
\rho & = \rho_{1}^{(0)}\phi + \rho_{2}^{(0)}(1-\phi), \\
u & =U^{(0)}\sin \bigg(\frac{x}{L}\bigg)\cos \bigg(\frac{y}{L}\bigg)\cos \bigg(\frac{z}{L}\bigg), \\
v & =-U^{(0)}\cos \bigg(\frac{x}{L}\bigg)\sin \bigg(\frac{y}{L}\bigg)\cos \bigg(\frac{z}{L}\bigg), \\
w & =0,\\
p & =p^{(0)}+\frac{\rho_{0} U_{0}}{16}\Bigg[\cos \bigg(\frac{2x}{L}\bigg)+\cos \bigg(\frac{2y}{L}\bigg)\Bigg] \Bigg[\cos \bigg(\frac{2z}{L}\bigg)+2 \Bigg]. 
\end{aligned}
\right.
\end{equation} 
with $x$, $y$ and $z$ the spatial coordinates, $u$, $v$, $w$ the velocity components along $x$, $y$ and $z$. The parameters $U^{(0)}$, $P^{(0)}$ and $\rho_{1,2}^{(0)}$ are used in such a way to obtain a certain characteristic Mach number of the flow field. 

As a first test we considered the single phase case but simulated using the proposed five equation model. In other words, we selected $\rho_{1,2}^{(0)}=1$ and $U^{(0)}$, $P^{(0)}$ such that the initial flow Mach number is equal to $0.05$. The Reynolds number has been set to $500$. This set-up is popular in the community of Taylor-Green Vortex simulations where DNS is feasible. Finally, regarding the parameters determining the equation of state of the two phases we simply put $\gamma_{1,2}=1.4$ and $p^{\infty}_{1,2}=0$. In this way the system exactly coincides with the standard one-phase version of the TGV case for an ideal gas. The resolution is fixed at $96^{3}$ total number of DoF. Three orders of approximation have been considered: $3^{\mathrm{rd}}$, $4^{\mathrm{th}}$ and $5^{\mathrm{th}}$. Consequently, the grids employed are made of $32^{3}$, $24^{3}$ and $19^{3}$ quadratics elements, respectively. 

Secondly, we decided to start modifying the initial density of the first phase $\rho_{1}$ to $2.0$, $0.5$ and $0.25$.

\begin{figure}[h!]
\centering
\subfigure[Kinetic energy.]{\includegraphics[width=0.49\textwidth]{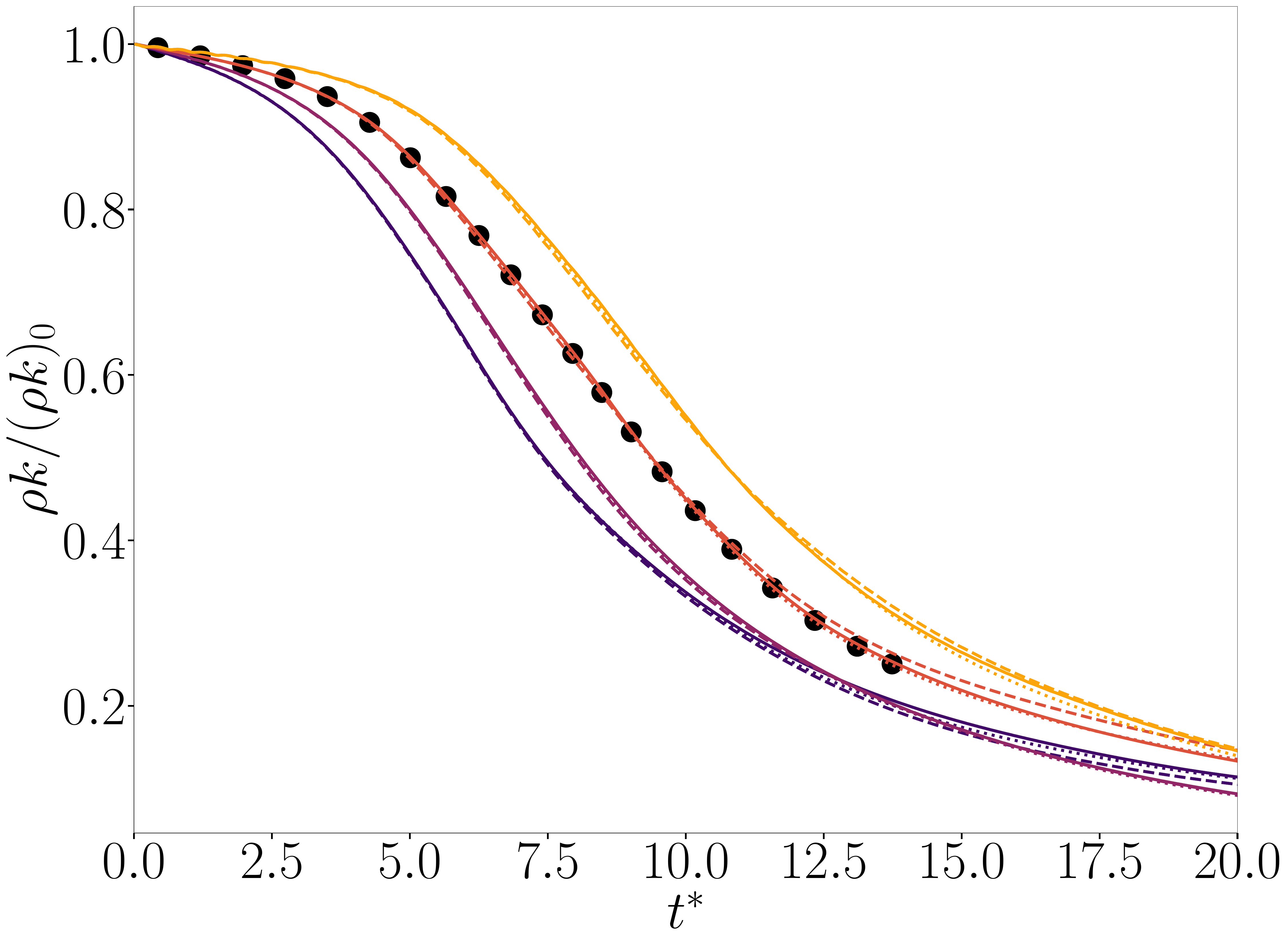}}
\subfigure[Viscous dissipation.]{\includegraphics[width=0.49\textwidth]{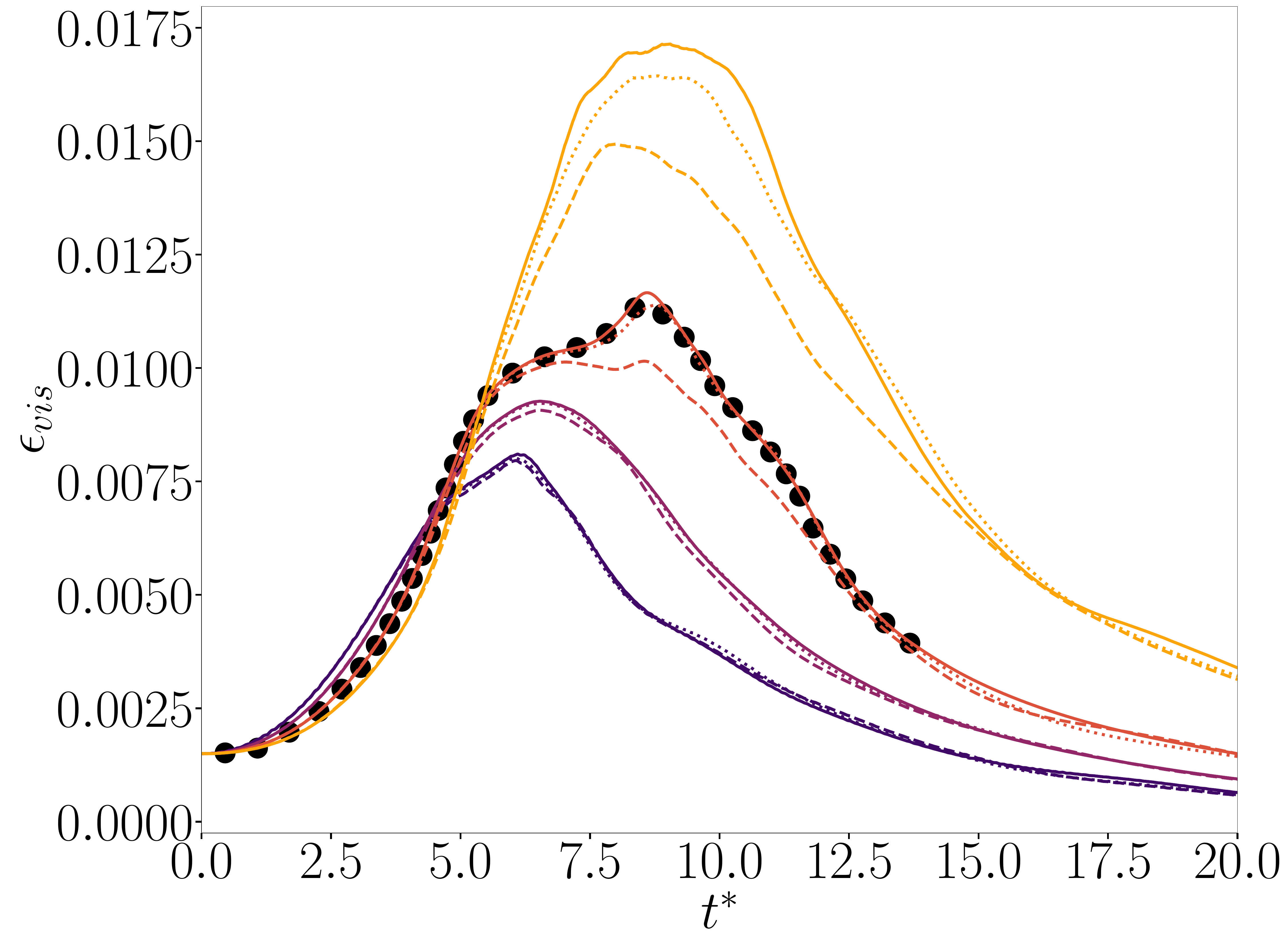}}
\caption{Averaged kinetic energy (left) and viscous dissipation (right) as function of the time. Colour gradient from light orange to dark purple indicates smaller values of $\rho_{1}$. Namely, $\rho_{2}/\rho_{1}=0.5,1,2,4$. Dashed line, $3^{\mathrm{rd}}$-order; dotted line, $4^{\mathrm{th}}$-order; solid line, $5^{\mathrm{th}}$-order. The black dots represent the DNS data for the single phase case~\cite{chapelier2012inviscid}. The kinetic energy is normalised by its initial value.}
\label{fig:TGV_ke}
\end{figure}

In figure \ref{fig:TGV_ke} we show the averaged kinetic energy and viscous dissipation for the cases herein considered. 
First, let us comment the general behaviour of the single phase case where data from the literature are available: we can observe trends that are in good agreement with respect to the reference work by Chapelier et al.~\cite{chapelier2012inviscid}. Also referring to the single phase case, we can observe similar trends with respect to~\cite{chapelier2012inviscid} in terms of approximation order influence: the $3^{\mathrm{rd}}$-order discretisation is characterised by stronger numerical dissipation, leading to an under-estimation of the viscous dissipation. On the other hand, $4^{\mathrm{th}}$-order and $5^{\mathrm{th}}$-order computations are very similar. The levels of numerical dissipation are negligible in these cases. 

In the two-phase simulations, as $\rho_{2}/\rho_{1}$ increases, the differences between the orders of approximation tend to decrease. This behaviour extends also in the case of $\rho_{1}<\rho_{2}$. In fact, for the case $\rho_{1}/\rho_{2}=0.5$ we notice a significant influence of the order of approximation on the viscous dissipation field.

\begin{figure}[h!]
\centering
\subfigure[$t^{*}=0$.]{\includegraphics[width=0.32\textwidth]{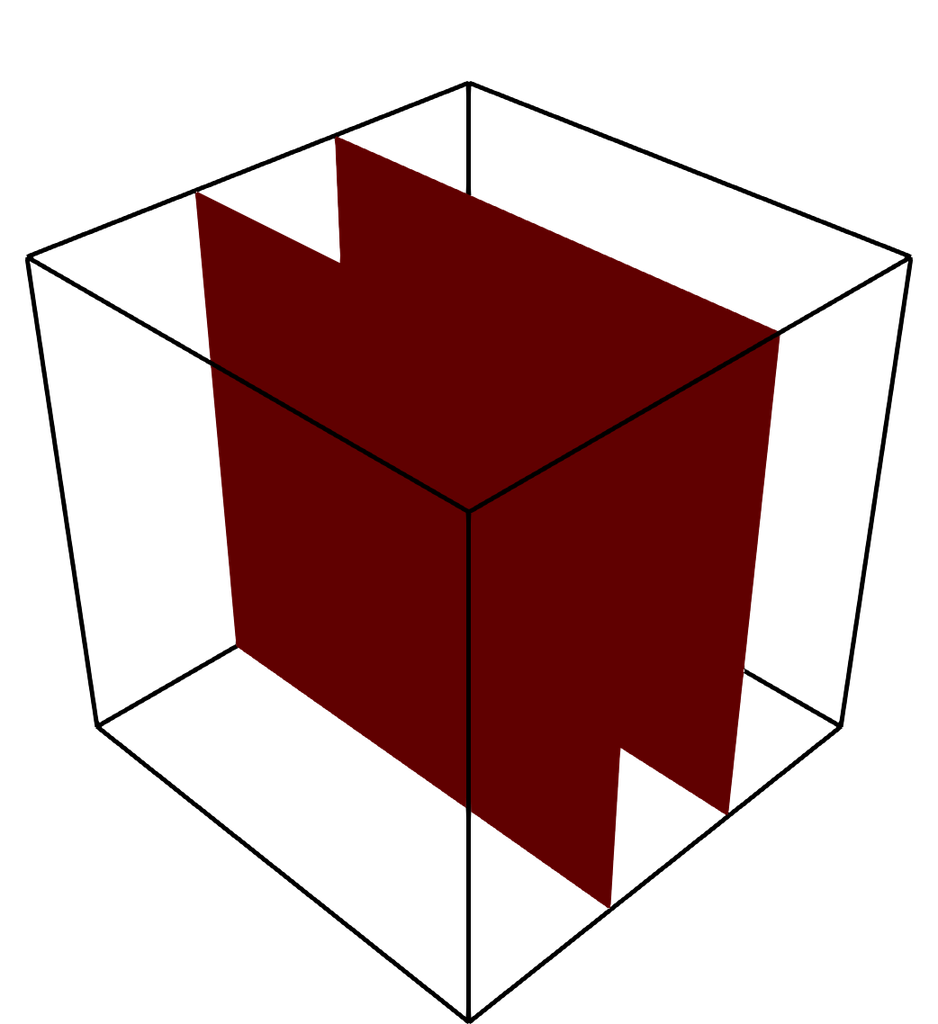}}
\subfigure[$t^{*}=2$.]{\includegraphics[width=0.32\textwidth]{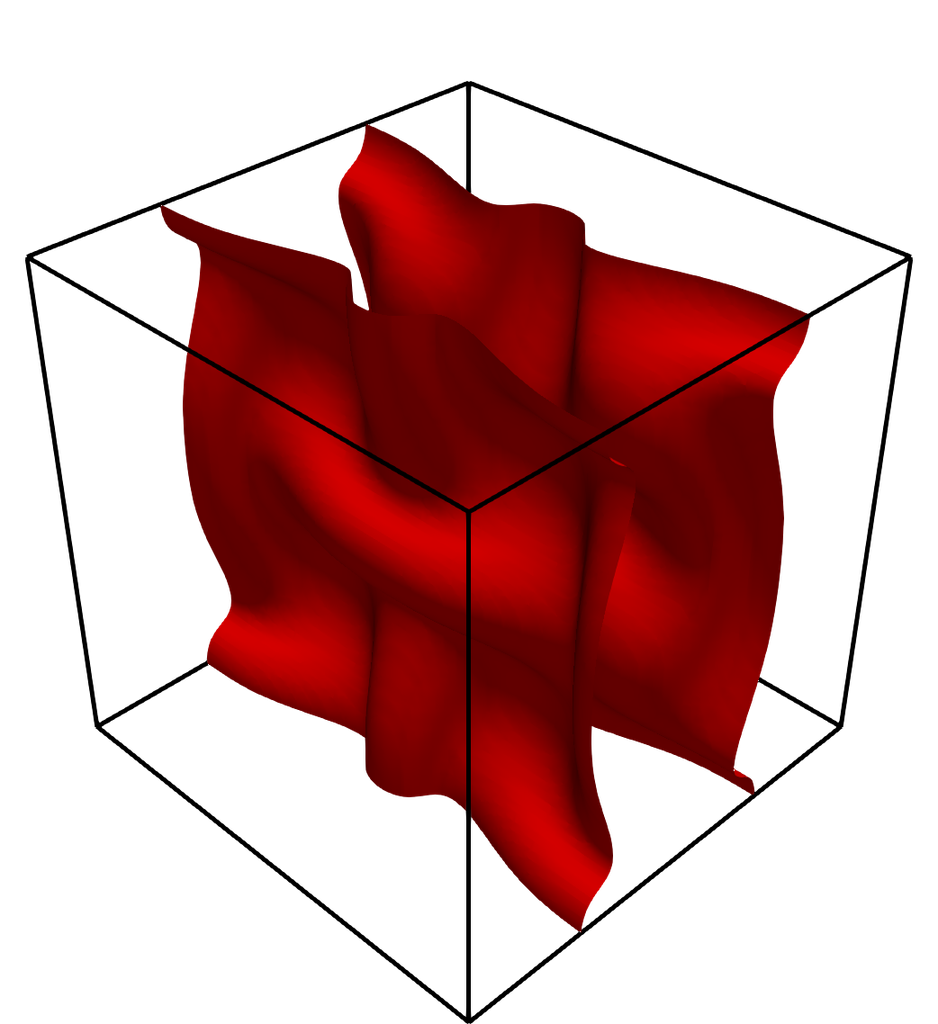}}
\subfigure[$t^{*}=4$.]{\includegraphics[width=0.32\textwidth]{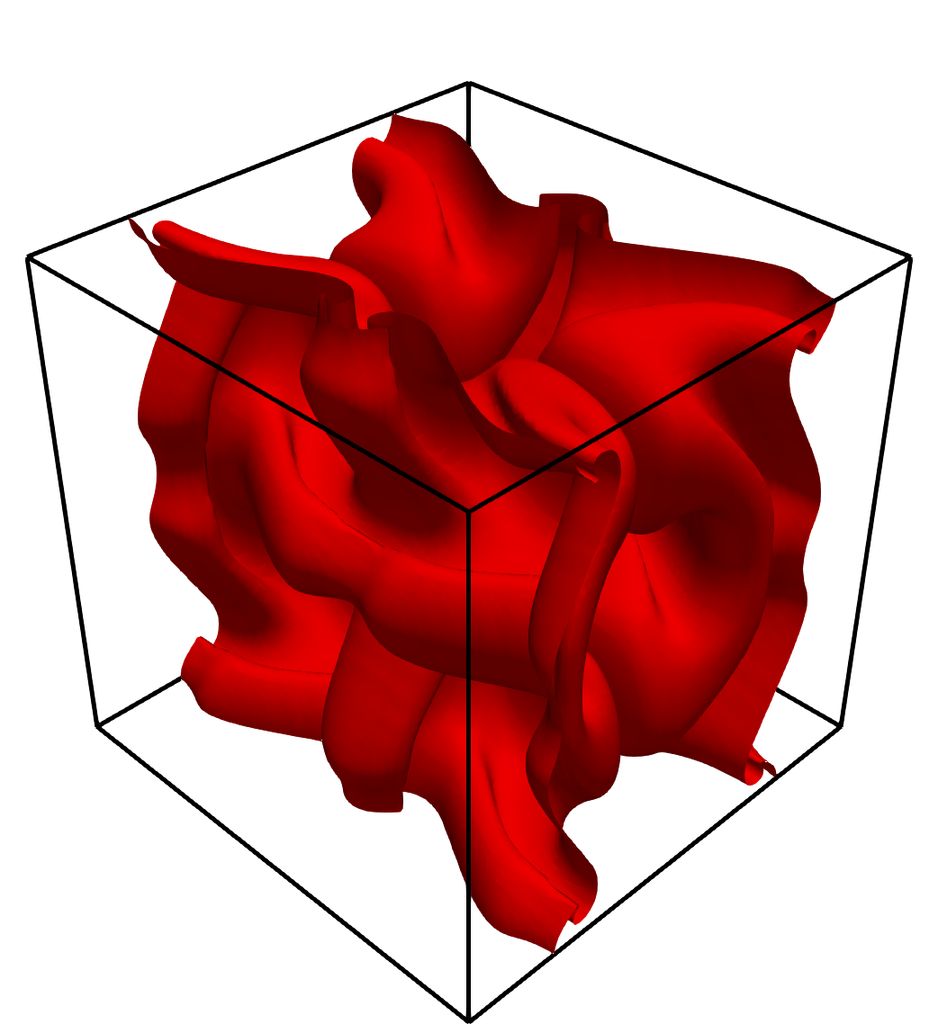}}
\subfigure[$t^{*}=6$.]{\includegraphics[width=0.32\textwidth]{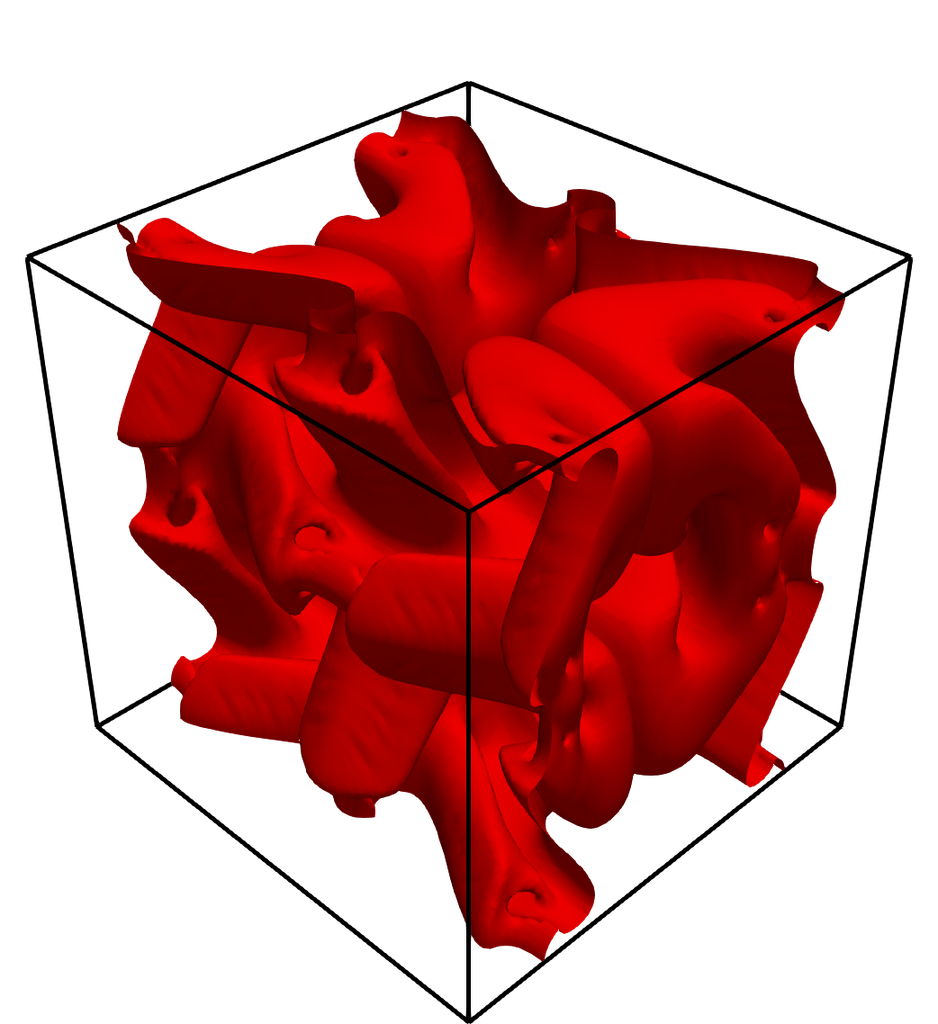}}
\subfigure[$t^{*}=8$.]{\includegraphics[width=0.32\textwidth]{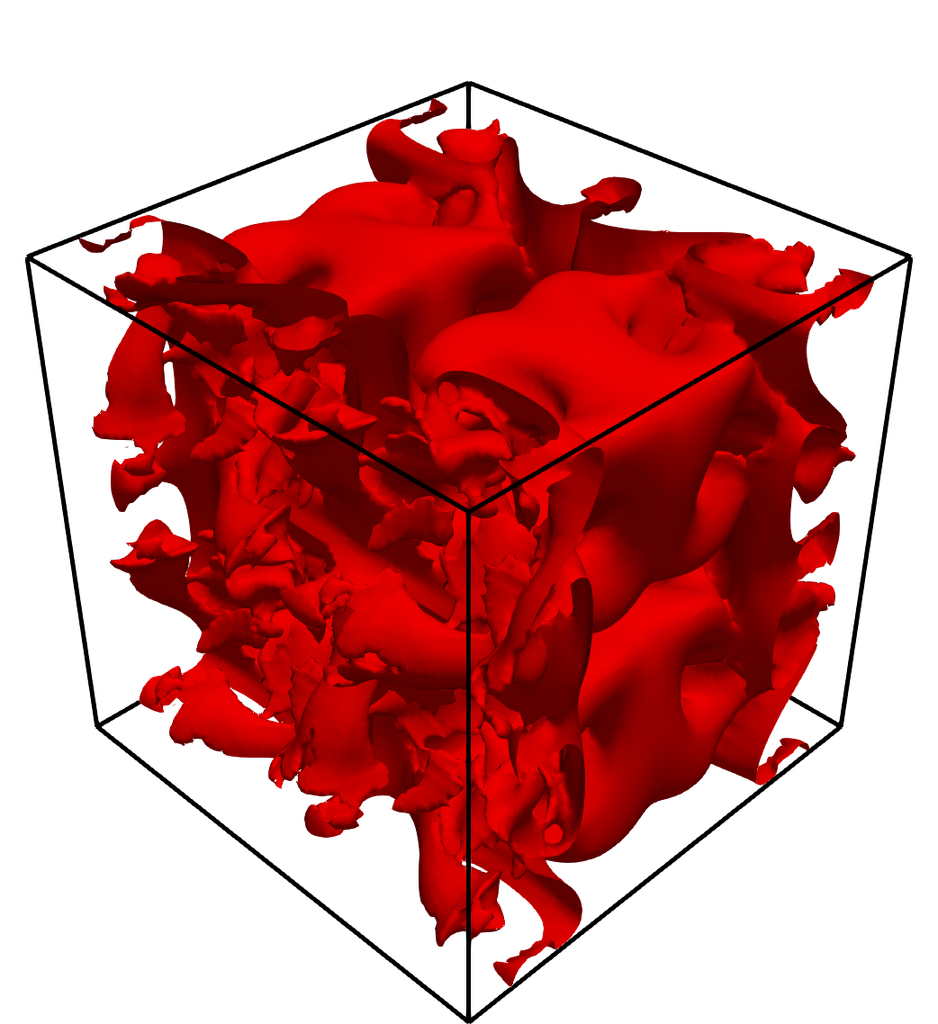}}
\subfigure[$t^{*}=10$.]{\includegraphics[width=0.32\textwidth]{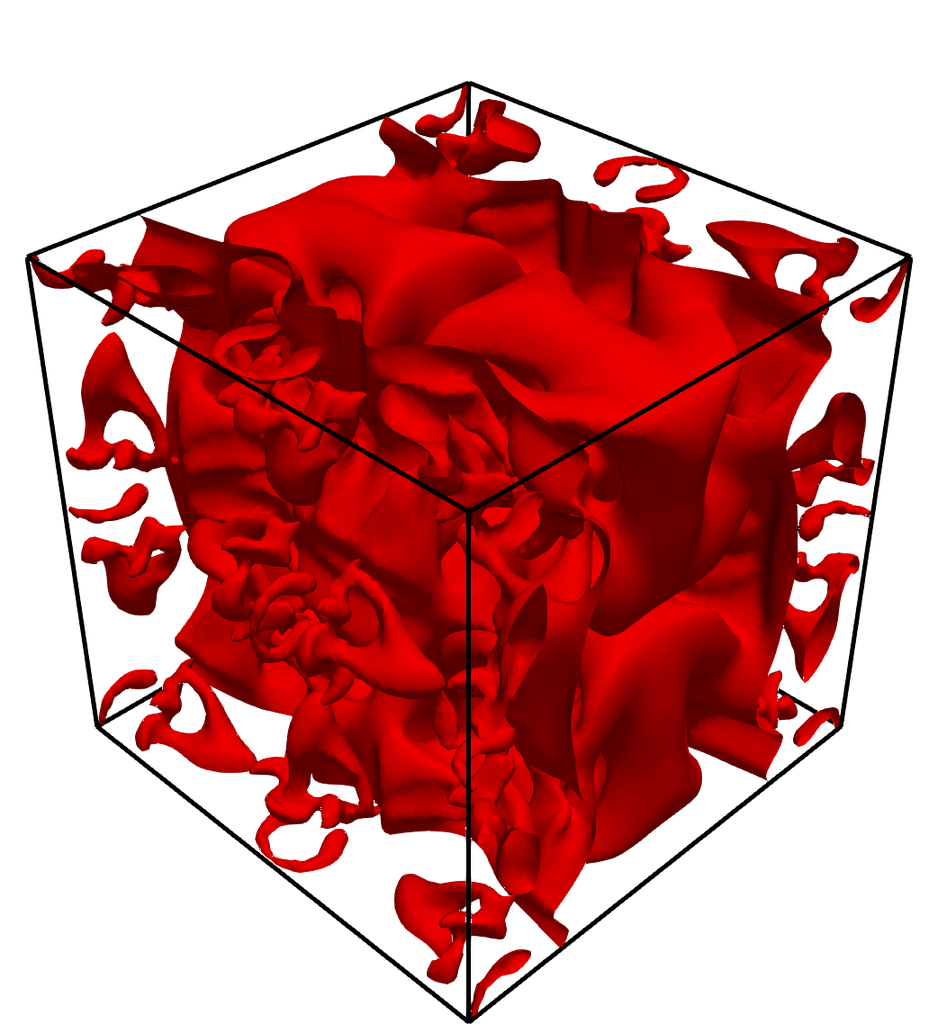}}
\subfigure[$t^{*}=12$.]{\includegraphics[width=0.32\textwidth]{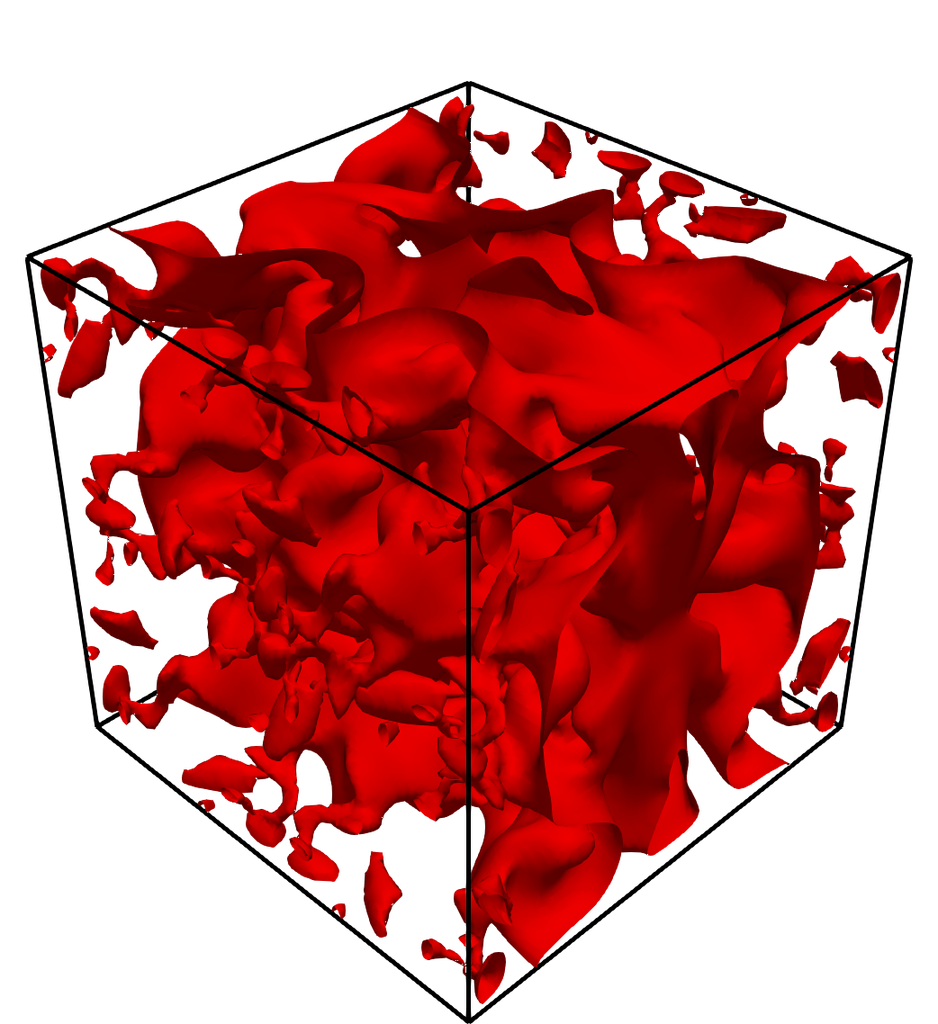}}
\subfigure[$t^{*}=14$.]{\includegraphics[width=0.32\textwidth]{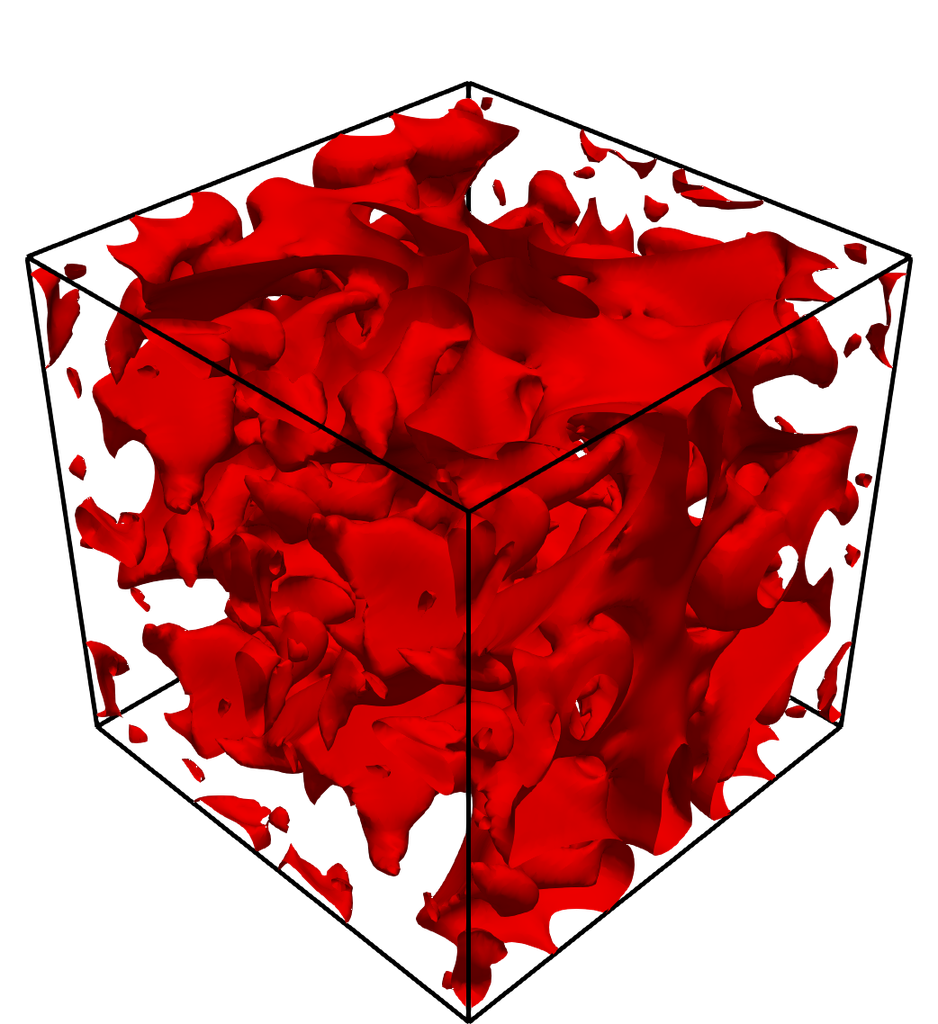}}
\subfigure[$t^{*}=20$.]{\includegraphics[width=0.32\textwidth]{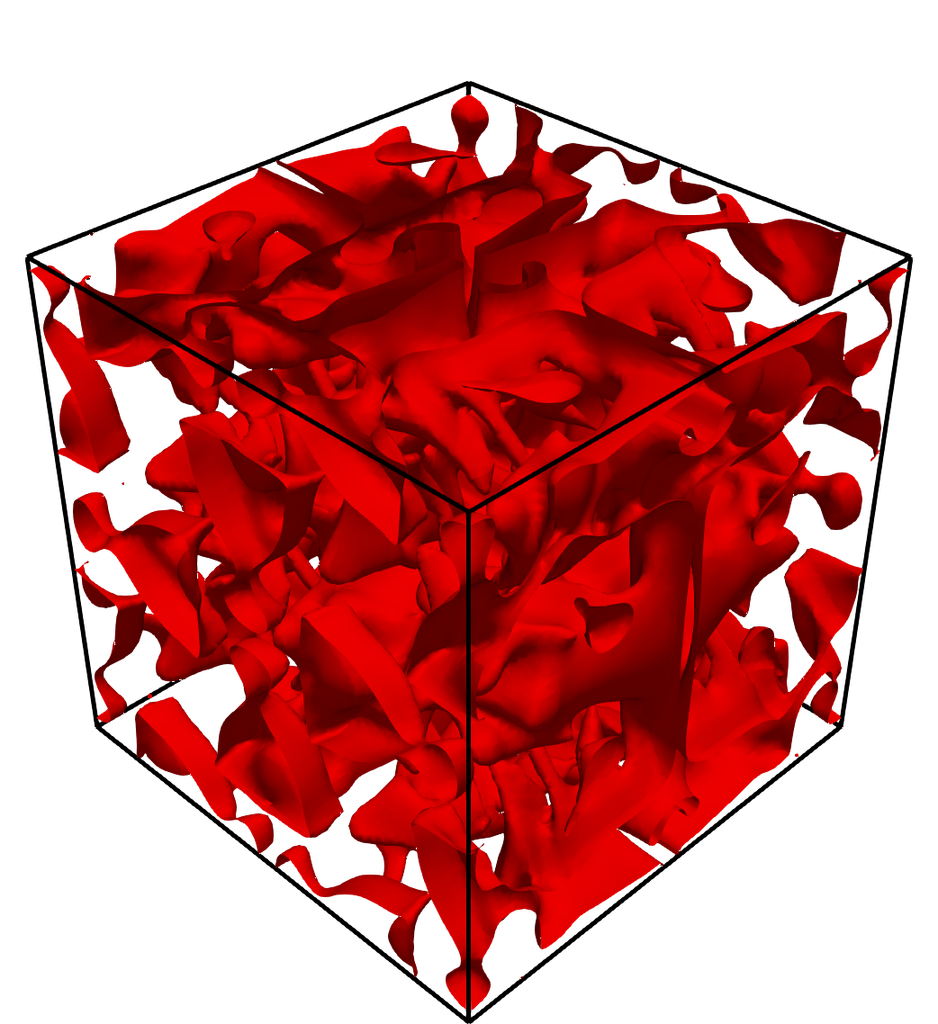}}
\caption{Interface snapshots at different time steps. Case $\rho_{1}/\rho_{2}=2$ with $4^{\mathrm{th}}$-order discretisation.}
\label{fig:TGV3}
\end{figure}

\begin{figure}[h!]
\centering
\includegraphics[width=.95\textwidth]{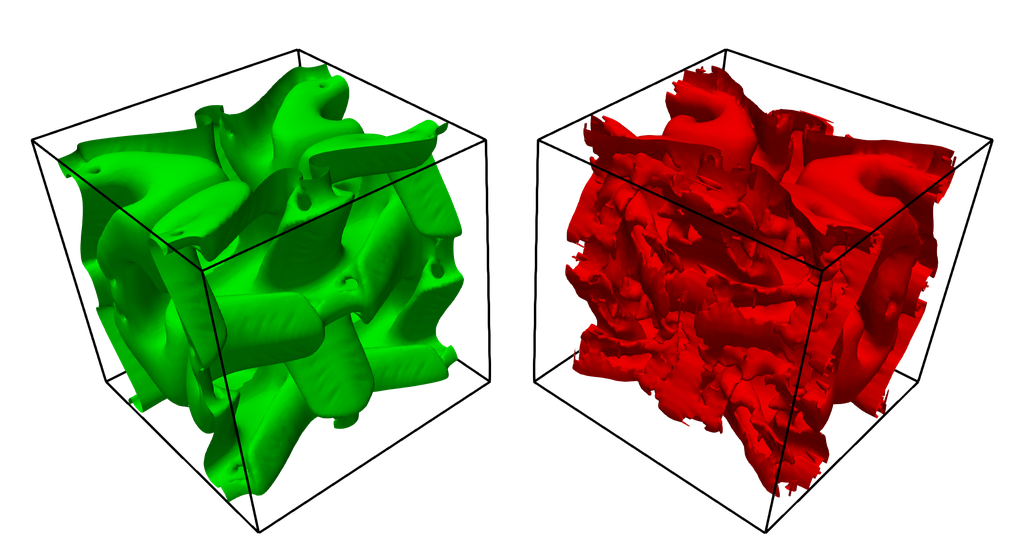}
\caption{Interface location at $t^{*}=5$ for the $4^{\mathrm{th}}$-order (left) and $3^{\mathrm{rd}}$-order (right) simulations for $\rho_{1}/\rho_{2}=2.0$.}
\label{fig:TGV4}
\end{figure}

\begin{figure}[h!]
\centering
\subfigure[$t^{*}=0$.]{\includegraphics[width=0.32\textwidth]{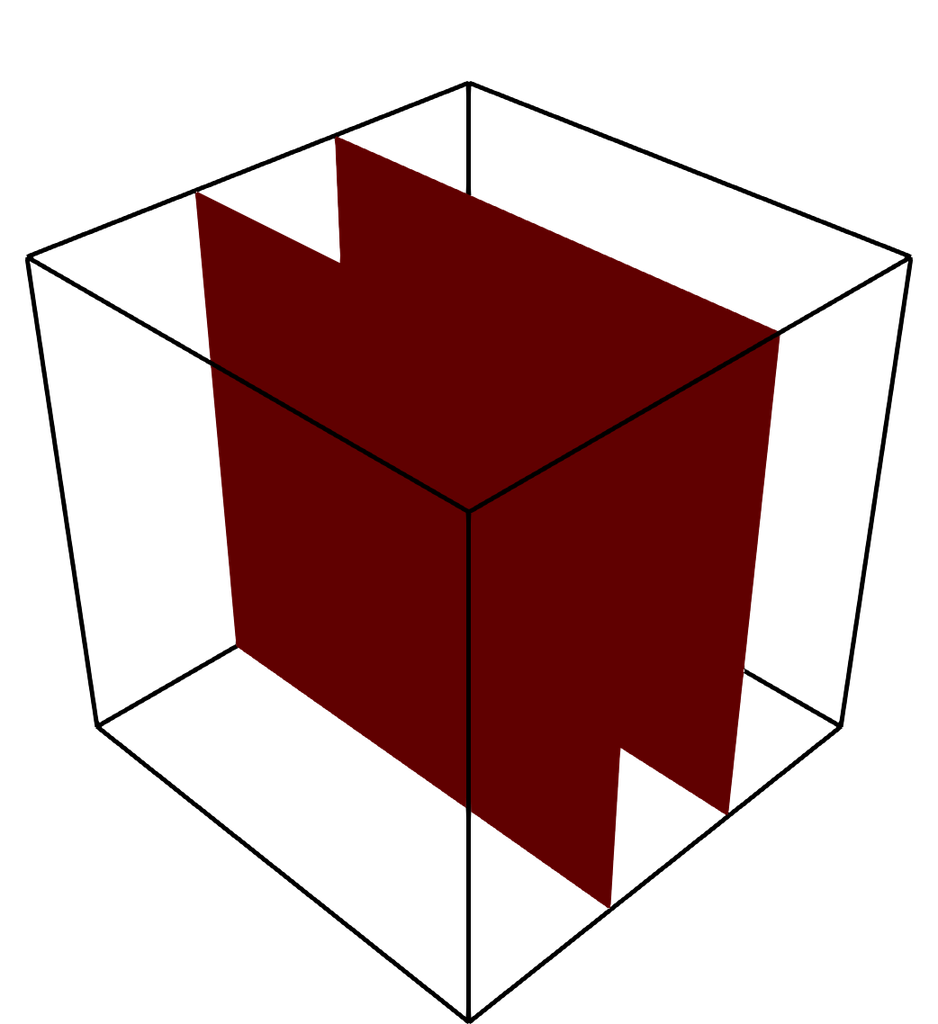}}
\subfigure[$t^{*}=2$.]{\includegraphics[width=0.32\textwidth]{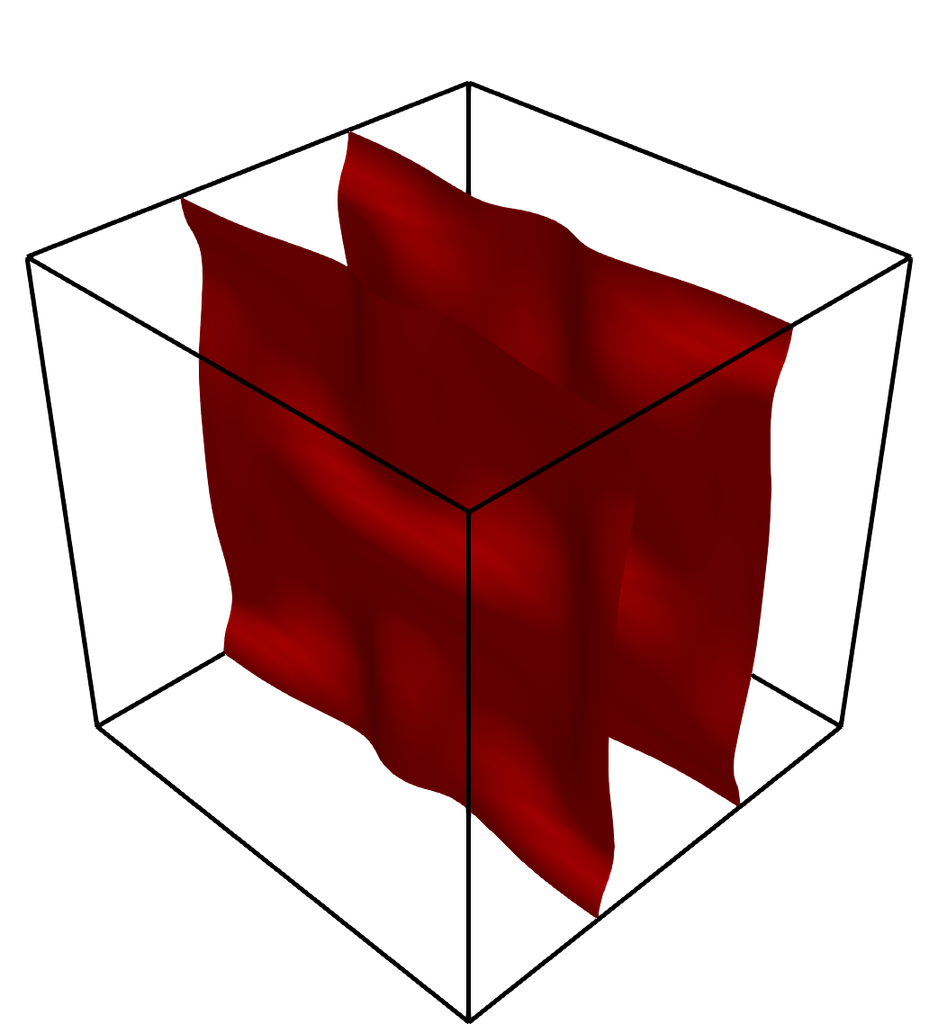}}
\subfigure[$t^{*}=4$.]{\includegraphics[width=0.32\textwidth]{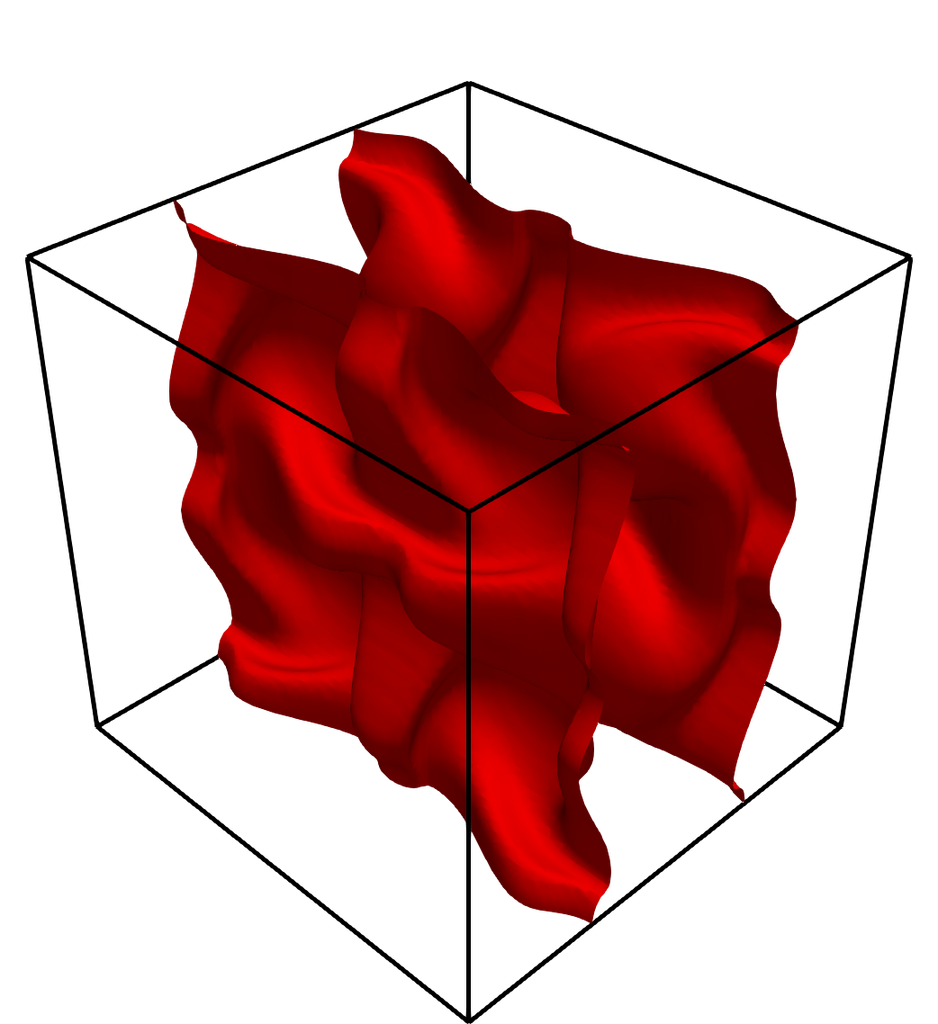}}
\subfigure[$t^{*}=6$.]{\includegraphics[width=0.32\textwidth]{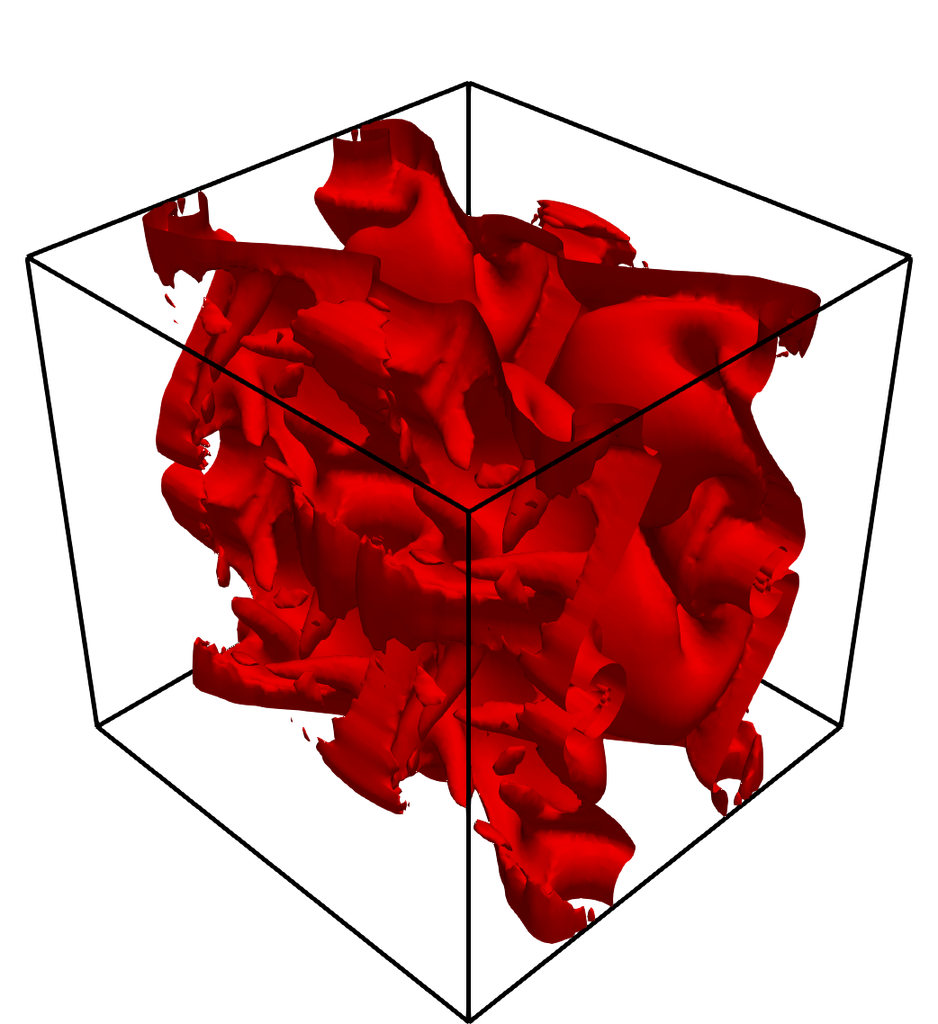}}
\subfigure[$t^{*}=8$.]{\includegraphics[width=0.32\textwidth]{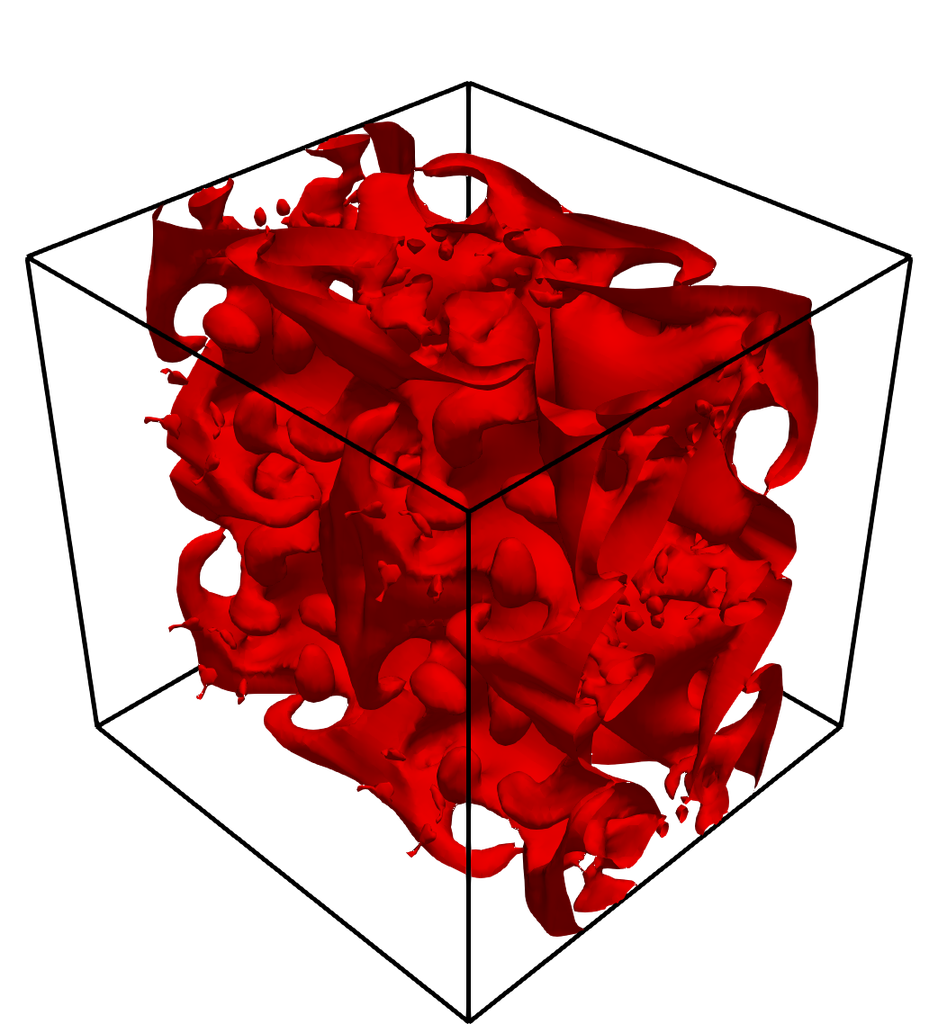}}
\subfigure[$t^{*}=10$.]{\includegraphics[width=0.32\textwidth]{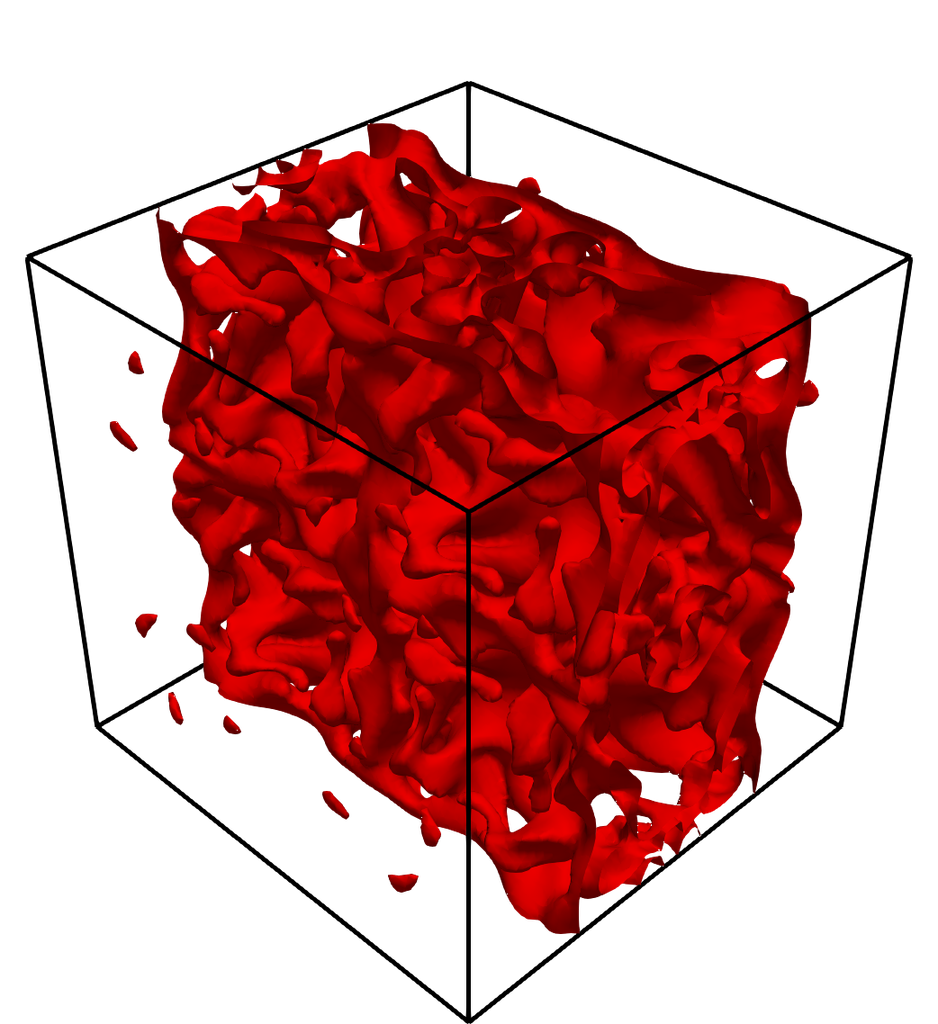}}
\subfigure[$t^{*}=12$.]{\includegraphics[width=0.32\textwidth]{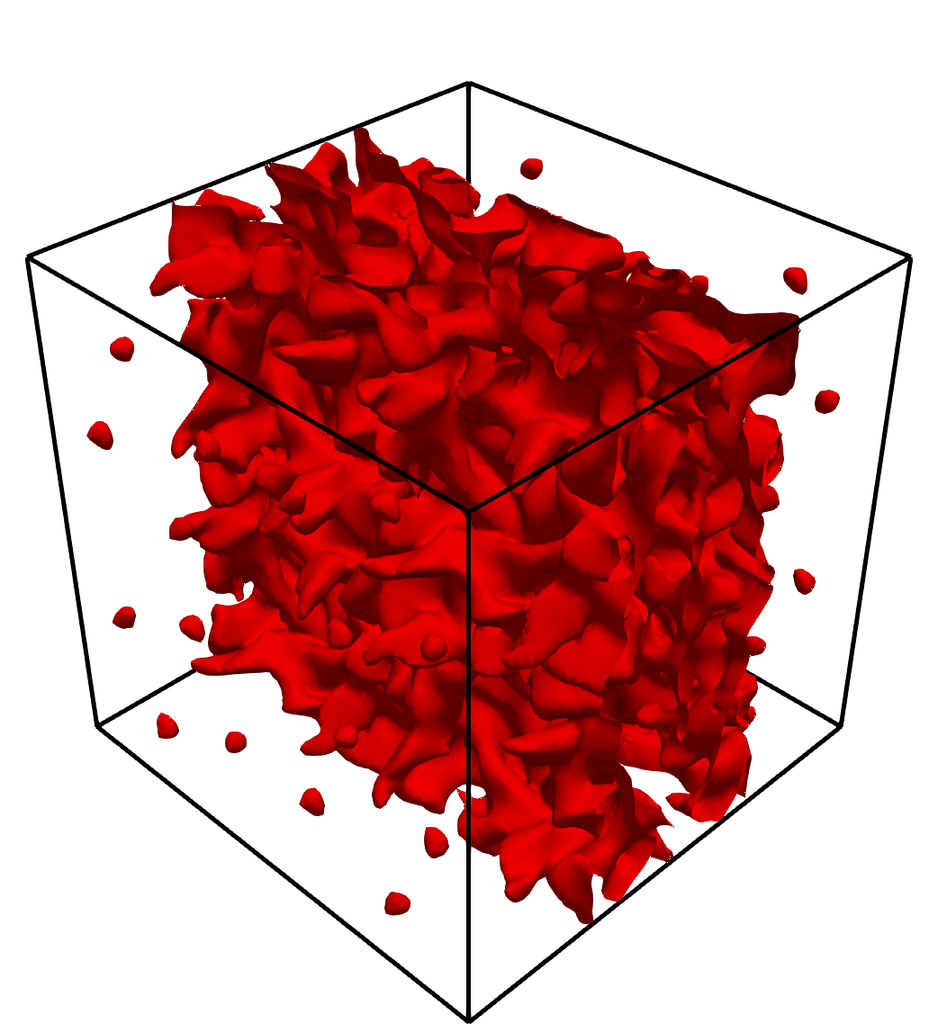}}
\subfigure[$t^{*}=14$.]{\includegraphics[width=0.32\textwidth]{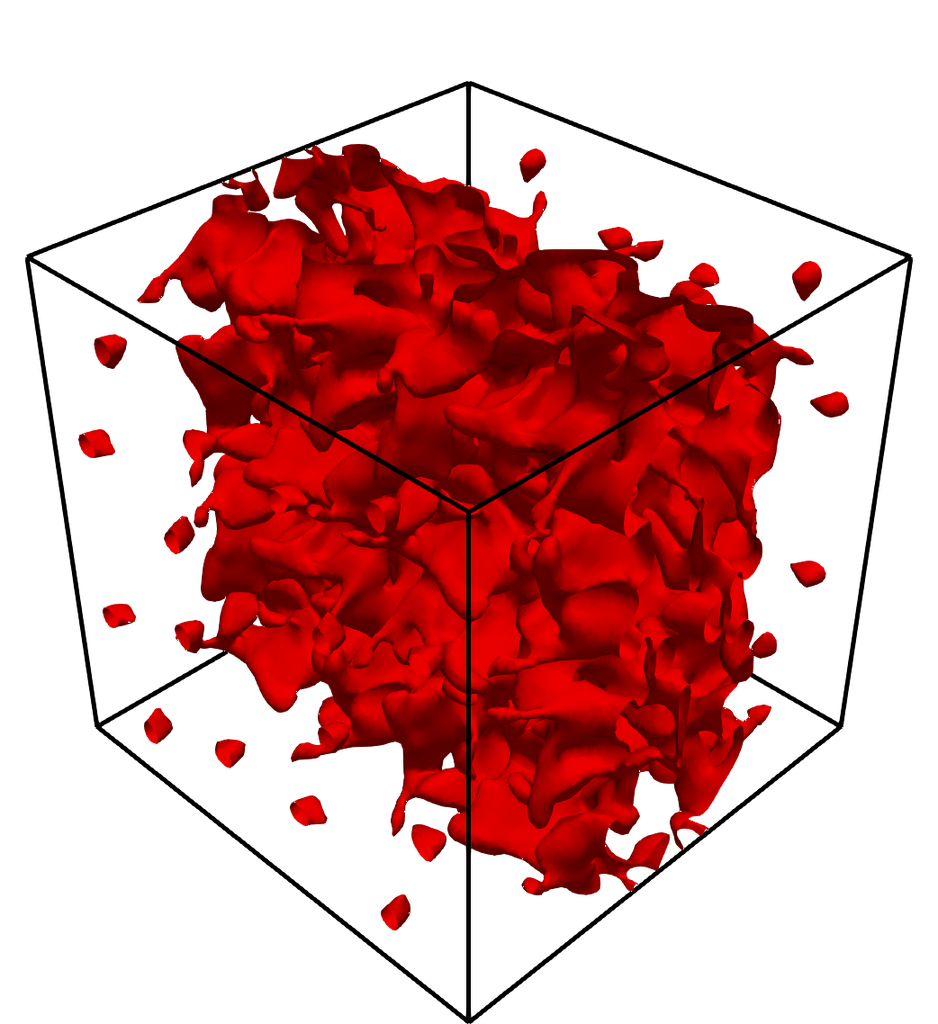}}
\subfigure[$t^{*}=20$.]{\includegraphics[width=0.32\textwidth]{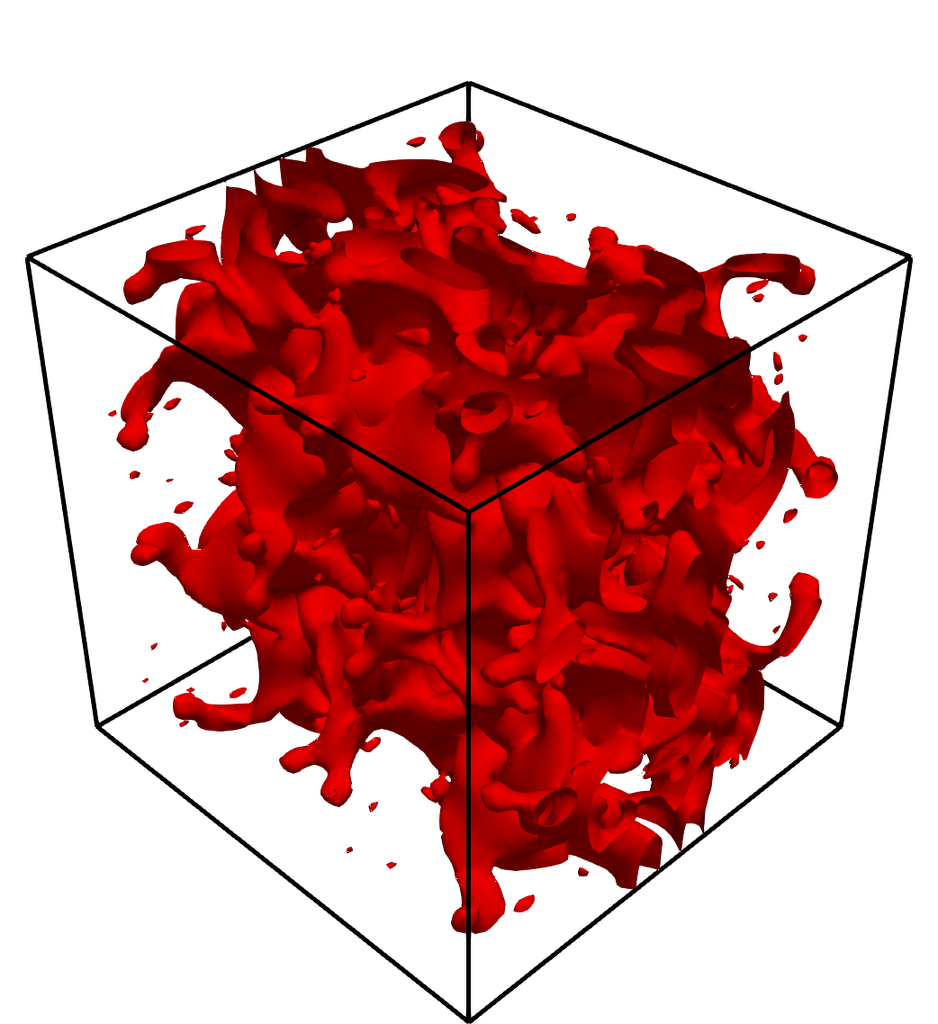}}
\caption{Interface snapshots at different time steps. Case $\rho_{1}/\rho_{2}=0.5$ with $4^{\mathrm{th}}$-order discretisation.}
\label{fig:TGV7}
\end{figure}

As visual evolution of the interface over time for this problem, we show in figure \ref{fig:TGV3} a series of snapshots of the $\phi=0.5$ iso-contour over time for the case $\rho_{1}/\rho_{2}=2$ and $4^{\mathrm{th}}$-order. 
In figure \ref{fig:TGV4} a comparison between different orders of approximation is shown. Here we can observe similar trends as those that were observed before: the lower order simulation tends to postpone break-up events, whereas smaller structure develops faster using a high-order discretisation.
In figure \ref{fig:TGV7} the same snapshots are shown for the case $\rho_{1}/\rho_{2}=0.5$. It is interesting to notice that the dynamics of the interface is completely different. In the case $\rho_{1}/\rho_{2}=2$ the inner layer of the first phase was characterised by relatively smooth curves whereas primary break-up was expanding from this layer towards the exterior. In the case $\rho_{1}/\rho_{2}=0.5$, instead, the trend is the opposite as more and more complex structures develops as the system evolves but they stay primarily close to the initial layer of the first phase.
%
\subsubsection{Two-phase Taylor-Green vortex: $\mathrm{Re}=5000$}
In this section we consider the same Taylor-Green vortex case for a higher Reynolds number (i.e. $\mathrm{Re}=5000$). With this case, in fact, we seek to assess the present formulation in dealing with under-resolved turbulent flows.
Consequently, for the same resolution of the previous test case ($96^{3}$ DoF), the problem is severely under-resolved and sub-grid scale (SGS) modelling is required. In particular, due to its previous success within the spectral difference scheme, we consider the Spectral Element Dynamic Model (SEDM) developed by Chapelier \& Lodato~\cite{chapelier2016spectral} as SGS model.

\begin{figure}[h!]
\centering
\subfigure[Viscous dissipation.]{\includegraphics[width=0.49\textwidth]{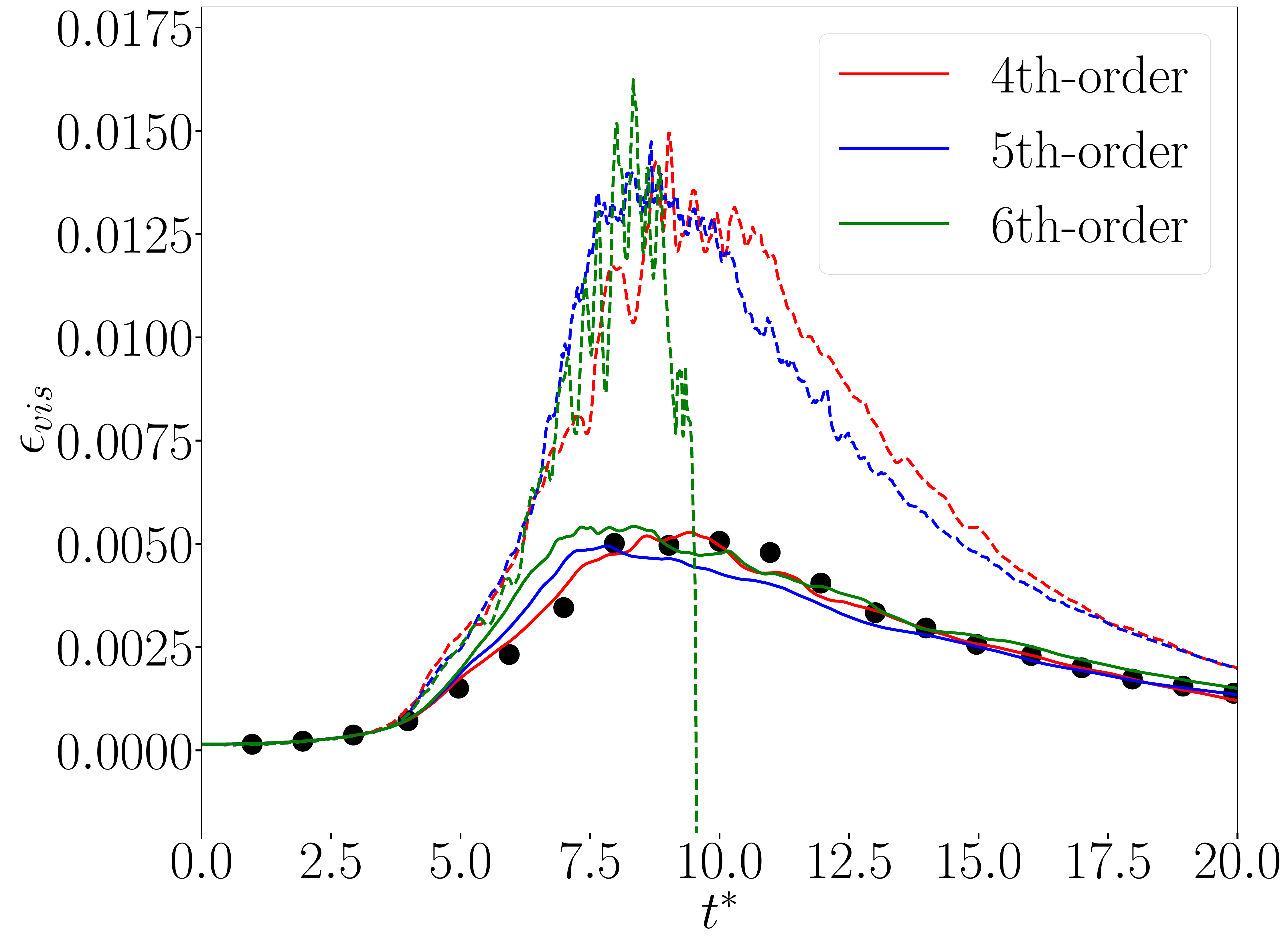}}
\subfigure[Total SGS dissipation.]{\includegraphics[width=0.49\textwidth]{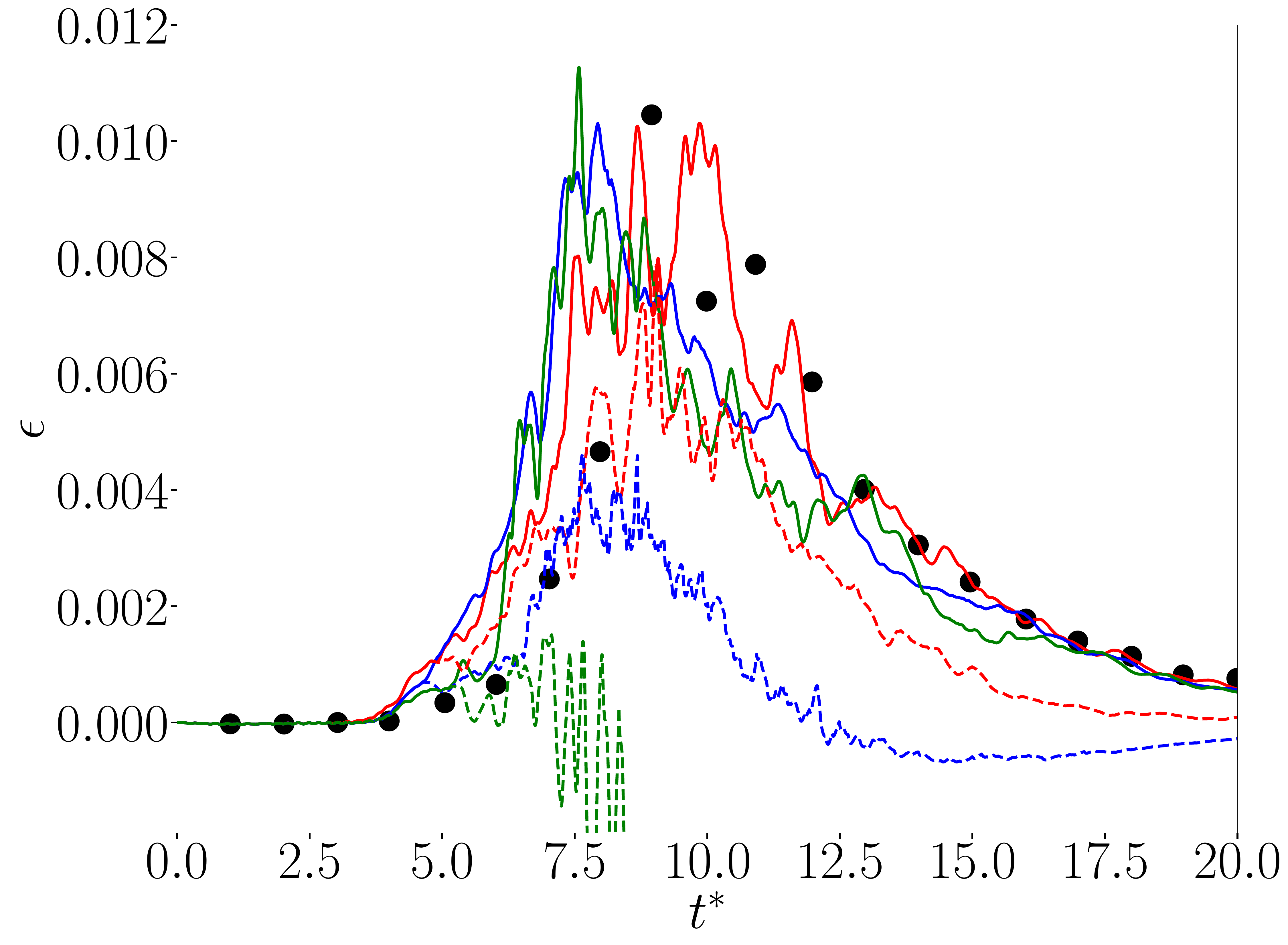}}
\caption{Averaged viscous dissipation (left) and total SGS dissipation (right) over time. The black dots represent the DNS data for the single phase case~\cite{chapelier2016spectral}. Solid lines, with SEDM; dashed lines, without SEDM.}
\label{fig:TGV_SEDM}
\end{figure}

As in the low-Reynolds TGV case, we first consider the single-phase formulation. We compare in figure \ref{fig:TGV_SEDM} the averaged viscous and total SGS dissipation for different orders of approximation with and without the SEDM. The total SGS dissipation is defined as the residual of the averaged kinetic energy balance without considering SGS terms:
\begin{equation}
\underbrace{-\frac{d}{dt}\bigg(\frac{1}{2}\int \rho u_{i} u_{i} dV \bigg)}_{\text{Total dissipation}}=\underbrace{\int \bigg(2\mu S_{ij} S_{ij}-\frac{2}{3}\mu \bigg(\pd{u_{j}}{x_{j}}\bigg)^{2}\bigg)dV}_{\text{Viscous}}-\underbrace{\int p\bigg(\pd{u_{j}}{x_{j}}\bigg)dV}_{\text{Dilatation}}+\varepsilon
\end{equation}
where $\varepsilon$ includes both explicit (SEDM) and implicit (numerical dissipation) SGS terms. 

We can observe that explicit SGS modelling greatly improves the results. It is worthwhile underlying that the SEDM was developed to be coupled with a Roe Riemann solver. In this work, we are employing a Lax–Friedrichs flux. Furthermore, the diffusive numerical fluxes are different with respect to the single-phase counterparts due to the interface regularisation terms. In fact, even if the treatment of the interface is potentially redundant for this formulation, for consistency, we still solve the full system of equations, and, consequently, the penalty terms in the diffusive fluxes depend also on the parameters $\Gamma$ and $\epsilon$. 

Based on these considerations, and knowing that the Lax–Friedrichs flux should be low-dissipative in the low-Mach regime (see \cite{moura2017eddy, mengaldo2018spatial,manzanero2018role,tonicello2021comparative}), we increased the SEDM constant to $0.45$ instead of the nominal value of $0.23$.

\begin{figure}[h!]
\centering
\subfigure[SEDM.]{\includegraphics[width=0.49\textwidth]{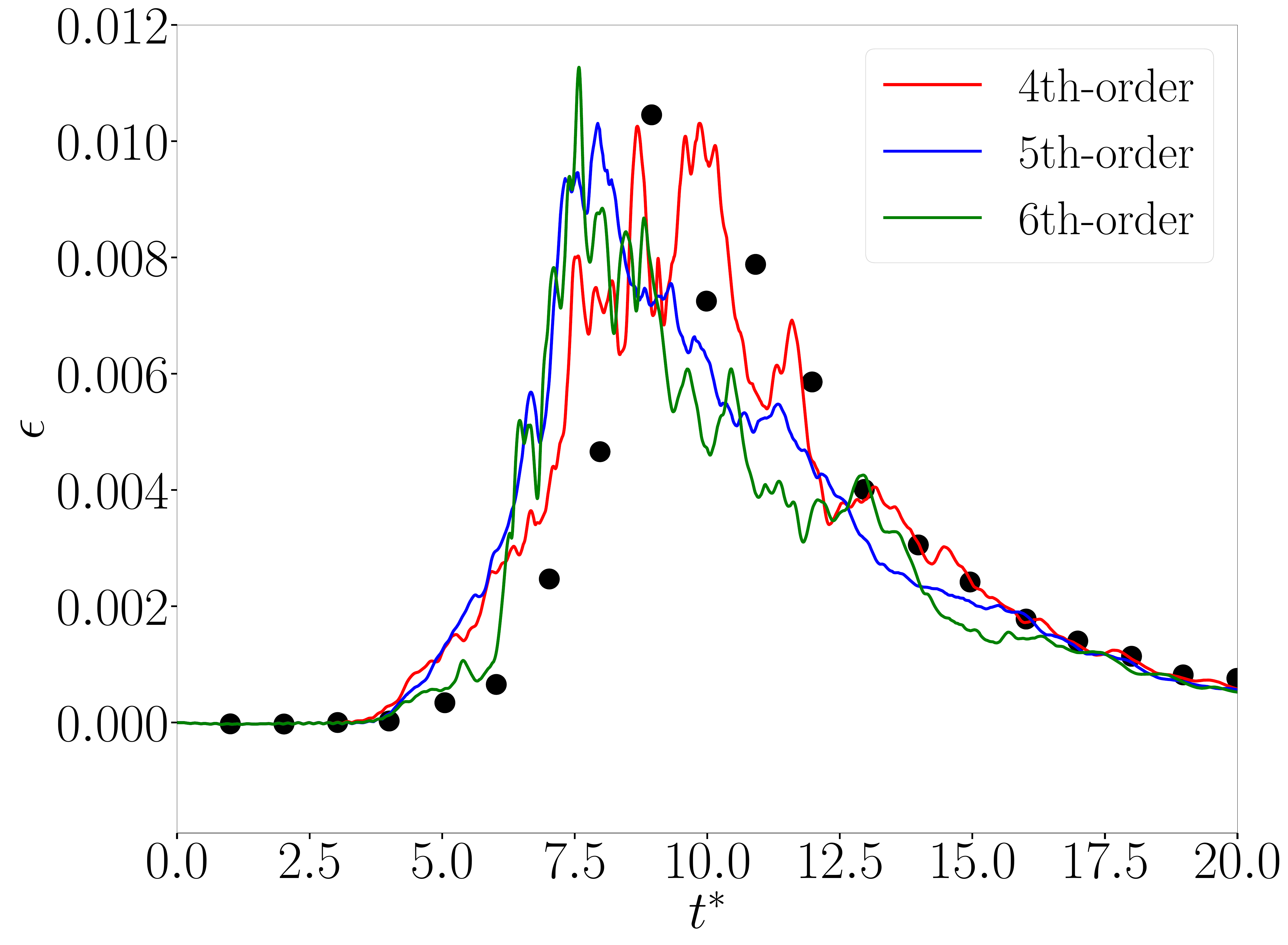}}
\subfigure[ILES.]{\includegraphics[width=0.49\textwidth]{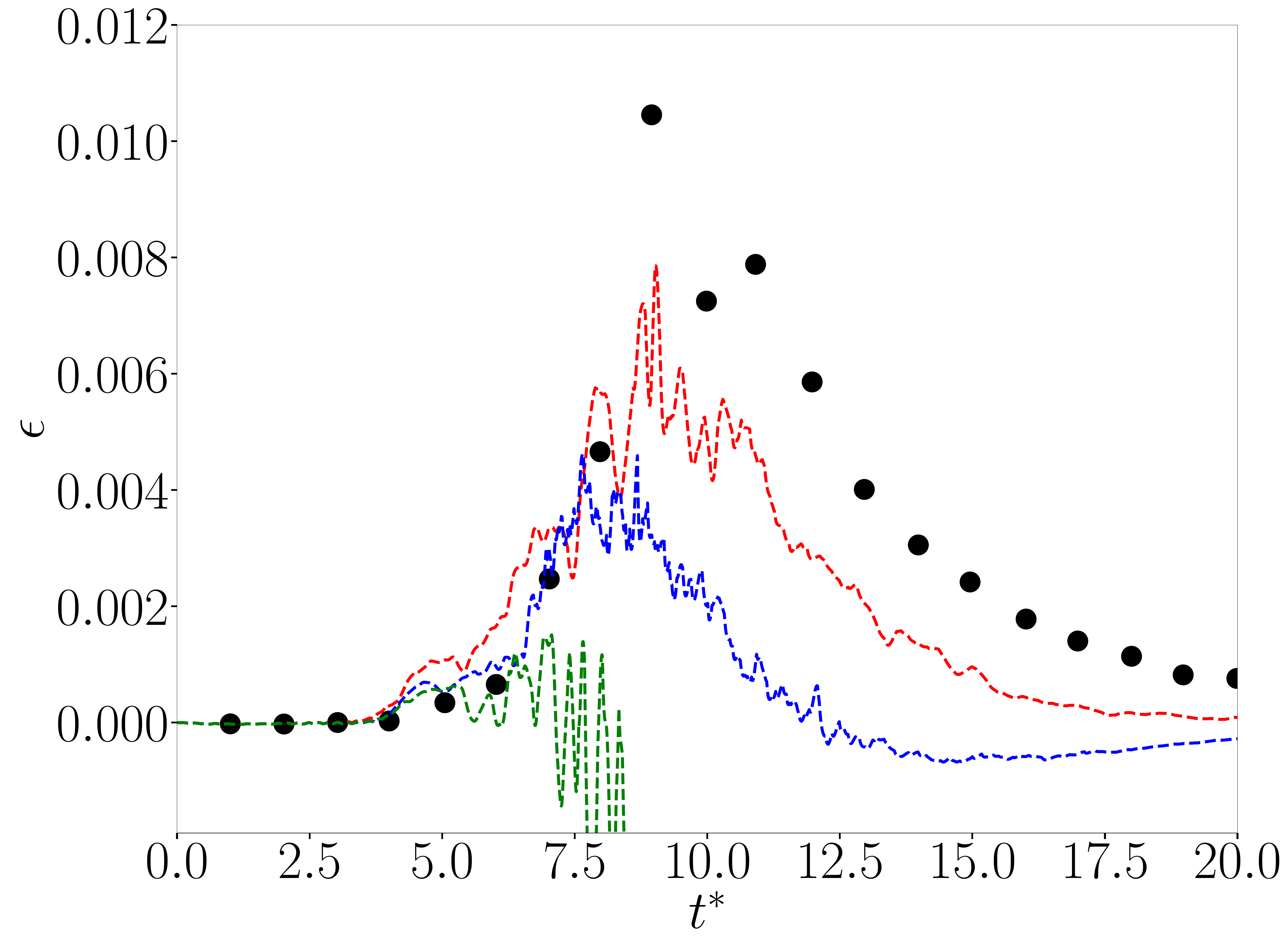}}
\caption{Total SGS dissipation as a function of time with (left) and without (right) SEDM. The black dots represent the DNS data for the single phase case~\cite{chapelier2016spectral}.}
\label{fig:TGV_SEDM2}
\end{figure}

From figure \ref{fig:TGV_SEDM} it is evident that the implicit LES are characterised by very different numerical results: with increasing orders of approximation, the viscous dissipation continues to excessively grow (as expected by the inverse cascade of vorticity). For the $6^{\mathrm{th}}$-order simulation, vorticity grows at such a rapid rate that the simulation rapidly becomes unstable. Instead, the addition of the SEDM counterbalances such behaviour, providing the same trend of overall dissipation, successfully matching the expected filtered DNS data. Notice that we did not consider $3^{\mathrm{rd}}$-order simulations in this case since the SEDM is based on a modal decay sensor which is not suitable for low approximation orders.
The flexibility and adaptivity of the SEDM can also be noticed in figure \ref{fig:TGV_SEDM2} where the total SGS dissipation is shown separately for explicit and implicit large eddy simulations. The total amount of SGS dissipation in the implicit LES (which is entirely due to numerical dissipation) varies greatly depending on the order of approximation: larger orders lead, in fact, to lower numerical dissipation. Instead, by adding explicit SGS modelling, we can obtain similar levels of total SGS dissipation for all the orders of approximation. 

\begin{figure}[h!]
\centering
\subfigure[Kinetic energy.]{\includegraphics[width=0.49\textwidth]{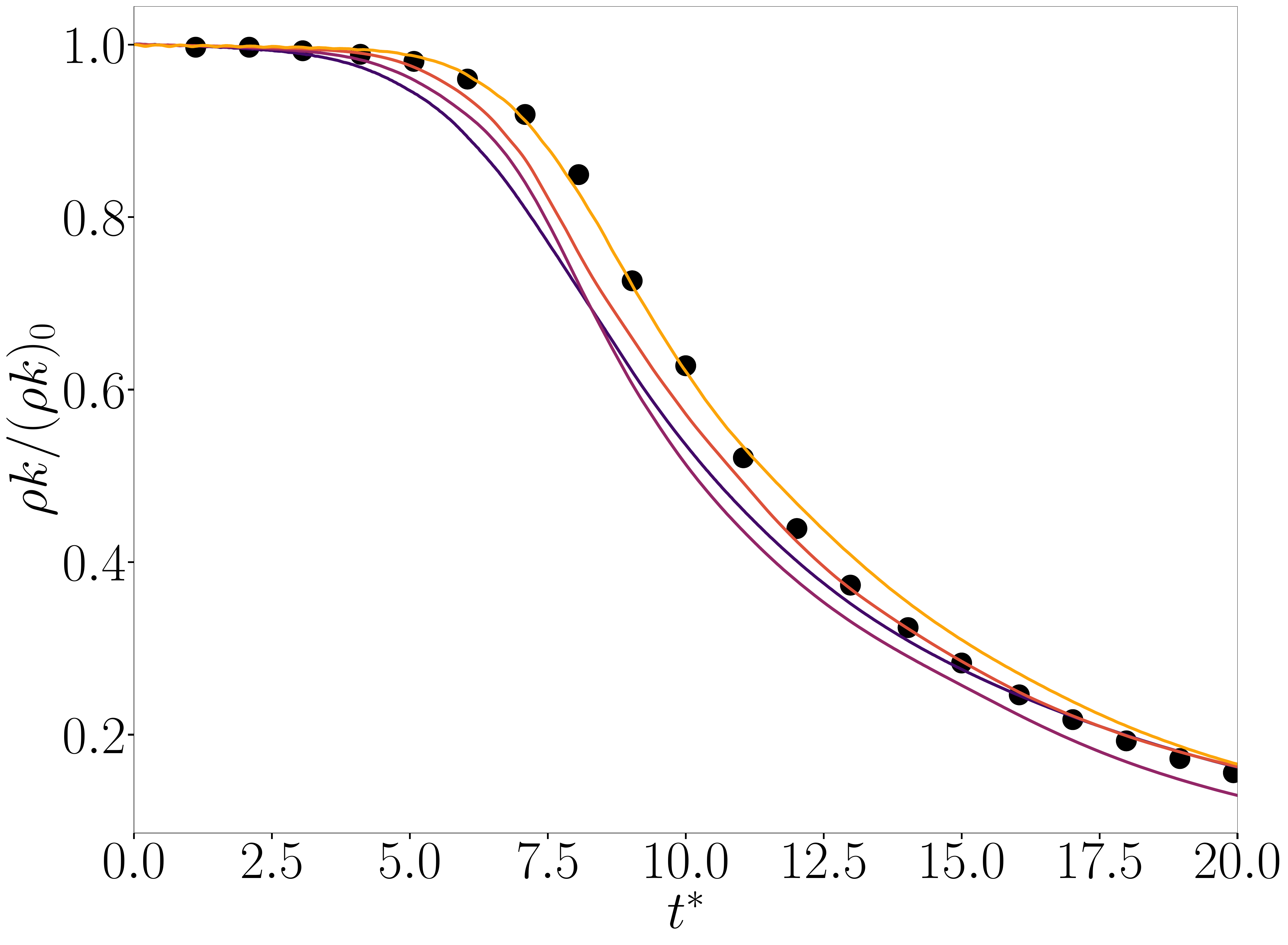}}
\subfigure[Viscous dissipation.]{\includegraphics[width=0.49\textwidth]{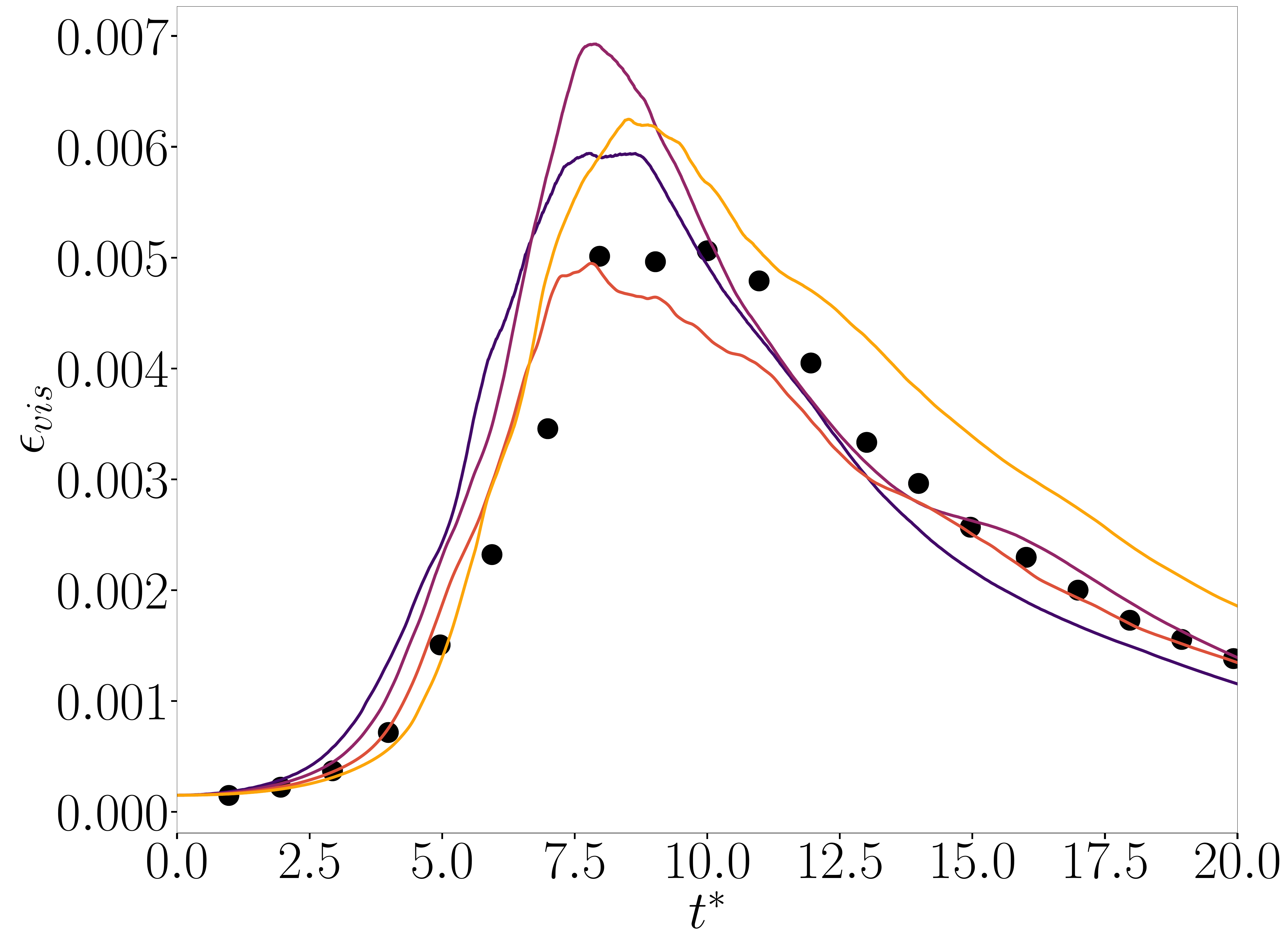}}
\caption{Averaged kinetic energy (left) and viscous dissipation(right) over time for the $5^{\mathrm{th}}$-order simulations. Colour gradient from light orange to dark purple indicates smaller values of $\rho_{1}$. Namely, $\rho_{2}/\rho_{1}=0.5,1,2,4$. The black dots represent the DNS data for the single phase case~\cite{chapelier2016spectral}. Values of kinetic energy are normalised by their initial value.}
\label{fig:TGV_ke2}
\end{figure}

Similarly to the previous case, we proceed in modifying the density ratio between the two fluids in order to assess the robustness of the proposed approach also in under-resolved turbulent flows. The same ratios of the previous case are considered (namely, $\rho_{2}/\rho_{1}=0.5,1,2,4$). The averaged kinetic energy and viscous dissipation are shown in figure \ref{fig:TGV_ke2}.

\begin{figure}[h!]
\centering
\includegraphics[width=.95\textwidth]{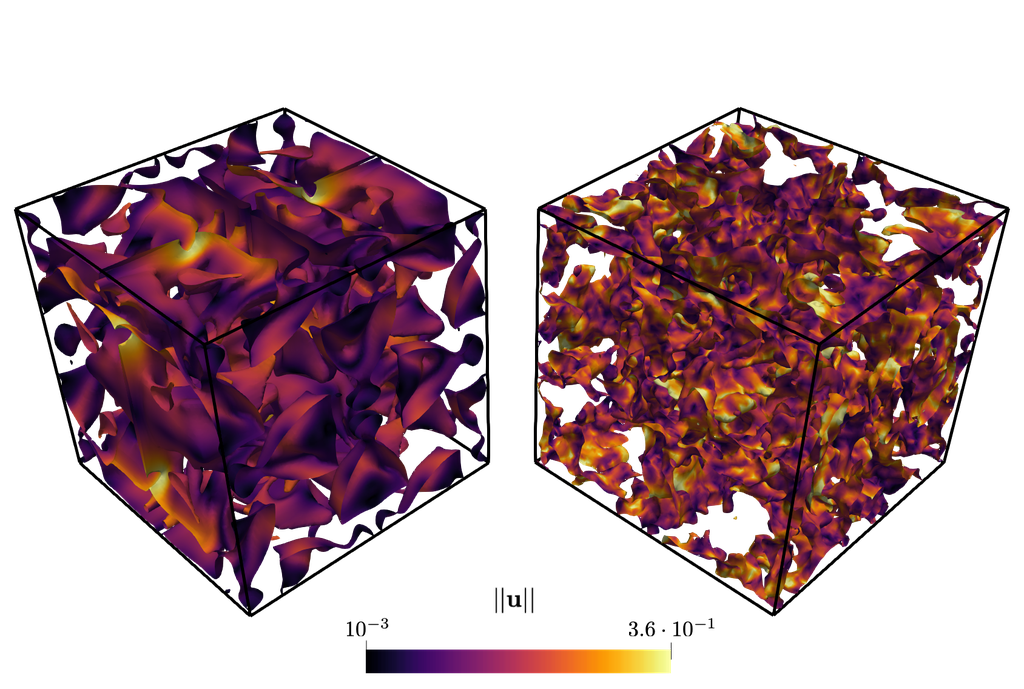}
\caption{Phase field isocontour at $t^{*}=20$ for the $\mathrm{Re}=500$ (left) and $\mathrm{Re}=5000$ (right) cases for $\rho_{1}/\rho_{2}=0.5$ (coloured by velocity magnitude).}
\label{fig:TGV8}
\end{figure}

First, it can be noticed that with respect to the previous case the deviations between different density ratios are smaller. This is related to the concept that even smaller variations of the density ratio can significantly change the dynamics of low Reynolds flows, whereas smaller differences are expected when the flow is fully turbulent. In other words, the Taylor-Green vortex is considerably different at $\mathrm{Re}=500$ and $\mathrm{Re}=250$ whereas the differences between $\mathrm{Re}=5000$ and $\mathrm{Re}=2500$ are significantly smaller. 
Also in terms of interface dynamics, the low and high Reynolds number cases can differ significantly. In figure \ref{fig:TGV8} the instantaneous iso-contour of the phase field is shown at the end of the simulation for the two Reynolds numbers. We can observe that much smaller structures arise for the $\mathrm{Re}=5000$ case: after fully transitioning to turbulence smaller scales are generated, leading to a widespread break-up of the initially layered phase field.
\section{Conclusions}
In the present work the spectral difference scheme was employed for the discretisation of the five equation model equipped with the additional Allen-Cahn regularisation terms for interface capturing purposes~\cite{chiu2011conservative}. Within this framework of the spectral difference scheme, in order to mitigate pressure oscillations in proximity of material interfaces in two-phase flows, a change of variables in the extrapolation to the flux points was used. In particular, instead of interpolating the conservative variables, as it is custom in the spectral difference scheme, the primitive variables (namely pressure and velocity) were used to alleviate the non-linearity of the stiffened-gas equation of state. 

A series of numerical tests of increasing complexity were considered in order to assess the robustness of the solver, including both kinematic and two-phase test cases. In particular, the Rider-Kothe vortex was studied in order to quantify convergence properties of the scheme and mass conservation errors. Regarding this last point, the use of a high-order discretisation showed significant improvements in  mass conservation. In the two-phase flow problems, the capability of the change of variables to fulfil the interface equilibrium condition was tested, showing increased stability. Finally, a series of more complex problems were considered. In particular, whereas previous cases were characterised by either imposed velocity field, or very simple ones (constant advection of a droplet), more complex flow dynamics was considered in the Rayleigh-Taylor instability, a shock-droplet interaction and a three-dimensional, two-phase version of the Taylor-Green Vortex problem. In all of these cases, the use of high orders of approximation showed improvements in the representation of the vortical dynamics of the flow for well-resolved flows. In particular, it was observed that the excessive amount of numerical dissipation introduced by low-order discretisations provided a less accurate description of the velocity field, and, consequently, an implicit pollution of the interface dynamics. The interface tends, in fact, to be characterised by a delayed break-up where larger droplets are promoted over the development of thin, small-scale structures.

Finally, in order to assess the robustness of the proposed approach in a relatively challenging scenario, an under-resolved TGV test case was studied where sub-grid scale modelling was handled using the Spectral Element Dynamic Model~\cite{chapelier2016spectral}.  
Overall, the proposed approach shows a good robustness in dealing with a large variety of classical test cases in the two-phase simulation community, advocating the advantages in using high-order discretisations. It is worthwhile mentioning that the essential modifications needed to adapt the single-phase spectral difference scheme to a two-phase framework are considerably localised, providing a simple and flexible approach for the simulation of compressible two-phase flows. Future research will focus on exploiting such flexibility by considering more sophisticated Riemann solvers on the numerical side and more complex fluid dynamics processes in terms of physics. Namely, surface tension, phase change, cavitation and more sophisticated SGS modelling represent the natural evolution of the present work.

\section*{Acknowledgments}
The use of the SD solver originally developed by Antony Jameson's group at Stanford University is gratefully acknowledged.
This study was funded by the European Union - NextGenerationEU, in the framework of the iNEST - Interconnected Nord-Est Innovation Ecosystem (iNEST ECS00000043 – CUP G93C22000610007). MI acknowledges financial support from NSF (Award 1909379) and Daikin. The views and
opinions expressed are solely those of the authors and do not necessarily reflect those of the European Union, nor can the European Union be held responsible for them.

\end{document}